\documentclass[prc,showpacs,amsmath,amssymb,superscriptaddress,floatfix,nofootinbib,10pt]{revtex4}

\usepackage{inputenc}
\usepackage{amstext,amsfonts}
\usepackage{mathrsfs}
\usepackage{bbm,bm}
\usepackage{slashed}
\usepackage{setspace}
\usepackage{graphicx}
\usepackage{caption}
\usepackage{subfig}
\usepackage{booktabs}
\usepackage{float}
\usepackage{array}
\usepackage{hyperref}

\DeclareUnicodeCharacter{2212}{-}

\usepackage{changes}
\newcommand{\trace}[1]{\ensuremath{\langle #1\rangle}}

\begin{document}
\title{Nonperturbative two-pion exchange contributions to the nucleon-nucleon interaction in covariant baryon chiral perturbation theory}

\author{Chun-Xuan Wang}
\affiliation{School of Physics, Beihang University, Beijing 102206, China}

\author{Jun-Xu Lu}
\affiliation{School of Space and Environment, Beihang University, Beijing 102206, China}
\affiliation{School of Physics, Beihang University, Beijing 102206, China}

\author{Yang Xiao}
\affiliation{School of Physics, Beihang University, Beijing 102206, China}
\affiliation{Université  Paris-Saclay, CNRS/IN2P3, IJCLab, 91405 Orsay, France}

\author{Li-Sheng Geng}
\email[E-mail: ]{lisheng.geng@buaa.edu.cn}
\affiliation{School of Physics, Beihang University, Beijing 102206, China}
\affiliation{Beijing Key Laboratory of Advanced Nuclear Materials and Physics, Beihang University, Beijing, 102206, China}
\affiliation{School of Physics and Microelectronics, Zhengzhou University, Zhengzhou, Henan 450001, China}

\begin{abstract}
We calculate the nonperturbative two-pion exchange (TPE) contributions to the $NN$ interaction in covariant baryon chiral perturbation theory. We study how the nonperturbative resummation affects the $NN$ phase shifts for partial waves with $J \geq 3$  and  $L \leq 6$. No significant differences are observed between the nonperturbative phase shifts and perturbative ones for most partial waves except for $^3D_3$, for which the nonperturbative resummation greatly improves  the description of the phase shifts. However, a significant cutoff dependence is found for this partial wave and a reasonable description of the  phase shifts can only be obtained with a particular  cutoff. Furthermore, we compare the so-obtained nonperturbative phase shifts with those obtained in the heavy baryon chiral perturbation theory. We show that the contributions from relativistic nonperturbative TPE are more moderate than those from the nonrelativistic TPE obtained in the dimensional regularization scheme. A proper convergence pattern is observed for most of the partial waves studied except for $^3F_3$, $^3F_4$, and $^3H_6$, for which the subleading TPE  contributions are a bit strong. We find that for $H$ and $I$ partial waves, the OPE alone can already describe the phase shifts reasonably well. 

\end{abstract}

\pacs{21.30.−x}

\maketitle

\section{Introduction}\label{sec:introduction}
Deriving the nucleon-nucleon ($NN$) interaction continues to be one of the most important topics in nuclear physics. Starting from the seminal idea of Yukawa~\cite{Yukawa:1935xg} to describe the nuclear force in a field-theoretical manner, a number of high-precision phenomenological $NN$ interactions have been constructed, such as Reid93~\cite{Stoks:1994wp}, Argonne $\textrm{V}_{18}$~\cite{Wiringa:1994wb}, and CD-Bonn~\cite{Machleidt:2000ge}. Although these interactions describe experimental data well, they are not directly linked to the underlying theory of the strong interaction, quantum chromodynamics (QCD). So far, the two most widely accepted model-independent approaches to describe the $NN$ interaction based on QCD are  lattice QCD and chiral effective field theory (ChEFT). In past decades, with the rapidly increasing computational resources and improvements in algorithms, lattice QCD simulations have been employed to derive the $NN$ interaction from first principles~\cite{Ishii:2006ec,Aoki:2009ji,Beane:2010em} and remarkable progress has been made~\cite{Barnea:2013uqa,Drischler:2019xuo,Aoki:2020bew}. However, most lattice QCD simulations of the $NN$ interaction were still performed with unphysically large pion masses~\cite{NPLQCD:2020lxg,McIlroy:2017ssf}. On the other hand, chiral effective field theory has been successfully applied to study the $NN$ interaction since the pioneering works of Weinberg~\cite{Weinberg:1990rz,Weinberg:1991um}. ChEFT is based on chiral symmetry of QCD and its spontaneous and explicit breakings. In ChEFT, the long-range interaction is provided by the exchange of the Nambu-Goldstone bosons (pions in the $u$ and $d$ two-flavor sector and the pseudoscalar nonet in the $u$, $d$, and $s$ three-flavor sector), and the short-range interaction is described by the so-called low-energy constants (LECs) that encode the effects of higher energy degs of freedom. Nowadays, the chiral nuclear force  has been constructed in the heavy-baryon scheme up to the fifth order~\cite{Epelbaum:2014sza,Reinert:2017usi,Entem:2017gor} and become the de facto standard in $ab$ $initial$  nuclear structure and reaction studies~\cite{Epelbaum:2008ga,Machleidt:2011zz,Hammer:2019poc,Tews:2020hgp}.

Recently, a covariant power counting approach  similar to the extended-on-mass-shell scheme in the one-baryon sector was proposed for the $NN$ interaction~\cite{Ren:2016jna,Xiao:2018jot}, with the full structure of the Dirac spinor retained. At leading order (LO), the covariant chiral nuclear force with five LECs can already provide a reasonably good description of the $NN$ phase shifts for $J=0$ and $1$ partial waves~\cite{Ren:2016jna}. In addition, the unique features of  the $^1S_0$ partial wave, i.e., the pole position of the virtual bound state and the zero amplitude at the scattering momentum~340 MeV~\cite{SanchezSanchez:2017tws},  can be well reproduced~\cite{Ren:2017yvw}. This framework has also been successfully applied to study lattice QCD $NN$ interactions~\cite{Bai:2020yml,Bai:2021uim}, hyperon-nucleon interactions~\cite{Li:2016paq,Li:2016mln,Li:2018tbt,Song:2018qqm,Liu:2020uxi,Song:2021yab}, and $\Lambda_{c} N$ interaction~\cite{Song:2020isu,Song:2021war}.  In Ref.~\cite{Wang:2020myr}, the renormalizibility of the LO covariant chiral $NN$ interaction was investigated in detail, which indeed shows some interesting improvements, e.g., in the puzzling $^3P_0$ channel.

Clearly, the LO covariant $NN$ interaction still cannot reach the accuracy required for realistic applications, e.g., for many-body nuclear structure studies. To further improve the description of the $NN$ phase shifts, higher order contributions need to be taken into account, which contain  contact terms and two-pion exchanges.  Up to next-to-leading order (NLO), there are 17 LECs  in the covariant scheme~\cite{Xiao:2018jot}, which contribute to all the partial waves with $J=0, 1, 2$. At next-to-next-to-leading order (NNLO), no further LECs appear.  In Ref.~\cite{Xiao:2020ozd}, the covariant TPE has been calculated perturbatively up to order $O(p^3)$ for higher partial waves with $L\ge2$. It was shown that the contributions of relativistic TPE are more moderate than those of the nonrelativistic TPE at the perturbative level. Since the nucleon-nucleon system is nonperturbative in nature, it is necessary to explicitly check how the nonperturbative effect affects the TPE contributions  in the covariant scheme, especially for the $D$ and $F$ partial waves, where the nonperturbative effects are expected to be relatively strong.~\footnote{In the present work, we refrain from studying the lower partial waves with $J=0,1,2$, because they involve unknown LECs up to NNLO and therefore a dedicated fitting to data is needed for a meaningful study.}

In this work, we start from the covariant chiral Lagrangians~\cite{Fettes:2000gb} and calculate the nonperturbative TPE up to order $O(p^3)$. To resum the potential, we solve the Blankenbecler and Sugar (BbS) equation~\cite{Blankenbecler:1965gx} in which  the two intermediate nucleons are equally off shell. Then, we calculate the phase shifts for higher partial waves of $J\ge3$ and $L\le6$  and compare the resulting phase shifts with the perturbative ones of Ref.~\cite{Xiao:2020ozd} and nonrelativistic nonperturbative ones of Ref.~\cite{Epelbaum:1999dj}.
 
This article is organized as follows. In Sec. II, we spell out the covariant $\pi$N Lagrangians needed for computing the OPE and TPE contributions. In Sec. III,  we explain how to construct  the leading and subleading nonperturbative TPE, paying particular attention to the difference between the nonperturbative treatment with the perturbative treatment of Ref.~\cite{Xiao:2020ozd}. The notations for the scattering equation and phase shifts are specified in Sec. IV. We show the partial wave phase shifts in Sec. V and discuss  the cutoff dependence and convergence, followed by a short summary and outlook in Sec. VI.

\section{Chiral Lagrangian}\label{sec:lagrangians}

\begin{figure}
\centering
\includegraphics[width=0.09\textwidth]{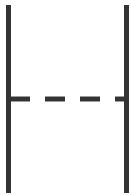}
\caption{One-pion exchange diagram at LO [$O(p^0)$]. The pion-nucleon vertices refer to vertices from $\mathcal{L}_{\pi N}^{(1)}$}.
\label{fig:ope}
\end{figure}

\begin{figure}
\centering
\includegraphics[width=0.63\textwidth]{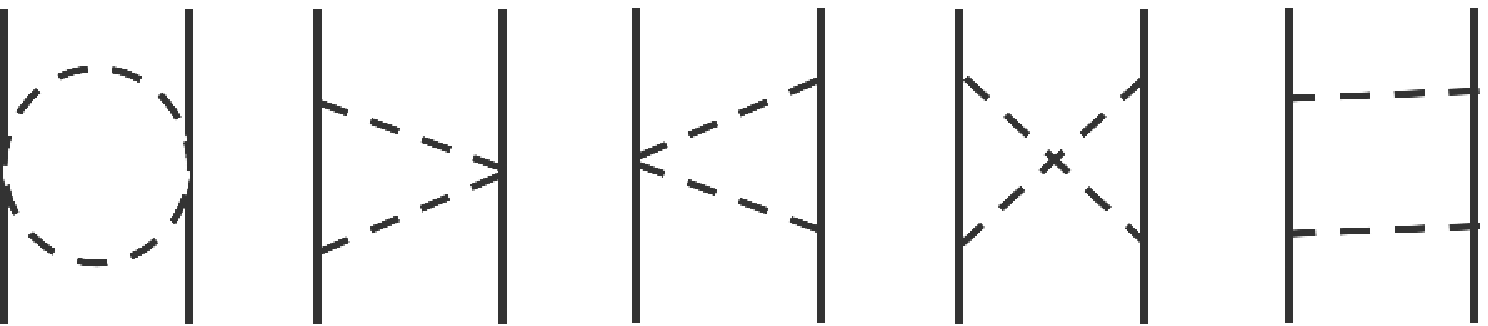}
\caption{Two-pion exchange diagrams at NLO [$O(p^2)$]. The pion-nucleon vertices refer to vertices from $\mathcal{L}_{\pi N}^{(1)}$}.
\label{fig:tpelo}
\end{figure}

\begin{figure}
\centering
\includegraphics[width=0.55\textwidth]{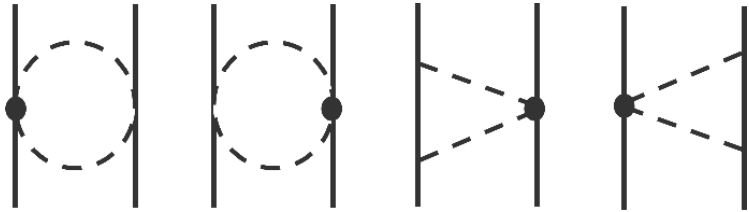}
\caption{Two-pion exchange diagrams at NNLO [$O(p^3)$]. The black dots denote vertices from $\mathcal{L}_{\pi N}^{(2)}$}.
\label{fig:tpenlo}
\end{figure}

For the purpose of including the contributions to the $NN$ interaction from the pion exchanges up to NNLO, one needs to compute the Feynman diagrams shown in Figs.~\ref{fig:ope},~\ref{fig:tpelo}, and~\ref{fig:tpenlo}. 
The relevant Lagrangians for the $\pi N$ vertices read~\cite{Fettes:2000gb}
\begin{align}\label{eq:lagpiN}
\mathcal{L}_{\pi N}^{(1)} &=\bar{\Psi}\left( {\rm{i}} \slashed{D}- M + \frac{g_A}{2} \slashed{u} \gamma_5 \right) \Psi ,\\
\mathcal{L}_{\pi N}^{(2)} &=c_{1}\trace{\chi_{+}}\bar{\Psi}\Psi- \frac{c_{2}}{4M^2}\trace{u^{\mu}u^\nu}\left(\bar{\Psi}D_{\mu}D_{\nu}\Psi +\text{H.c.}\right)+\frac{c_3}{2}\trace{u^2}\bar{\Psi} \Psi-\frac{c_4}{4}\bar{\Psi}\gamma^{\mu}\gamma^{\nu}\left[u_{\mu},u_{\nu}\right] \Psi,
\end{align}
with the axial vector coupling $g_A = 1.29$~\cite{Machleidt:2011zz}, the nucleon mass $M = 939$ MeV, the pion mass $m_{\pi} = 139 $ MeV~\cite{ParticleDataGroup:2018ovx}, the pion decay constant $f_{\pi} = 92.4$ MeV, and the low-energy constants $ c_1 =-1.39$, $c_2=4.01$, $c_3=-6.61$, $c_4=3.92$~\cite{Chen:2012nx} (all in units of GeV$^{-1}$). The SU(2) matrix $u = {\rm{exp}}\left(\frac{{\rm{i}}\Phi}{2f_\pi}\right)$, the pion and nucleon fields $\Phi$ and $\Psi$ read
\begin{align}
\Phi=\left(
 \begin{matrix}
   \pi^0 &  \sqrt{2} \pi^{+}\\
   \sqrt{2} \pi^{-} &  -\pi^0\\
  \end{matrix}
 \right),\quad
 \Psi = \left(
 \begin{matrix}
   p\\
   n\\
  \end{matrix}
 \right),
\end{align}
and $\chi_{+}=u^\dag \chi u + u \chi u^\dag$ with $\chi=\mathcal{M}=\text{diag}\left(m_\pi^2,m_\pi^2\right)$. 
The covariant derivative is defined as 
\begin{align}
D_{\mu}=&\partial_{\mu}+\Gamma_{\mu},\\
\Gamma_{\mu}=&\frac{1}{2}\left(u^\dag \partial_{\mu} u+ u \partial_{\mu}u^\dag\right),\\
u_{\mu}=&{\rm{i}}\left(u^\dag \partial_{\mu} u- u \partial_{\mu} u^\dag\right).
\end{align}
From the above Lagrangians, we can obtain the products of isospin and coupling factors for the TPE diagrams shown in Figs.~\ref{fig:ope},~\ref{fig:tpelo}, and~\ref{fig:tpenlo}. For details, we refer to Table~I of Ref.~\cite{Xiao:2020ozd}.

\section{Two-pion exchange contributions} \label{sec:oneloop}
We calculate the two-pion exchanges in the center-of-mass (c.m.) frame and in the isospin limit $m_u=m_d$. Because the large nonzero nucleon mass at the chiral limit leads to the so-called power counting breaking (PCB) problem, one should recover the power counting rule defined in Ref.~\cite{Xiao:2018jot}. Here, we adopt the  extended on mass shell (EOMS) scheme and extend it to the two-baryon sector to remove the PCB terms, which has been well established in the one-baryon~\cite{Gegelia:1999gf,Fuchs:2003qc,Geng:2013xn} and meson-baryon sectors~\cite{Chen:2012nx,Lu:2018zof}. We use FEYNCALC~\cite{Shtabovenko:2020gxv,Shtabovenko:2016sxi,Mertig:1990an} to decompose the full TPE potential into the scalar integrals A0, B0, C0, and D0 multiplied with certain polynomials and fermion bilinears, and then calculate the potential numerically with the help of ONELOOP~\cite{vanHameren:2009dr,vanHameren:2010cp}. 

Unlike Ref.~\cite{Xiao:2020ozd}, we  focus in the present work on the off-shell TPE contributions which are  necessary for the nonperturbative treatment. For an off-shell nucleon $u(p,M)$ with momentum $p=(p^0,\vec{p})$ $\left [p^2=p^2_0-\vec{p}^2 \right ]$, the equation of motion $\slashed{p}u(p,M)=Mu(p,M)$ is  no longer  applicable. Instead, one now has
\begin{equation}\label{offshellME}
  \slashed{p}u(p,M)=\left(
  \begin{array}{cc}
    p^0-E+M  &  0 \\
    0  & E-p^0+M  \\
  \end{array}
\right)u(p,M),
\end{equation}
where $E=\sqrt{|\bm{p}|^2+M^2}$ is the energy of the nucleon. For on-shell nucleons, one has $p^0=E$ and then the normal equation of motion is recovered. 

Next, we evaluate the various classes of Feynman diagrams one by one. In writing down the potential, the explicit expressions of the potential will  not be given  due to their complexity.~\footnote{They can be obtained as a MATHEMATICA notebook from the authors upon request}


\subsection{ Leading order $\left[O(p^0)\right]$ results }

At LO, we only have the one-pion exchange (OPE) diagram  shown in Fig.~\ref{fig:ope}. The OPE potential can be expressed as
\begin{align}
V_{\textrm{OPE}}=\frac{g^2_A}{4f^2_{\pi}}\frac{\bar{u}(p_3)\gamma^{\mu}\gamma_5(p_{1\mu}-p_{3\mu})u(p_1){\bar{u}(p_4)}\gamma^{\nu}\gamma_5(p_{4\nu}-p_{2\nu})u(p_2)}{(p_3-p_1)^2+m^2_{\pi}}.
\end{align}

\subsection{ Next-to-leading order $\left[O(p^2)\right]$ results }

At NLO, there are five TPE diagrams  as shown in Fig.~\ref{fig:tpelo}, which from left to right correspond to the football diagram, the two triangle diagrams, the crossed box diagram, and the planar box diagram, respectively. The corresponding TPE potentials at this order can be written as
\begin{align}
V_{\textrm{Box}}=&\frac{-i g^4_A}{16f^4_{\pi}}\int\frac{d^4l}{(2\pi)^4}\frac{\bar u(p_3)\gamma^{\nu}\gamma_5(l_{\nu}+p_{3\nu}-p_{1\nu})(\slashed{p}_1-\slashed{l}+M)\gamma^{\mu}\gamma_5l_{\mu}u(p_1)\bar u(p_4)\gamma^{\tau}\gamma_5(l_{\tau}+p_{2\tau}-p_{4\tau})}{[(p_2+l)^2+M^2][(p_1-l)^2+M^2]}\\
\nonumber &\times \frac{(\slashed{p}_2+\slashed{l}+M)\gamma^{\rho}\gamma_5l_{\rho}u(p_2)}{(l^2+m^2_{\pi})[(l+p_2-p_4)^2+m^2_{\pi}]},\\
V_{\textrm{Cross}}=&\frac{-i g^4_A}{16f^4_{\pi}}\int\frac{d^4l}{(2\pi)^4}\frac{\bar u(p_3)\gamma^{\nu}\gamma_5(l_{\nu}+p_{3\nu}-p_{1\nu})(\slashed{p}_1-\slashed{l}+M)\gamma^{\mu}\gamma_5l_{\mu}u(p_1)\bar u(p_4)\gamma^{\tau}\gamma_5l_{\tau}(\slashed{p}_4-\slashed{l}+M)}{[(p_1-l)^2+M^2][(p_4-l)^2+M^2]}\\
\nonumber &\times \frac{\gamma^{\rho}\gamma_5(l_{\rho}+p_{2\rho}-p_{4\rho})u(p_2)}{(l^2+m^2_{\pi})[(l+p_2-p_4)^2+m^2_{\pi}]},\\
V_{\textrm{Football}}=&\frac{1}{2}\frac{-i}{16f^4_{\pi}}\int\frac{d^4l}{(2\pi)^4}\frac{\bar{u}(p_3)\gamma^{\mu}(2l_{\mu}+p_{1\mu}-p_{3\mu}) u(p_1){\bar{u} (p_4)}\gamma^{\nu}(2l_{\nu}+p_{4\nu}-p_{2\nu})u(p_2)}{(l^2+m^2_{\pi})[(l+p_4-p_2)^2+m^2_{\pi}]},\\
V_{\textrm{TriangleL}}=&\frac{-i g^2_A}{16f^4_{\pi}}\int\frac{d^4l}{(2\pi)^4}\frac{\bar{u}(p_3)\gamma^{\mu}(2l_{\mu}+p_{4\mu}-p_{2\mu}) u(p_1)\bar u (p_4)\gamma^{\nu}\gamma_5(l_{\nu}+p_{4\nu}-p_{2\nu})(\slashed{p}_2-\slashed{l}+M)\gamma^{\rho}\gamma_5(-l_{\rho})u(p_2)}{[(p2-l)^2+M^2](l^2+m^2_{\pi})[(l+p_4-p_2)^2+m^2_{\pi}]},\label{NLO:TRIL}\\
V_{\textrm{TriangleR}}=&\frac{-i g^2_A}{16f^4_{\pi}}\int\frac{d^4l}{(2\pi)^4}\frac{\bar{u}(p_3)\gamma^{\nu}\gamma_5(l_{\mu}+p_{3\mu}-p_{1\mu})(\slashed{p}_1-\slashed{l}+M)\gamma^{\rho}\gamma_5(-l_{\rho})u(p_1)\bar u(p_4)\gamma^{\mu}(2l_{\mu}+p_{3\mu}-p_{1\mu})u(p_2)}{[(p1-l)^2+M^2](l^2+m^2_{\pi})[(l+p_3-p_1)^2+m^2_{\pi}]}.\label{NLO:TRIR}
\end{align}
Obviously, the potentials of the triangle left diagram and the triangle right diagram are the same. The corresponding PCB terms at NLO in helicity basis read
\begin{align}\label{eq:pcblo}
V^{\textrm{PCB}}_{\textrm{Football}} & = 0,\\
V^{\textrm{PCB}}_{\textrm{TriangleL}} & =  \frac{g^2_A}{4f^4_{\pi}}{H_1} M^2 \ln \left(\frac{\mu}{M}\right),\\
V^{\textrm{PCB}}_{\textrm{TriangleR}} &=  \frac{g^2_A}{4f^4_{\pi}}{H_1} M^2 \ln \left(\frac{\mu}{M}\right),\\
V^{\textrm{PCB}}_{\textrm{Cross}} & = -\frac{g^4_A}{4f^4_{\pi}}H_1 M^2 \left[3 \ln \left(\frac{\mu}{M}\right)-1\right],\\
V^{\textrm{PCB}}_{\textrm{Box}}   & = -\frac{g^4_A}{4f^4_{\pi}}H_1 M^2 \left[ \ln \left(\frac{\mu}{M}\right)+1\right], 
\end{align}
where $\mu$ refers to the renormalization scale and is set at 1 GeV in our numerical study. $H_{1}$ is the inner product of the initial and final helicity basis and reads
\begin{align}
H_1 = \left[|\bar{\lambda}_1 +\lambda_1| \cos\left(\frac{\theta}{2}\right) +|\bar{\lambda}_1 -\lambda_1| \sin\left(\frac{\theta}{2}\right)  \right] \left[ |\bar{\lambda}_2 +\lambda_2| \cos\left(\frac{\theta}{2}\right) - |\bar{\lambda}_2 -\lambda_2| \sin\left(\frac{\theta}{2}\right) \right],
\end{align}
where $\lambda_{1,2},\bar{\lambda}_{1,2}$ denote the helicities of incoming and outgoing particles, respectively, and $\theta$ is  the scattering angle.

\subsection{Next-to-next-to-leading order $\left[O(p^3)\right]$ results}

At NNLO, we have four diagrams as shown in Fig.~\ref{fig:tpenlo}. In general, we only need to replace one of the $\pi N$ vectices at $O(p)$ in the diagrams of Fig.~\ref{fig:tpelo} with the corresponding $O(p^2)$ vertex. Note that no box diagram and crossed box diagram appears at this order because there is no $\pi NN$ vertex at $O(p^2)$. Corresponding to the four terms $c_1$, $c_2$, $c_3$, $c_4$ in the $O(p^2)$ $\pi N$ vertices, the potential of the football left diagram in which the left $\pi N$ vertex is replaced reads
\begin{align}
V^{c_1}_{\textrm{FootballL}}=&\frac{1}{2}\frac{-i{c_1} m^2_{\pi}}{16f^4_{\pi}}\int\frac{d^4l}{(2\pi)^4}\frac{\bar u(p_3)u(p_1)\bar u(p_4)\gamma^{\mu}(2l_{\mu}+p_{4\mu}-p_{2\mu})u(p_2)}{(l^2+m^2_{\pi})[(l+p_1-p_3)^2+m^2_{\pi}]},\\
V^{c_2}_{\textrm{FootballL}}=&\frac{1}{8M^2}\frac{-i{c_2} }{16f^4_{\pi}}\int\frac{d^4l}{(2\pi)^4}\frac{\bar u(p_3)\big{(}l\cdot p_1(l\cdot p_1+p_1\cdot p_1-p_3\cdot p_1)+l\cdot p_3(l\cdot p_3+p_1\cdot p_3-p_3\cdot p_3)\big{)}u(p_1)}{(l^2+m^2_{\pi})[(l+p_1-p_3)^2+m^2_{\pi}]}\\
\nonumber &\times \bar u(p_4)\gamma^{\mu}(2l_{\mu}+p_{4\mu}-p_{2\mu})u(p_2),\\
V^{c_3}_{\textrm{FootballL}}=&\frac{1}{2}\frac{-i{c_3} }{16f^4_{\pi}}\int\frac{d^4l}{(2\pi)^4}\frac{\bar u(p_3)(l\cdot l+l\cdot p_1-l\cdot p_3)u(p_1)\bar u(p_4)\gamma^{\mu}(2l_{\mu}+p_{4\mu}-p_{2\mu})u(p_2)}{(l^2+m^2_{\pi})[(l+p_1-p_3)^2+m^2_{\pi}]},\\
V^{c_4}_{\textrm{FootballL}}=&\frac{1}{2}\frac{-i{c_4} }{16f^4_{\pi}}\int\frac{d^4l}{(2\pi)^4}\frac{\bar u(p_3)\big{(}-\gamma^{\mu}(l_{\mu}+p_{1\mu}-p_{3\mu})\gamma^{\nu}l_{\nu}+\gamma^{\mu}l_{\mu}\gamma^{\nu}(l_{\nu}+p_{1\nu}-p_{3\nu})\big{)}u(p_1)}{(l^2+m^2_{\pi})[(l+p_1-p_3)^2+m^2_{\pi}]}\\
\nonumber &\times \bar u(p_4)\gamma^{\mu}(2l_{\mu}+p_{4\mu}-p_{2\mu})u(p_2),
\end{align}
and the complete football left potential can be expressed as
\begin{align}
{V_{\textrm{FootballL}} =  V^{c_1}_{\textrm{FootballL}} + V^{c_2}_{\textrm{FootballL}}+ V^{c_3}_{\textrm{FootballL}}+ V^{c_4}_{\textrm{FootballL}}.}
\end{align}

Similarly, the potential of the triangle left diagram  reads, 
\begin{align}
{V_{\textrm{TriangleL}} =  V^{c_1}_{\textrm{TriangleL}} + V^{c_2}_{\textrm{TriangleL}}+ V^{c_3}_{\textrm{TriangleL}}+ V^{c_4}_{\textrm{TriangleL}},}
\end{align}
where
\begin{align}
V^{c_1}_{\textrm{TriangleL}}=&\frac{1}{2}\frac{-i{c_1} m^2_{\pi}g^2_A}{16f^4_{\pi}}\int\frac{d^4l}{(2\pi)^4}\frac{\bar u(p_3)u(p_1)\bar u(p_4)\gamma^{\rho}\gamma_5(l_{\rho}+p_{4\rho}-p_{2\rho})(\slashed{p}_2-\slashed{l}+M)\gamma^{\tau}\gamma_5(-l_{\tau})u(p_2)}{(l^2+m^2_{\pi})[(l+p_1-p_3)^2+m^2_{\pi}][(p_2-l)^2+M^2]},\\
V^{c_2}_{\textrm{TriangleL}}=&\frac{1}{8M^2}\frac{-i{c_2} g^2_A}{16f^4_{\pi}}\int\frac{d^4l}{(2\pi)^4}\frac{\bar u(p_3)\big{(}l\cdot p_1(l\cdot p_1+p_1\cdot p_1-p_3\cdot p_1)+l\cdot p_3(l\cdot p_3+p_1\cdot p_3-p_3\cdot p_3)\big{)}u(p_1)}{(l^2+m^2_{\pi})[(l+p_1-p_3)^2+m^2_{\pi}][(p_2-l)^2+M^2]}\\
\nonumber &\times u(p_4)\gamma^{\rho}\gamma_5(l_{\rho}+p_{4\rho}-p_{2\rho})(\slashed{p}_2-\slashed{l}+M)\gamma^{\tau}\gamma_5(-l_{\tau})u(p_2),\\
V^{c_3}_{\textrm{TriangleL}}=&\frac{1}{2}\frac{-i{c_3} g^2_A}{16f^4_{\pi}}\int\frac{d^4l}{(2\pi)^4}\frac{\bar u(p_3)(l\cdot l+l\cdot p_1-l\cdot p_3)u(p_1)\bar u(p_4)\gamma^{\rho}\gamma_5(l_{\rho}+p_{4\rho}-p_{2\rho})(\slashed{p}_2-\slashed{l}+M)\gamma^{\tau}\gamma_5(-l_{\tau})u(p_2)}{(l^2+m^2_{\pi})[(l+p_1-p_3)^2+m^2_{\pi}][(p_2-l)^2+M^2]},\\
V^{c_4}_{\textrm{TriangleL}}=&\frac{1}{2}\frac{-i{c_4} g^2_A}{16f^4_{\pi}}\int\frac{d^4l}{(2\pi)^4}\frac{\bar u(p_3)\big{(}-\gamma^{\mu}(l_{\mu}+p_{1\mu}-p_{3\mu})\gamma^{\nu}l_{\nu}+\gamma^{\mu}l_{\mu}\gamma^{\nu}(l_{\nu}+p_{1\nu}-p_{3\nu})\big{)}u(p_1)}{(l^2+m^2_{\pi})[(l+p_1-p_3)^2+m^2_{\pi}][(p_2-l)^2+M^2]}\\
\nonumber &\times u(p_4)\gamma^{\rho}\gamma_5(l_{\rho}+p_{4\rho}-p_{2\rho})(\slashed{p}_2-\slashed{l}+M)\gamma^{\tau}\gamma_5(-l_{\tau})u(p_2).
\end{align}

The potentials for the football right diagram and triangle right diagram can then be obtained via simple exchanges of $p_1\leftrightarrow p_2$ and $p_3\leftrightarrow p_4$ {and are the same as the corresponding football and triangle left diagrams. The PCB terms at this order for the football left (right) diagrams are vanishing and those for the triangle left diagram in helicity basis read,   
\begin{align}
V^{\textrm{PCB}}_{{\textrm{TriangleL}}}=V^{c_1,\textrm{PCB}}_{{\textrm{TriangleL}}}+V^{c_2,\textrm{PCB}}_{{\textrm{TriangleL}}}+V^{c_3,\textrm{PCB}}_{{\textrm{TriangleL}}}+V^{c_4,\textrm{PCB}}_{{\textrm{TriangleL}}},
 \end{align}}
 {where}
{
 \begin{align}
   V^{c_1,\textrm{PCB}}_{{\textrm{TriangleL}}} &= \frac{3c_1g^2_A}{8f^4_{\pi}}   H_1 M m_\pi^2 \left[ 2\ln \left( \frac{\mu}{M}\right) + 1\right],\\
  V^{c_2,\textrm{PCB}}_{{\textrm{TriangleL}}} &=\frac{3c_2g^2_A{H_1} M}{2304 f^4_{\pi}} \bigg{[}6\textrm{ln} \left(\frac{\mu }{M}\right)  \left(4\left(\left(-6 {\lambda_1} {\bar{\lambda}_1} +10 {\lambda_2} {\bar{\lambda}_2} \right){p} {p'}{\text{cos}\left(\theta\right)}+6 M^2+6 {m^2_{\pi}}+3 {p}^2+3 {p'}^2\right)+9 \left({s-4M^2}\right)\right)\nonumber\\ 
   & -8\left(-16 \left({\lambda_2} {\bar{\lambda}_2} +6\right){p} {p'}\text{cos}\left(\theta\right)+3 {p}^2+3 {p'}^2\right)\bigg{]},\\ 
   V^{c_3,\textrm{PCB}}_{{\textrm{TriangleL}}} &=\frac{3c_3g^2_A{H_1} M}{32 f^4_{\pi}} \bigg{[}2\textrm{ln} \left(\frac{\mu }{M}\right) \left(-4 ({\lambda_1} {\bar{\lambda}_1}+{\lambda_2} {\bar{\lambda}_2}){p} {p'}{\text{cos}\left(\theta\right)}+4 M^2+{p}^2+{p'}^2\right)+\left(4 \textrm{ln} \left(\frac{\mu}{M}\right)+2\right) A(p,p')\bigg{]},\\
   V^{c_4,\textrm{PCB}}_{{\textrm{TriangleL}}} &=\frac{3c_4g^2_AM}{16 f^4_{\pi}} \bigg{[}\left(-2\ln \left( \frac{\mu}{M}\right) -2 \right)B(p,p',\lambda_{1,2},\bar{\lambda}_{1,2})-2{H_1} \textrm{ln} \left(\frac{\mu }{M}\right) \left(\left(-16 {\lambda_1} {\bar{\lambda}_1} +2\right){p} {p'}\text{cos}\left(\theta\right)+{p}^2+{p'}^2\right)\bigg{]},
 \end{align}}
{
\begin{align}
  A(p,p')=&\left(4 {m^2_{\pi}}+{p}^2-2 {pp'}\text{cos}\left(\theta\right)+{p'}^2\right),\\
  B(p,p',\lambda_{1,2},\bar{\lambda}_{1,2})=&\left(4 {H_2} ({\lambda_1} {p}+{\bar{\lambda}_1} {p'}) ({\lambda_2} {p}+{\bar{\lambda}_2} {p'})+{H_1} \left(\left(8 \left({\lambda_1} {\bar{\lambda}_1}+{\lambda_2} {\bar{\lambda}_2}\right)-2\right){p} {p'}\text{cos}\left(\theta\right) +{p}^2+{p'}^2\right)\right),
 \end{align}}
{ and}
 {\begin{align}
 H_2 =& -\left[\left(\bar{\lambda}_1 +\lambda_1\right) \sin\left(\frac{\theta}{2}\right) +|\bar{\lambda}_1 -\lambda_1| \cos\left(\frac{\theta}{2}\right)  \right] \left[ \left(\bar{\lambda}_2 +\lambda_2\right) \sin\left(\frac{\theta}{2}\right) - |\bar{\lambda}_2 -\lambda_2| \cos\left(\frac{\theta}{2}\right) \right]\nonumber\\
 &-\left[|\bar{\lambda}_1 +\lambda_1| \sin\left(\frac{\theta}{2}\right) -\left(\bar{\lambda}_1 -\lambda_1\right) \cos\left(\frac{\theta}{2}\right)  \right] \left[ |\bar{\lambda}_2 +\lambda_2| \sin\left(\frac{\theta}{2}\right) + \left(\bar{\lambda}_2 -\lambda_2\right) \cos\left(\frac{\theta}{2}\right) \right]\nonumber\\
 &-\left[\left(\bar{\lambda}_1 +\lambda_1 \right)\cos\left(\frac{\theta}{2}\right) -|\bar{\lambda}_1 -\lambda_1| \sin\left(\frac{\theta}{2}\right)  \right] \left[ \left(\bar{\lambda}_2 +\lambda_2\right) \cos\left(\frac{\theta}{2}\right) + |\bar{\lambda}_2 -\lambda_2| \sin\left(\frac{\theta}{2}\right) \right].
  \end{align}}
{ 
The PCB terms for the triangle right diagram can then be obtained via the exchanges $\lambda_1\leftrightarrow \lambda_2$ and $\bar{\lambda}_1\leftrightarrow \bar{\lambda}_2$.}
{We note that all the PCB terms and the ultraviolet divergences in the chiral loops up to NNLO can be absorbed by LO and NLO contact terms in the covariant framework.}

In deriving the potential, to be consistent, we must adopt the same approximation as that of the BbS equation, i.e., the two nucleons are equally off the mass shell~\cite{Blankenbecler:1965gx,Woloshyn:1973mce}. In this approximation, the incoming momenta in the center-of mass frame are expressed as $p_1=(\sqrt{s}/2,\bm{p})$ and $p_2=(\sqrt{s}/2,-\bm{p})$, the outgoing momenta as $p_3=(\sqrt{s}/2,\bm{p'})$ and  $p_4=(\sqrt{s}/2,-\bm{p'})$,  and the transferred momentum as $q=(0,\bm{p'}-\bm{p})$. 

\subsection{Iterated OPE}
To avoid double counting the OPE contribution in the nonperturbative treatment of TPE contributions, one needs to subtract the contribution of iterated OPE (IOPE) which appears in the box diagram in the case of on-shell intermediate nucleons. We calculate the IOPE contribution {in partial wave matrix elements} numerically, i.e.,
{\begin{align}
V_{\textrm{IOPE}}^{l'l,sj}({p}',{p}|\sqrt{s}) &=\sum_{l^{''}}\int \frac{k^2d{k}}{(2\pi)^3} V_{\textrm{OPE}}^{l'l'',sj}({p}',{k}|\sqrt{s})G_{\textrm{BbS}}({k}|\sqrt{s})V_{\textrm{OPE}}^{l''l,sj}({k},{p}|\sqrt{s}),\label{IOPE}
\end{align}}
where $p = |\bm{p}|$, $p' = |\bm{p'}|$, $\sqrt{s}$ is the total energy of the two-nucleon system in the c.m. frame, {$V_{\textrm{OPE}}^{l'l,sj}$} is the OPE potential in the $LSJ$ representation, and $G_{\textrm{BbS}}$ is the propagator adopted in the BbS equation {and reads}
{\begin{align}
G_{\textrm{BbS}}({k}|\sqrt{s}) = \frac{M^2}{E_{k}}\frac{1}{p_{\text{cm}}^2-k^2}, 
\end{align}}
{where $|\bm{p}_{\text{cm}}|=\sqrt{s/4-M^2}$ is the nucleon momentum on the mass shell in the c.m. frame.} The integral above suffers from the same ultraviolet divergence as the scattering equation and thus needs to be properly regularized. We introduce an exponential regulator function $f^{\Lambda}$ as
\begin{align}
f^{\Lambda}(k) = \textrm{exp}[-\left(k/\Lambda\right)^{2n}]
\end{align}
and choose $n=3$ following Ref.~\cite{Epelbaum:2003xx}. In this way, the OPE potential $V_{\textrm{OPE}}$ in Eq.~(\ref{IOPE}) should be substituted with
\begin{align}
V_{\textrm{OPE}}({p}',{k}|\sqrt{s})\rightarrow&f^{\Lambda}(k)V_{\textrm{OPE}}({p}',{k}|\sqrt{s}),\\\nonumber
V_{\textrm{OPE}}({k},{p}|\sqrt{s})\rightarrow&f^{\Lambda}(k)V_{\textrm{OPE}}({k},{p}|\sqrt{s}).
\end{align}

%With the regulator function, the IOPE can be calculated by standard Gauss-Legendre quadrature and principal value integral. Therefore, we're not going into further discussion. 

\section{Scattering equation}

{To calculate the nonperturbative contributions of TPE, one should resum the above potential up to infinite order via a covariant scattering equation. In this work, we choose the one proposed by Blankenbecler and Sugar~\cite{Blankenbecler:1965gx} (BbS) which has the  practical advantage that the potential becomes energy independent~\cite{Entem:2001cg}. The BbS equation projected into the $LSJ$ representation reads
\begin{align}
T_{l'l}^{sj}({p}',{p}|\sqrt{s})=V_{l'l}^{sj}({p}',{p}|\sqrt{s})+\sum_{l^{''}}\int\frac{k^2dk}{(2\pi)^3}V_{l'l^{''}}^{sj}({p}',{k}|\sqrt{s})\frac{M^2}{E_{k}}\frac{1}{p^2-k^2+i\varepsilon}T_{l^{''}l}^{sj}({k},{p}|\sqrt{s}),
\end{align}
where the potential takes the following form
\begin{equation}\label{NNForce}
  V=V_{\mathrm{CT}}^{\mathrm{LO}}+V_{\mathrm{CT}}^{\mathrm{NLO}}+V_{\mathrm{OPE}}+V_{\mathrm{TPE}}^{\mathrm{NLO}}+V_{\mathrm{TPE}}^{\mathrm{NNLO}}-V_{\mathrm{IOPE}},
\end{equation} 
in which the first two terms refer to the contact terms at LO [${O}(p^0)$] and NLO [${O}(p^2)$], while the next three terms denote the contributions from OPE and TPE at NLO and NNLO. The last term represents the iterated OPE contribution, i.e., $V^{\mathrm{OPE}}\cdot g \cdot V^{\mathrm{OPE}}$.}

The iteration of $V$ in the BbS equation requires cutting $V$ off for high momenta to avoid ultraviolet divergence. This is achieved in the following way:
\begin{align}
V_{l'l}^{sj}({p}',{p}|\sqrt{s})=f_{R}({p})V({p}',{p}|\sqrt{s})f_{R}({p'}),
\end{align}
where the regulator function is chosen as,
\begin{align}
f_{R}({p})=f_{R}^{\textrm{sharp}}({p})=\theta(\Lambda^2-p^2).
\end{align}
It should be stressed that we use the same cutoff in the scattering equation as the one in the IOPE contribution. In fact, the phase shifts are basically unchanged if we only vary the cutoff in the IOPE piece but fix the cutoff in the scattering equation since it is known that the IOPE contributes little to scattering amplitudes for most partial waves~\cite{Epelbaum:2004fk}. 

The partial wave $S$ matrix is related to the on-shell $T$ matrix by
\begin{align}
S_{l'l}^{sj}({p}_{\text{cm}}) = \delta_{l'l}^{sj} + 2\pi i \rho T_{l'l}^{sj}({p}_{\text{cm}}),\quad \rho=-\frac{|{p}_{\text{cm}}|M^2}{16\pi^2E_{\text{cm}}}
\end{align}
where {${p}_{\text{cm}}=\sqrt{T_{\text{lab}}M/2}$}. The phase space factor can be determined by the elastic unitarity of the BbS equation. 

Phase shifts in the single channel case can be obtained from the $S$ matrix via
\begin{align}
S_{jj}^{0j}=\textrm{exp}{\left(2i\delta_{j}^{0j}\right)},\quad S_{jj}^{1j}=\textrm{exp}{\left(2i\delta_{j}^{1j}\right)}.
\end{align}

To calculate the phase shifts in coupled channels, we adopt the Stapp  parametrization~\cite{Stapp:1956mz} of the $S$ matrix, which can be written as
\begin{align}
S=&\begin{pmatrix} S_{--}^{1j} & S_{-+}^{1j} \\ S_{+-}^{1j} & S_{++}^{1j} \end{pmatrix}\nonumber\\
=&\begin{pmatrix} \textrm{exp}{\left(i\delta_{-}^{1j}\right)} & 0 \\ 0 & \textrm{exp}{\left(i\delta_{+}^{1j}\right)} \end{pmatrix}\begin{pmatrix} \textrm{cos}(2\varepsilon) & i\textrm{sin}(2\varepsilon) \\ i\textrm{sin}(2\varepsilon) & \textrm{cos}(2\varepsilon) \end{pmatrix}\begin{pmatrix} \textrm{exp}{\left(i\delta_{-}^{1j}\right)} & 0 \\ 0 & \textrm{exp}{\left(i\delta_{+}^{1j}\right)} \end{pmatrix}\nonumber\\
=&\begin{pmatrix} \textrm{cos}(2\varepsilon)\textrm{exp}{\left(2i\delta_{-}^{1j}\right)} & i\textrm{sin}(2\varepsilon)\textrm{exp}{\left(i\delta_{-}^{1j}+i\delta_{+}^{1j}\right)} \\
i\textrm{sin}(2\varepsilon)\textrm{exp}{\left(i\delta_{-}^{1j}+i\delta_{+}^{1j}\right)}  &
\textrm{cos}(2\varepsilon)\textrm{exp}{\left(2i\delta_{+}^{1j}\right)}
\end{pmatrix}
\end{align}
where the subscript ``$+$'' refers to $j+1$ and ``$-$'' to $j-1$. 

\section{Results and discussions} \label{sec:results}
In this section, the $\textrm{LO}$, $\textrm{NLO}$, and $\textrm{NNLO}$ relativistic nonperturbative phase shifts with $J \geq 3$ and  $L \leq 6$ are presented and compared with those of the perturbative counterparts~\cite{Xiao:2020ozd}, the nonrelativistic counterparts obtained in the dimensional regularization (DR) scheme~\cite{Epelbaum:1999dj}, and the Nijmegen partial wave phase shifts~\cite{Stoks:1993tb}.  For simplification, we use ``NNLO-Per," ``NNLO-NR-DR," and ``PWA93" to denote these results respectively in the following.

In the present work, we concentrate on the partial waves to which up to NNLO contact terms do not contribute. Therefore, the cutoff is the only free parameter. Following the treatment of Ref.~\cite{Epelbaum:1999dj}, we vary the cutoff between 600 and 1000 MeV in the present work and find that the best description of the phase shifts is obtained with a cutoff around 900 MeV. The fact that the cutoff used for NNLO studies is relatively larger than the usual value for LO or NLO studies has been noticed in the nonrelativistic schemes~\cite{Epelbaum:1999dj}, where it was pointed out that  the large contribution of subleading TPE shifts the cutoff to a larger value.
In the nonrelativistic potential obtained in the DR scheme~\cite{Epelbaum:1999dj}, a sharp cutoff with $\Lambda=875$ MeV is used in solving the scattering equation. In the following, we plot the phase shifts for each partial wave obtained with solely OPE and TPE contributions up to $T_{\textrm{lab}}=280$ MeV where the first inelastic channel opens. 
%~\footnote{With $R=1$ fm one obtains the smallest $\tilde{\chi}^2$ for all $J \geq 3$ and $L \leq 6$ partial waves among the choices of $R=0.8$, 0.9, 1.0, 1.1, 1.2 fm}

\subsection{$D$ wave}

\begin{figure}[htbp]
\centering
\subfloat{
\includegraphics[width=0.45\textwidth]{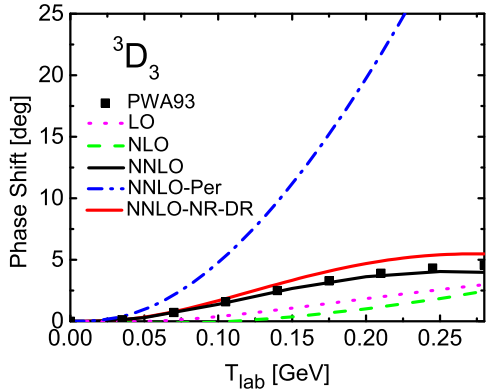}
}
\caption{$^3D_3$ phase shifts as a function of $T_{\text{lab}}$. The black dots denote the PWA93 results~\cite{Stoks:1993tb}. The magenta dotted, green dashed, and black solid lines represent the relativistic nonperturbative LO, NLO, and  NNLO phase shifts, respectively. For comparison, we also show the nonrelativistic counterparts (NNLO-NR-DR, red solid line)~\cite{Epelbaum:1999dj} and relativistic perturbative results at NNLO~\cite{Xiao:2020ozd} (NNLO-Per, magenta dash-dotted line).}
\label{tb:Dwave}
\end{figure}

In $D$ wave, only $^3D_3$ is free of contact contributions up to NNLO in our covariant framework. The phase shifts are shown in Fig.~\ref{tb:Dwave}. The NNLO nonperturbative phase shifts are in good agreement with PWA93 up to $T_{\textrm{lab}} = 250 $ MeV, while the perturbative results keep increasing rapidly in the whole energy region. This indicates that the nonperturbative effect plays a crucial role for this partial wave. Similar conclusions have been drawn in Refs.~\cite{Xiao:2020ozd,Kaiser:1997mw}. Compared with the NNLO-NR-DR scheme~\cite{Epelbaum:1999dj}, we find that both the nonrelativistic and our relativistic results can describe the phase shifts quite well below $T_{\textrm{lab}} = 100 $ MeV. However, above 100 MeV, the relativistic correction is of visible size and helps to improve the description of phase shifts.~\footnote{Note that the nonrelativistic result for the $^3D_3$ partial wave is also dependent on the  cutoff. The  cutoff of 875 MeV is the optimum value for all the partial waves. A better description of  the $^3D_3$ channel is possible if one fine-tunes the cutoff. } Compared with the LO and NLO results for this partial wave, the NNLO results describe the phase shifts significantly better. However, we notice that the differences between the NNLO and NLO results are much larger than those between the NLO and LO results. This indicates a quite large contribution from the subleading TPE, which  implies that the convergence of our covariant nuclear force is questionable up to NNLO. Actually this is not surprising since the nonrelativistic studies also suffer from such an extremely large subleading TPE contribution~\cite{Epelbaum:1999dj}. The spectral function regularization (SFR) method was proposed in Ref.~\cite{Epelbaum:2003gr} to deal with this problem by introducing a cutoff for the  chiral loops. In our covariant framework based on the EOMS scheme, such a cutoff method is not practical. Higher order contributions may also be helpful to balance the large subleading TPE contribution as was shown in Refs.~\cite{Entem:2002sf,Entem:2003ft}. This should be explicitly checked in the future.

It should be noted that for $^3D_3$ the phase shifts are sensitive to the choice of the cutoff (which is shown in Fig.~\ref{tb:cutoff-dependence} in the appendix) when we vary the cutoff from 800 to 1000 MeV, the phase shifts at $T_{\textrm{lab}} = 280$ MeV change approximately from $10$ to $-5$ deg. This comes from the strong cutoff dependence of subleading TPE. Higher order contact terms may be of great help to weaken the cutoff dependence, as was done in Ref.~\cite{Epelbaum:1999dj} where the strong cutoff dependence of the NNLO-NR-DR results for the $^1D_2$ partial wave is significantly reduced once the contact terms of N$^3$LO were introduced. 

\begin{figure}[htbp]
\centering

\subfloat{
\includegraphics[width=0.45\textwidth]{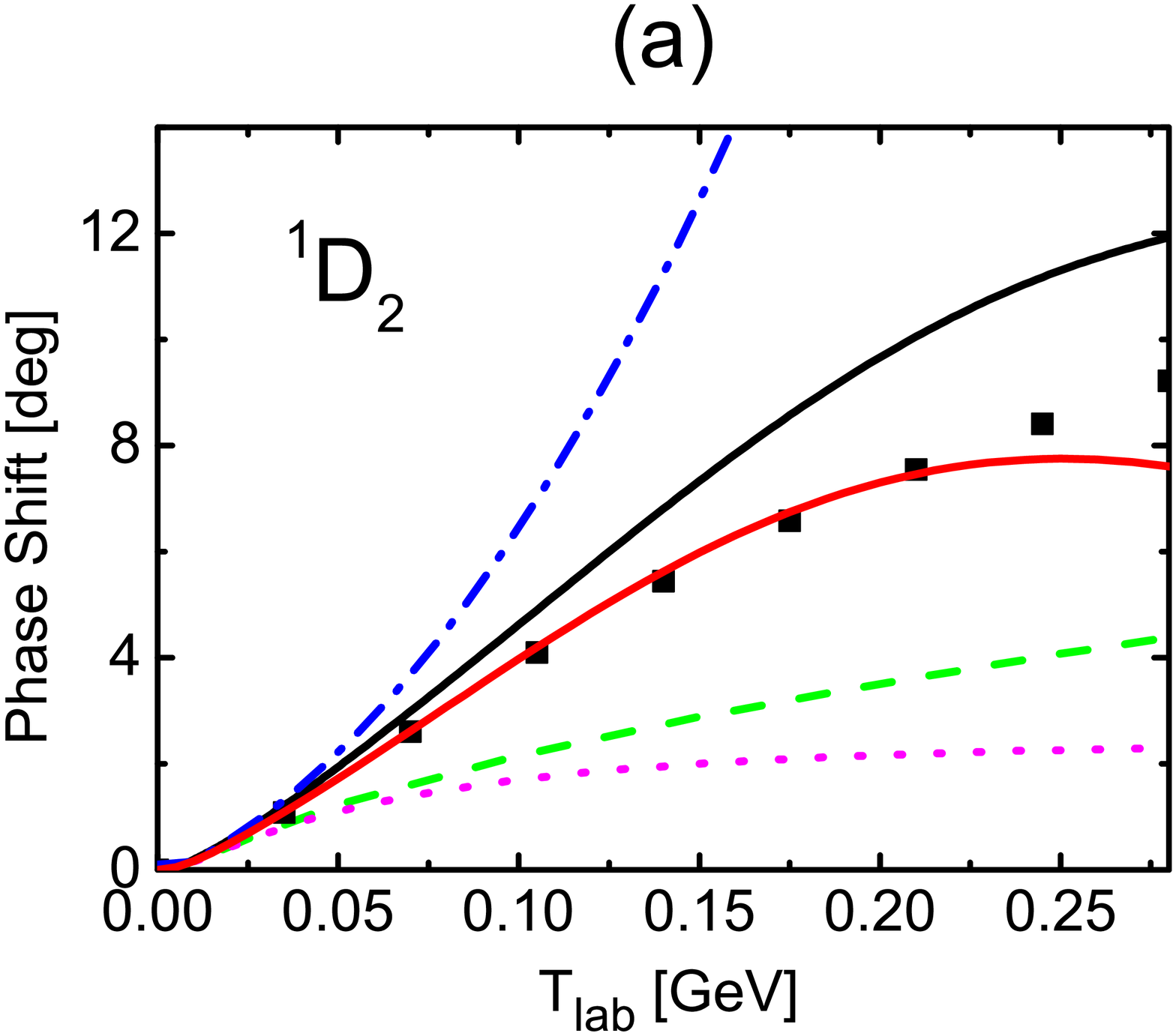}
} \hspace{-7mm}
\subfloat{
\includegraphics[width=0.45\textwidth]{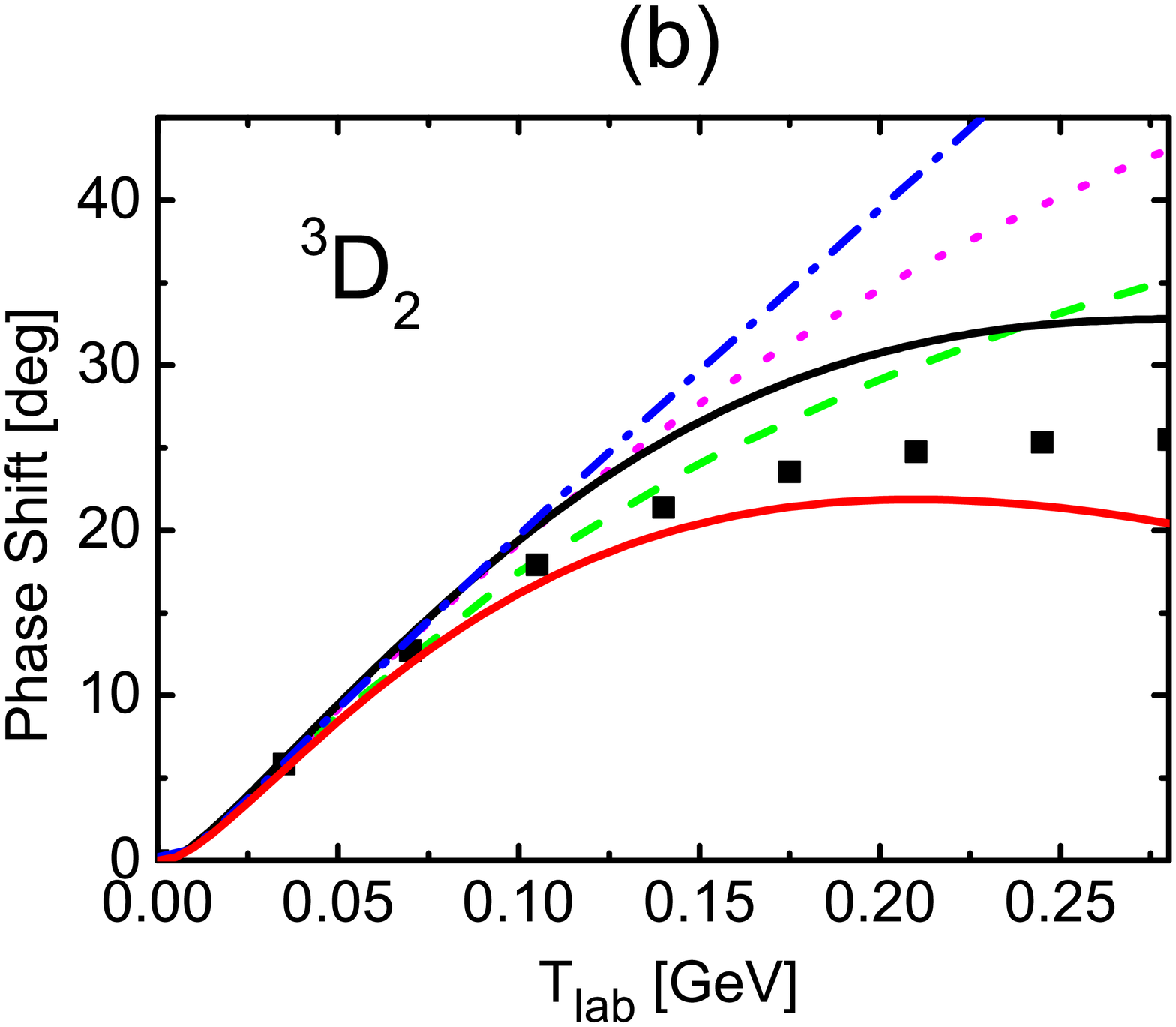}
}
\caption{Same as Fig.~\ref{tb:Dwave}, but for the $^1D_2$ and $^3D_2$ phase shifts.}
\label{tb:singleDwave}
\end{figure}

{To better understand the nonperturbative effects on the uncoupled $D$ waves, i.e.,  $^1D_2$ and $^3D_2$,  we plot in Fig.~\ref{tb:singleDwave} their phase shifts obtained by considering  only OPE and TPE contributions.~\footnote{{In the relativistic framework, the contact contributions to $^1D_2$ and $^3D_2$ up to NNLO have the following form: $V_{\text{1D2}}=C_{\text{1D2}}p^2p'^2/M^4$ and $V_{\text{3D2}}=C_{\text{3D2}}p^2p'^2/M^4$~\cite{Lu:2021gsb}.}} Unlike the perturbative results that increase rapidly in the whole energy region, the NNLO nonperturbative phase shifts seem more reasonable and are closer to their PWA93 counterparts. This indicates that the nonperturbative effects also play a crucial role for these two partial waves. Compared with the nonrelativistic results, the relativistic corrections are large and cause the phase shifts to overshoot the PWA93 phase shifts. However, the large corrections can be absorbed by contact terms  in the relativistic framework as shown in Ref.~\cite{Lu:2021gsb}. (The cutoff dependencies of the OPE and TPE contributions to $^1D_2$ and $^3D_2$ are plotted in Fig.~\ref{tb:cutoff-dependence} in the appendix.) One can see that the phase shifts are also cutoff dependent and therefore contact terms are needed to absorb the cutoff dependence.}\\

\subsection{$F$ wave}
\begin{figure}[htbp]
\centering
\subfloat{
\includegraphics[width=0.45\textwidth]{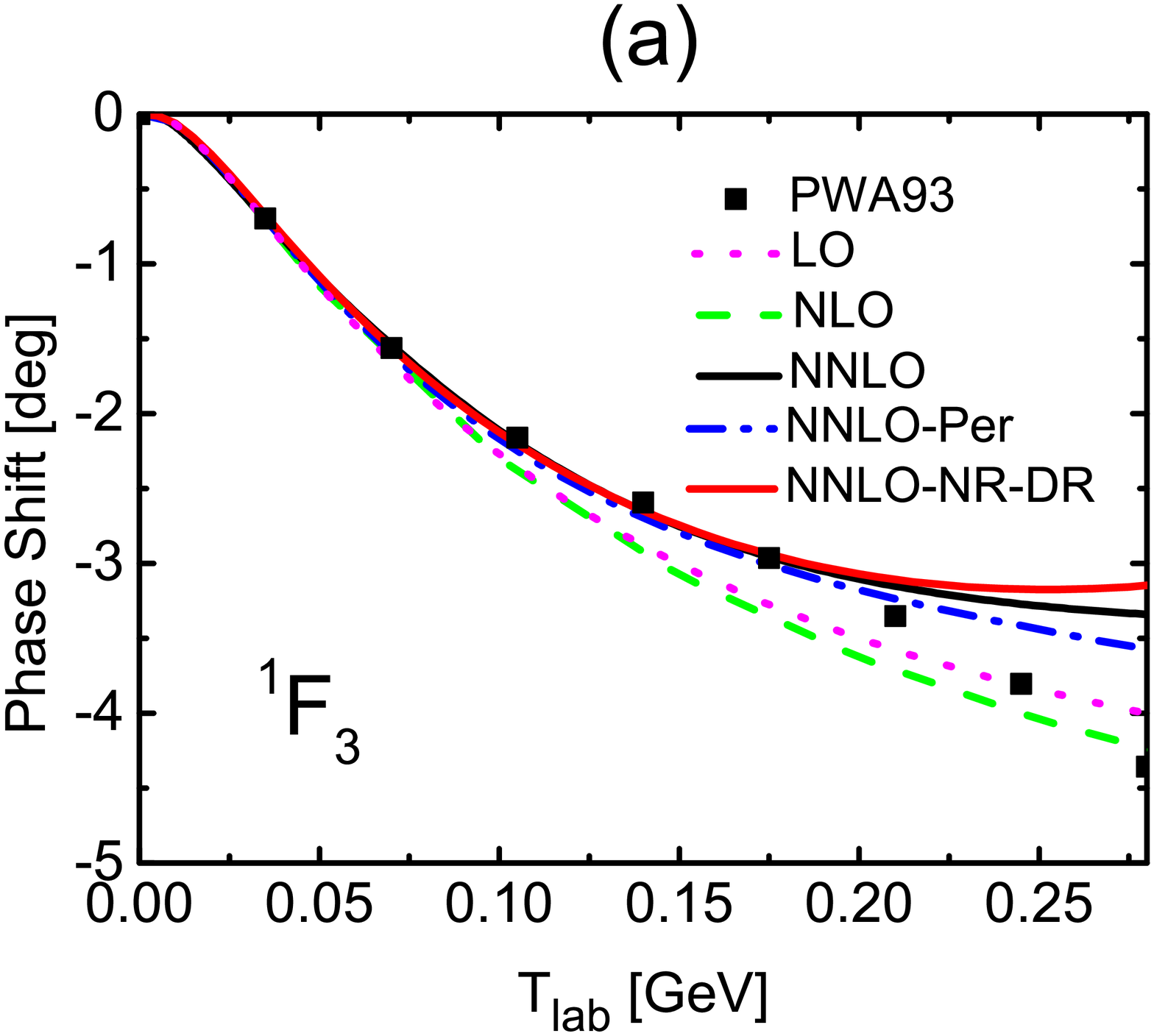}
}\hspace{-7mm}
\subfloat{
\includegraphics[width=0.45\textwidth]{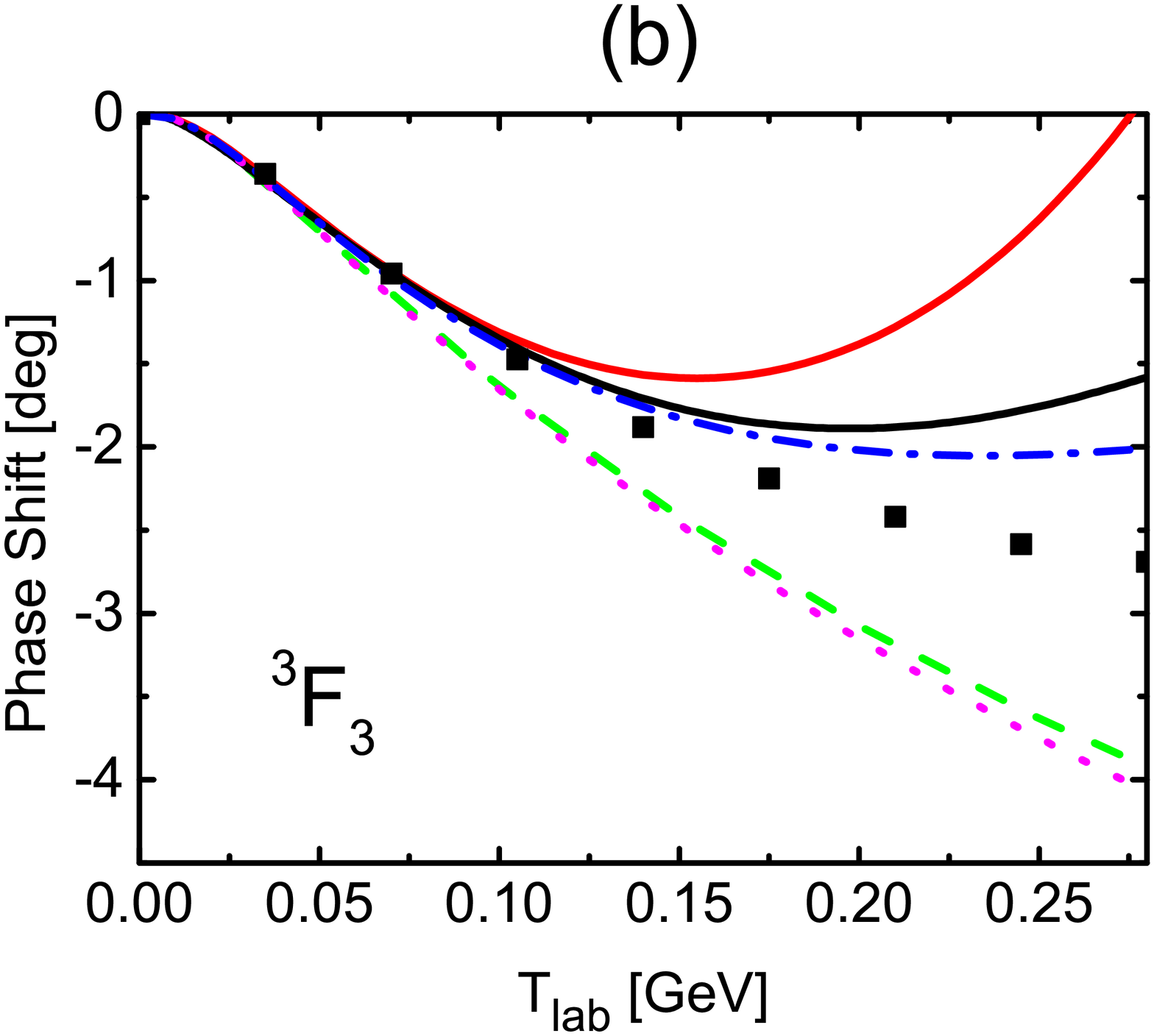}
}\\ \vspace{-2mm}
\subfloat{
\includegraphics[width=0.45\textwidth]{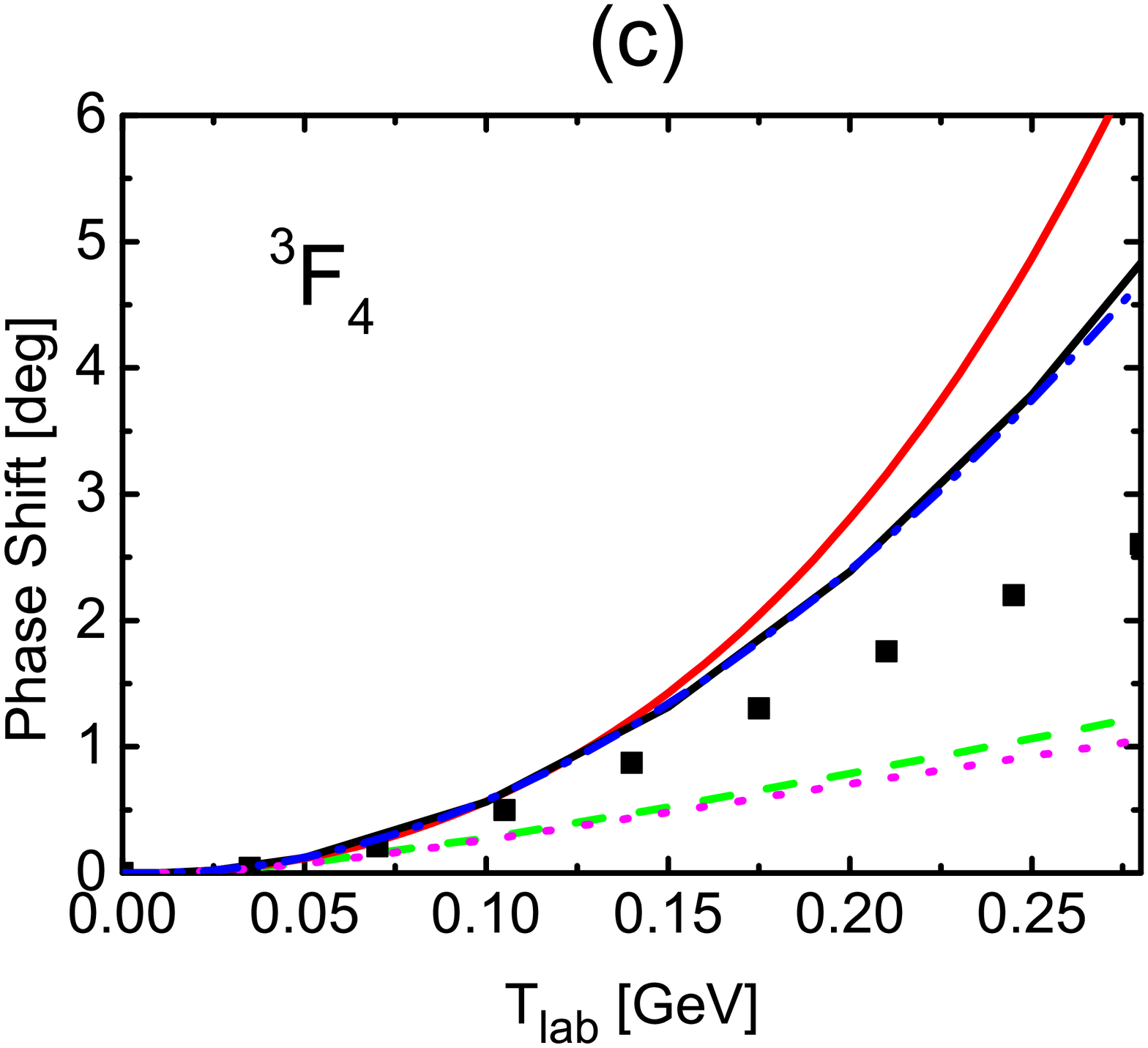}
}\hspace{-7mm}
\subfloat{
\includegraphics[width=0.45\textwidth]{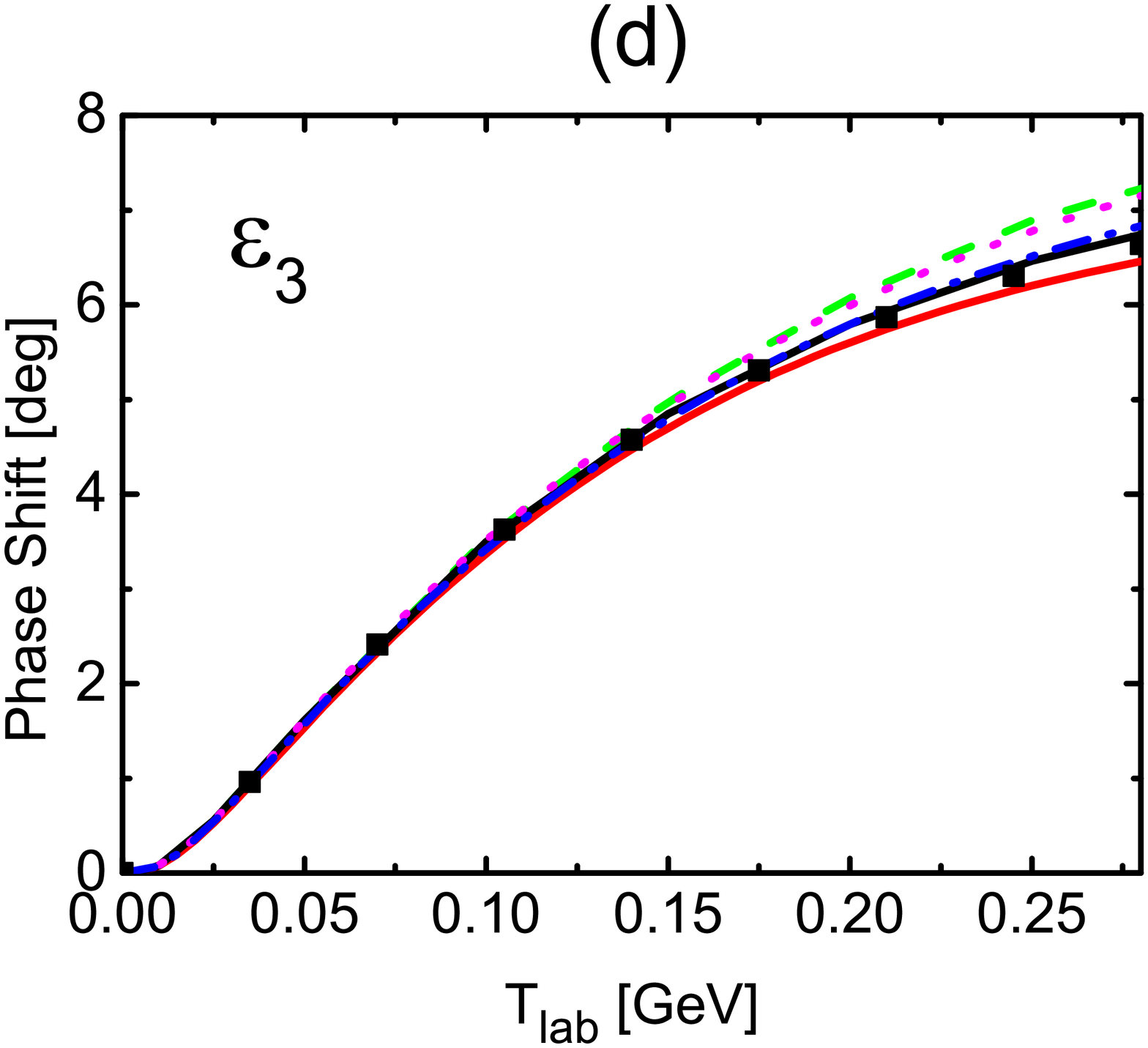}
}
\caption{ Same as Fig.~\ref{tb:Dwave}, but for the $F$-wave phase shifts and mixing angle $\varepsilon_3$.}
\label{tb:Fwave}
\end{figure}

The $F$-wave phase shifts and mixing angle $\epsilon_3$ are depicted in Fig.~\ref{tb:Fwave}. We do not show the phase shifts of $^3F_2$ because in our covariant scheme; this partial wave involves NLO contact terms. In general, the nonperturbative phase shifts for all the $F$ waves are compatible with the perturbative results. For the $^1F_3$, $^3F_3$,  and $^3F_4$ partial waves, the nonperturbative phase shifts are slightly worse than the perturbative results while the nonperturbative treatment slightly improves the description of $\varepsilon_3$. This indicates that for $F$ waves, the nonperturbative effect plays an insignificant role. Compared with the nonrelativistic results, both our covariant NNLO results and NNLO-NR-DR results could describe the PWA93 data for $T_{\textrm{lab}}\leq 100$ MeV equally well. However, in the higher energy region, the relativistic results are  better, especially for the $^3F_3$ and $^3F_4$ partial waves. The relativistic correction plays a positive and non-negligible role for these two partial waves. On the other hand, the larger differences between the NNLO and NLO results (than those between the NLO and LO results) indicate that these two partial waves receive  large subleading TPE contributions originated from the $c_3$ and $c_4$ terms.~\footnote{For the $^3F_3$ and $^3F_4$ partial waves, the NNLO nonrelativistic nonperturbative results obtained with the spectral function regularization (SFR) scheme describe the data better, because the SFR scheme suppresses the strong short-range contribution from subleading TPE~\cite{Epelbaum:2003gr}. However, the SFR scheme worsens the description of some other channels, such as $^3G_5$. }  
%Notice that the large numerical values of $c_3$ can be partially explained by the fact that the LECs $c_3$ are to a large extent saturated by the $\Delta$–excitation. This means a new smaller scale, $m_{\Delta}\sim293$ MeV, enters the values of these constants in EFT without explicit $\Delta$ and influence the convergence of the two channels. In principal, this can be cured by higher orders contact interactions%
The mixing angle $\varepsilon_3$ is  well described up to $T_{\textrm{lab}} = 250$ MeV in all the calculations. Both the higher order contributions and relativistic corrections do not have a visible effect on this mixing angle. Actually the OPE contribution alone can already describe the mixing angle well.

The $F$-wave phase shifts  and mixing angle $\epsilon_3$ obtained with different cutoffs are plotted in Fig.~\ref{tb:cutoff-dependence} in the appendix. We can see that the results obtained with cutoff values of 800, 900, and 1000 MeV  almost overlap with each other, which implies a quite weak cutoff dependence.
%For the subleading TPE, the strong attraction can be related to the very strong short-range components since the phase shifts changes thirty five percent with the cutoff in pion loop varys from 500 MeV to 800 MeV~\cite{Epelbaum:2003gr}

\subsection{$G$ wave and higher partial waves}

\begin{figure}[htbp]
\centering
\subfloat{
\includegraphics[width=0.45\textwidth]{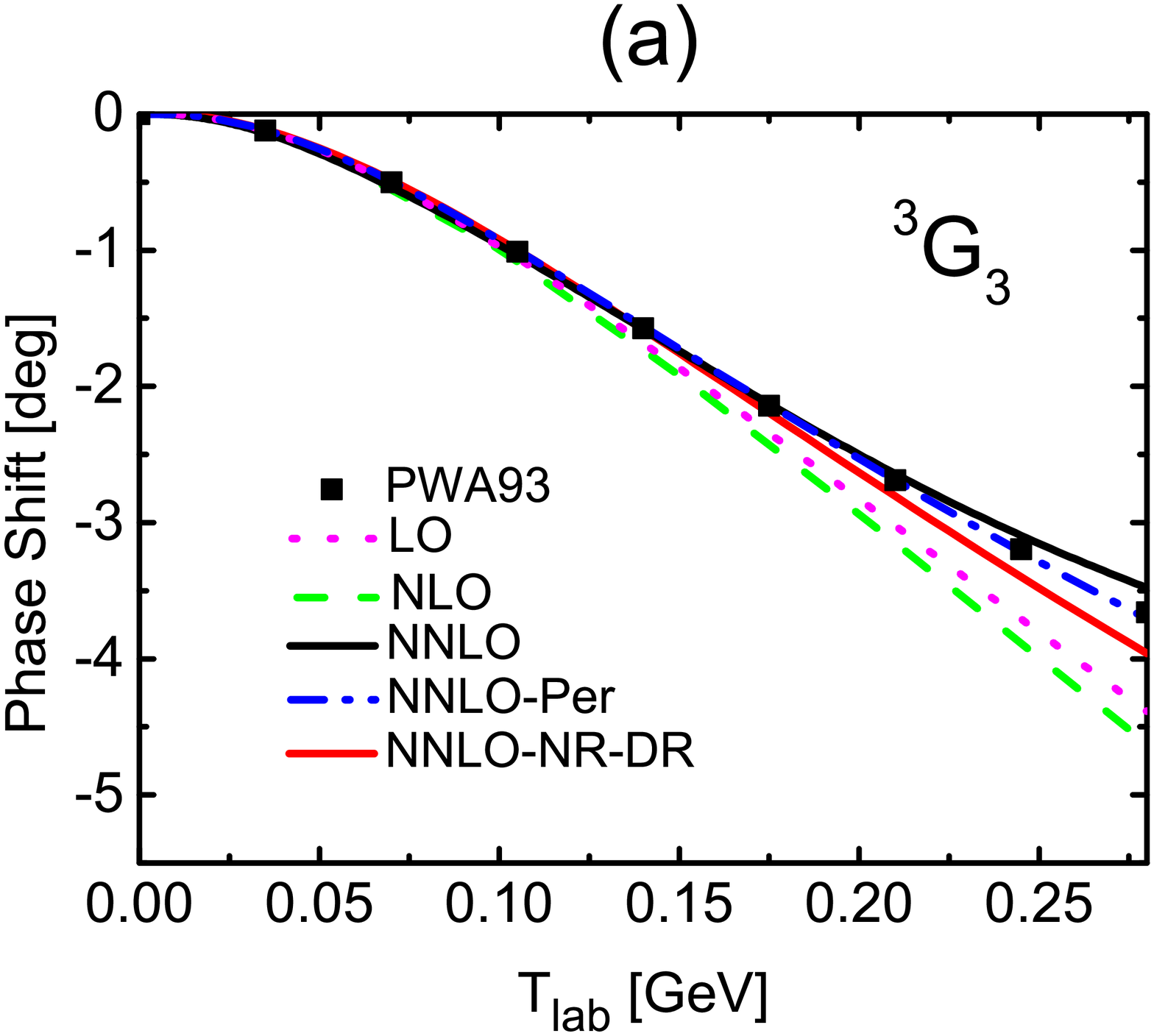}
}\hspace{-7mm}
\subfloat{
\includegraphics[width=0.45\textwidth]{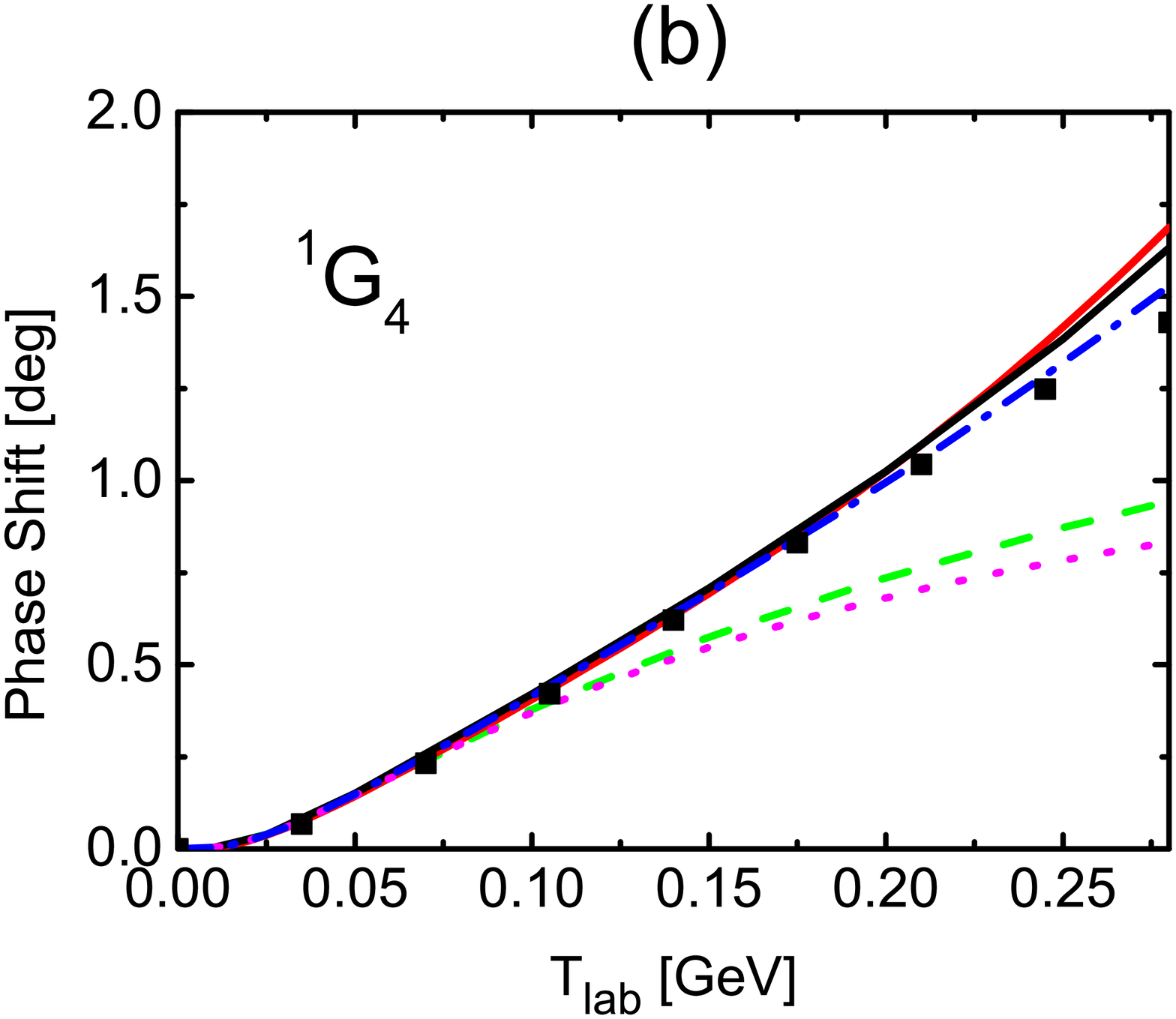}
}\\ \vspace{-2mm}
\subfloat{
\includegraphics[width=0.45\textwidth]{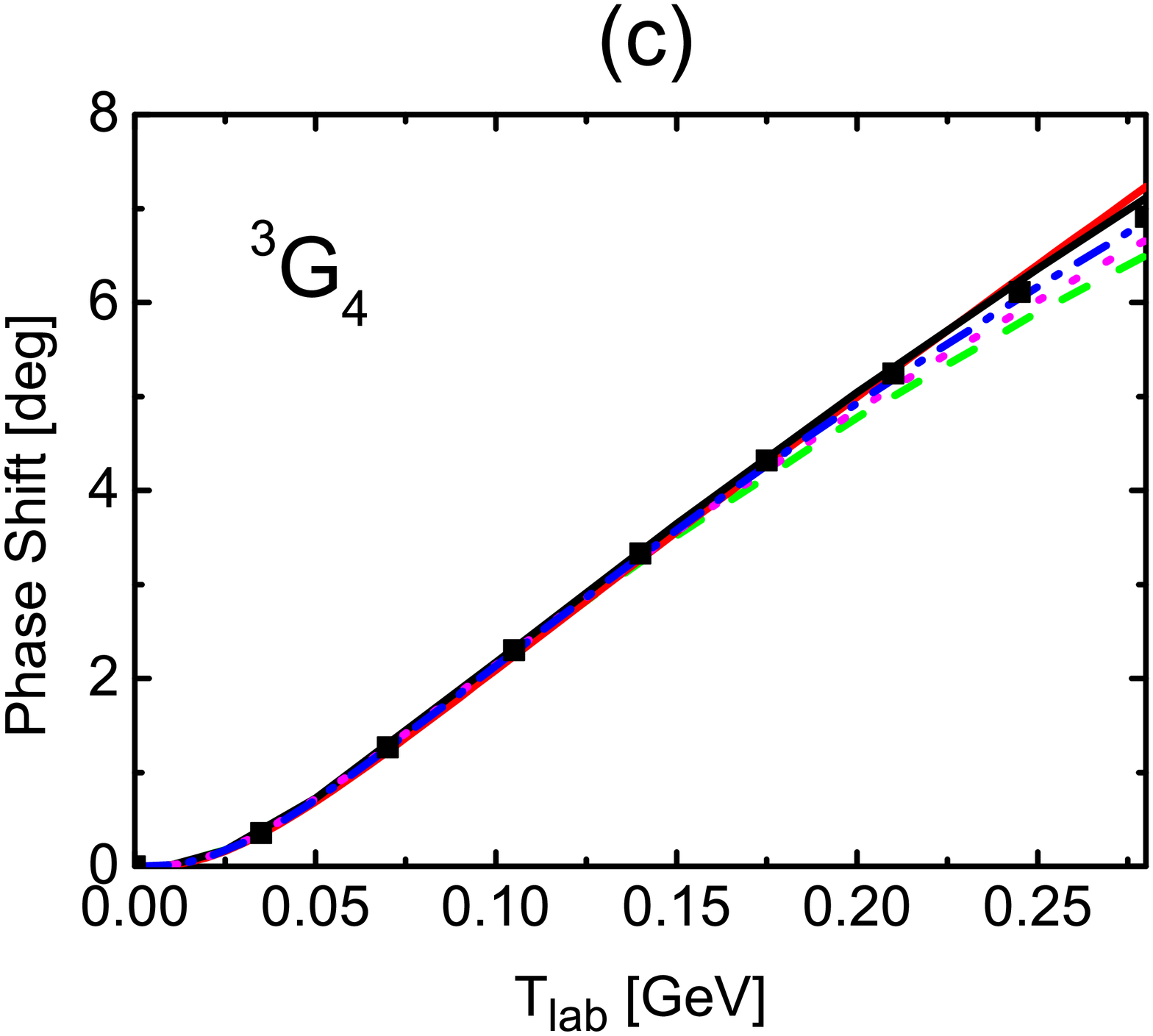}
}\hspace{-7mm}
\subfloat{
\includegraphics[width=0.45\textwidth]{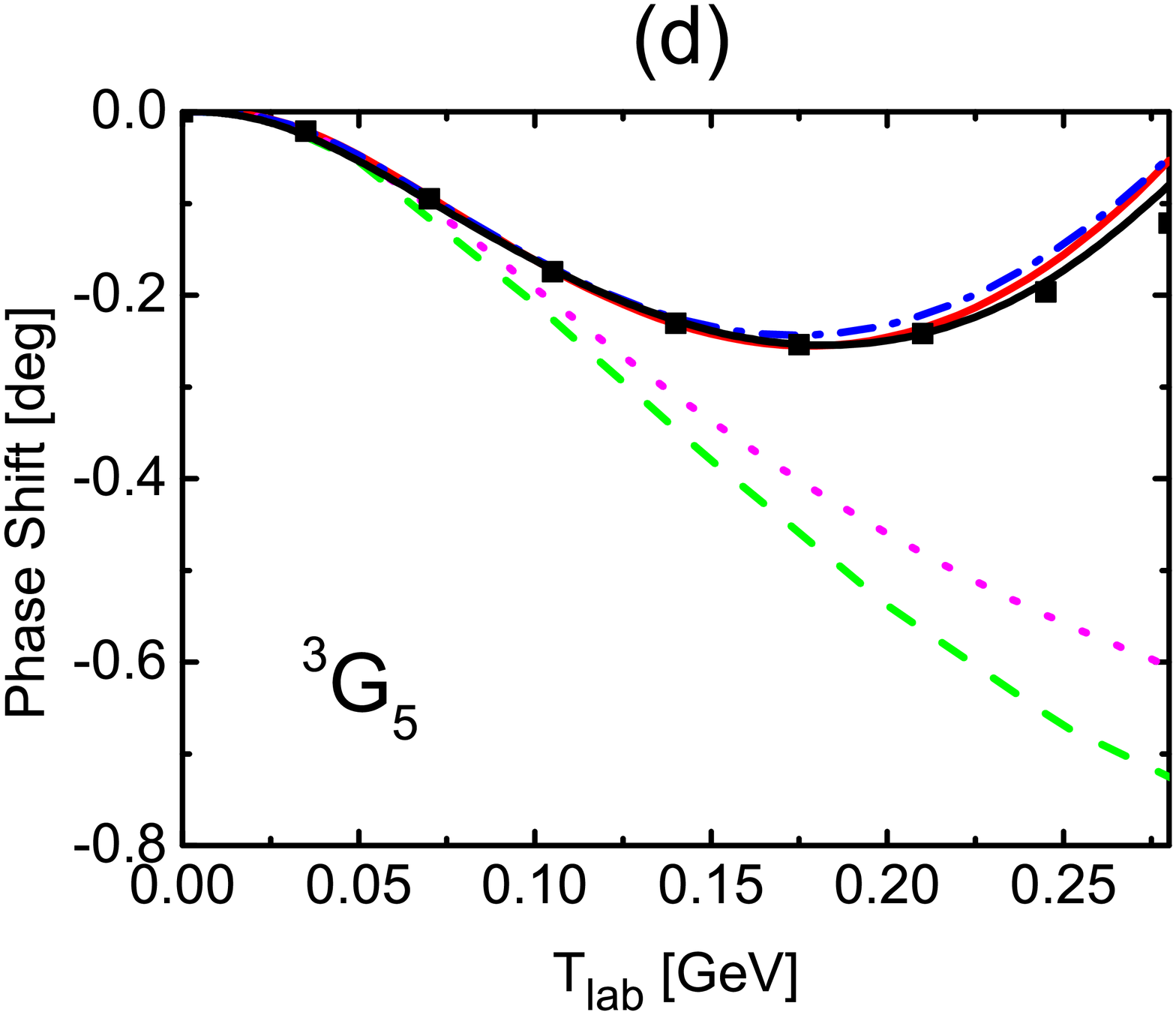}
}\\ \vspace{-2mm}
\subfloat{
\includegraphics[width=0.45\textwidth]{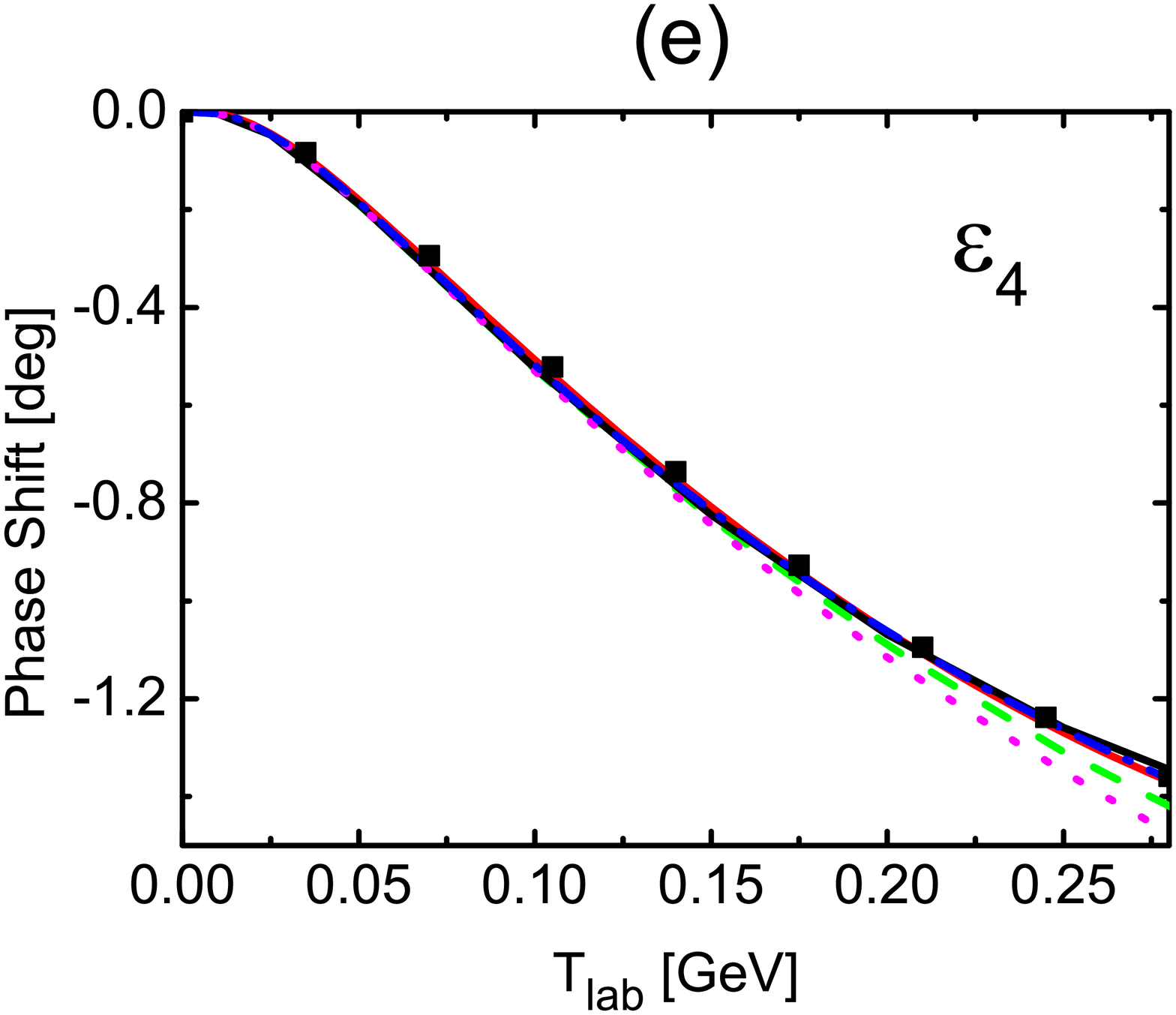}
}
\caption{Same as Fig.~\ref{tb:Dwave}, but for the $G$-wave phase shifts and mixing angle $\epsilon_4$.}
\label{tb:Gwave}
\end{figure}

\begin{figure}[htbp]
\centering
\subfloat{
\includegraphics[width=0.45\textwidth]{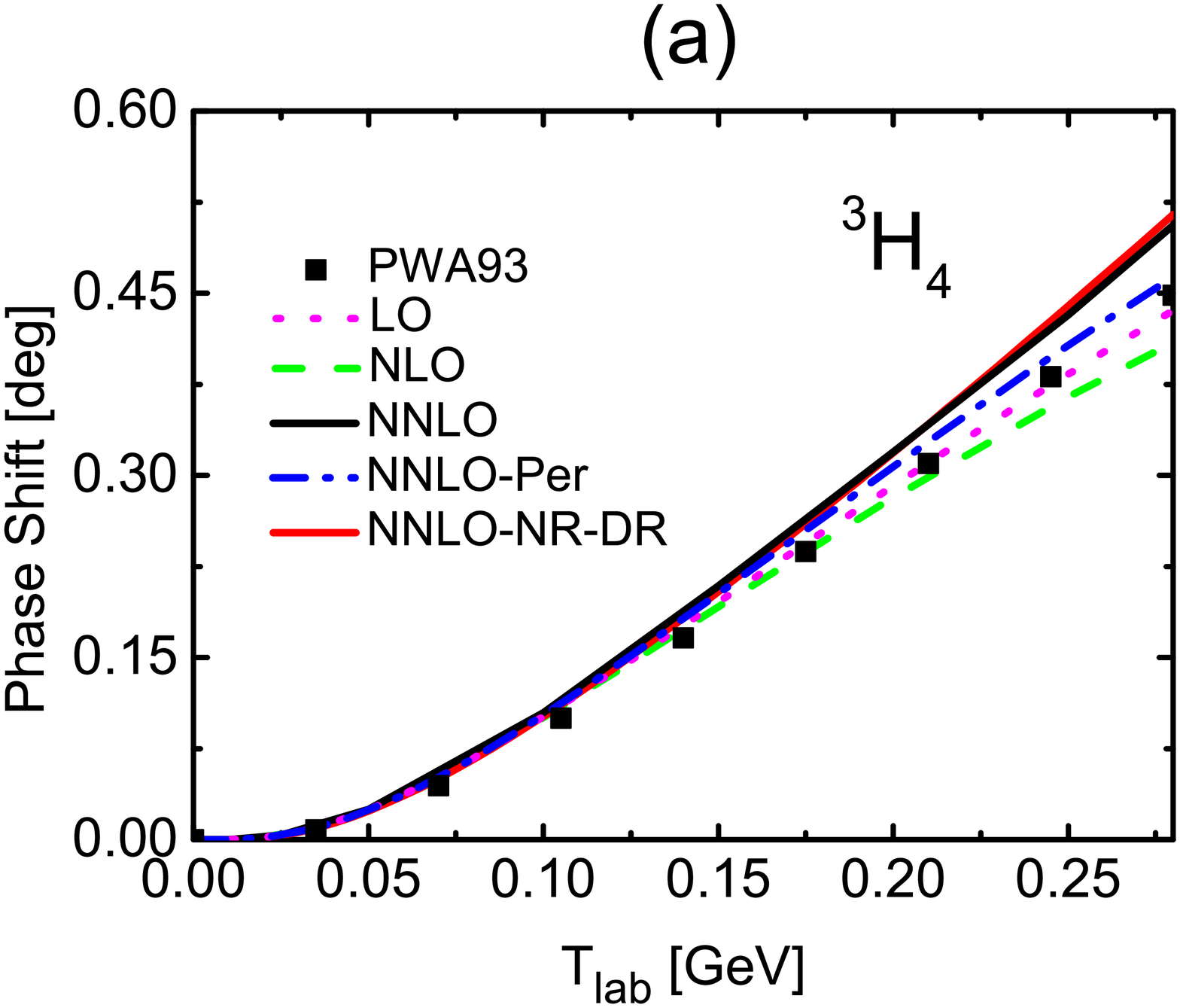}
}\hspace{-7mm}
\subfloat{
\includegraphics[width=0.45\textwidth]{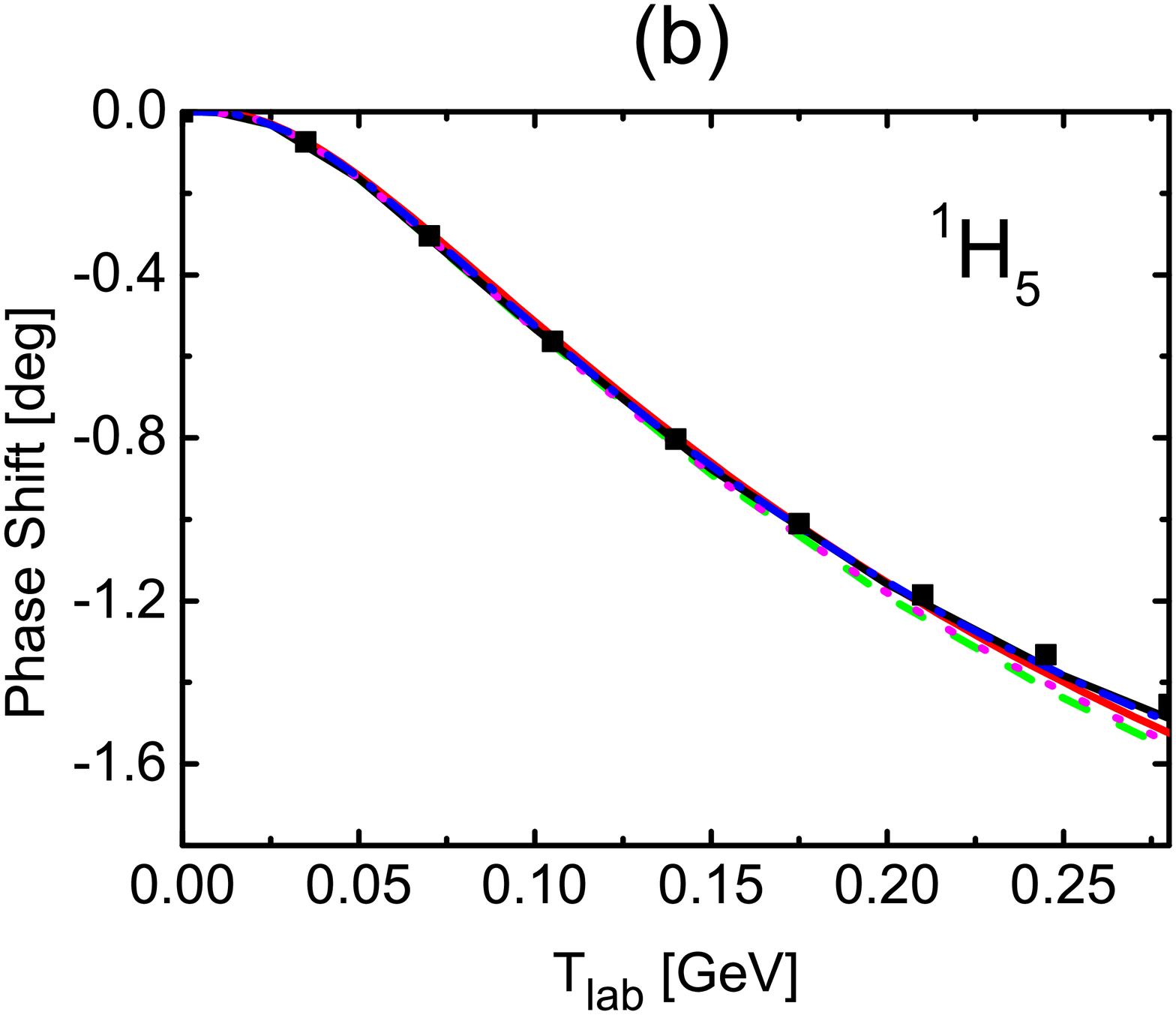}
}\\ \vspace{-2mm}
\subfloat{
\includegraphics[width=0.45\textwidth]{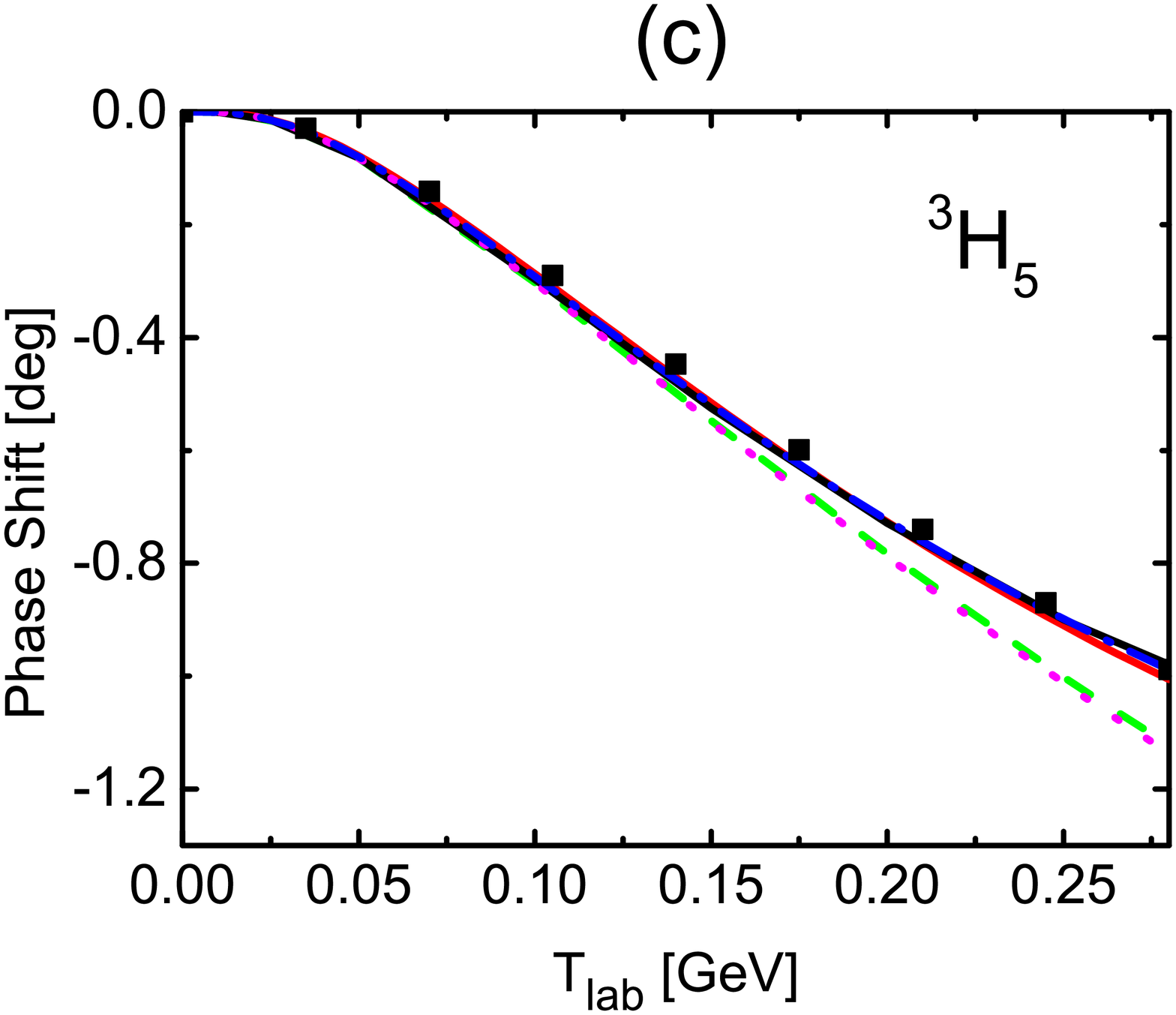}
}\hspace{-7mm}
\subfloat{
\includegraphics[width=0.45\textwidth]{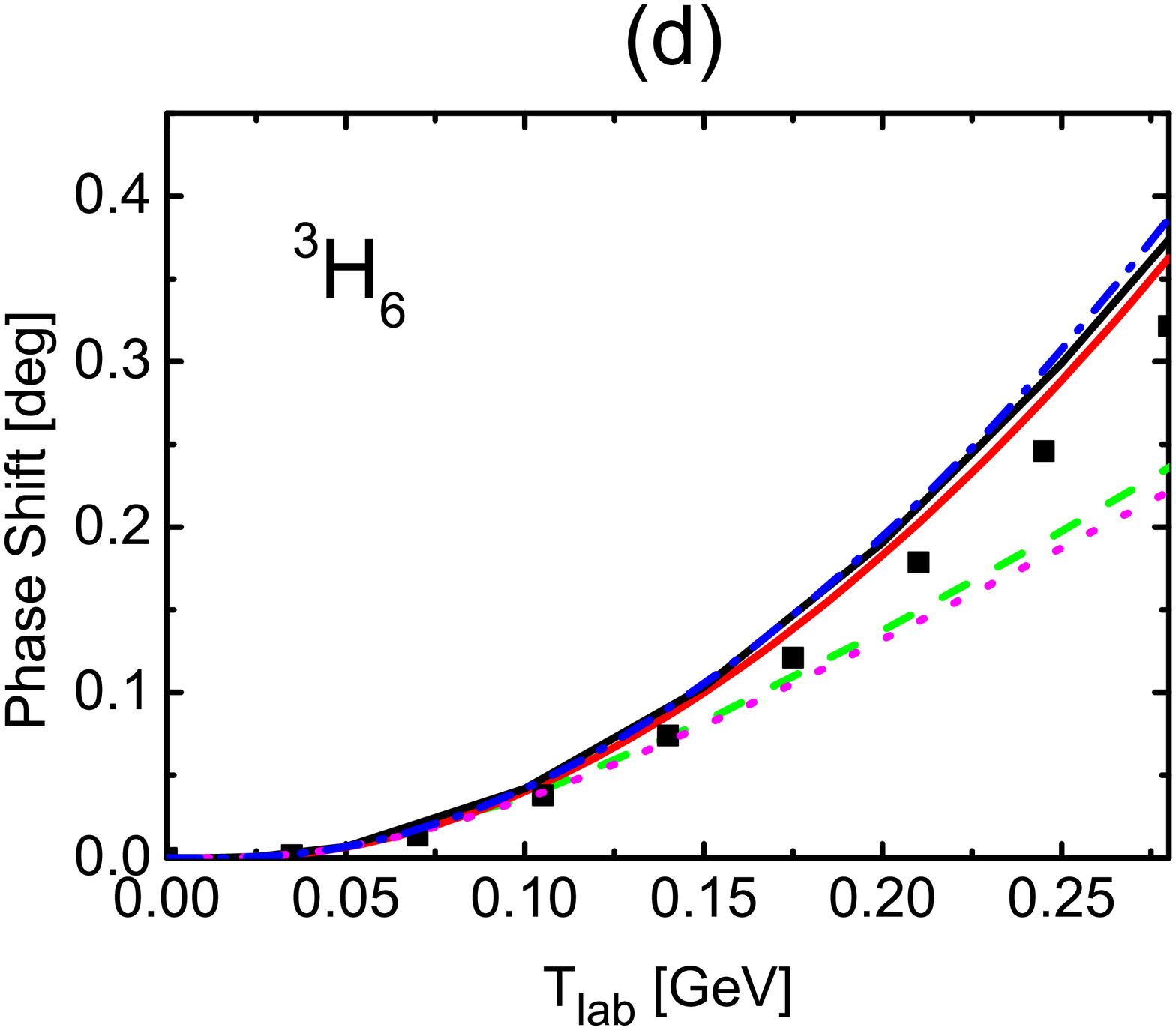}
}\\ \vspace{-2mm}
\subfloat{
\includegraphics[width=0.45\textwidth]{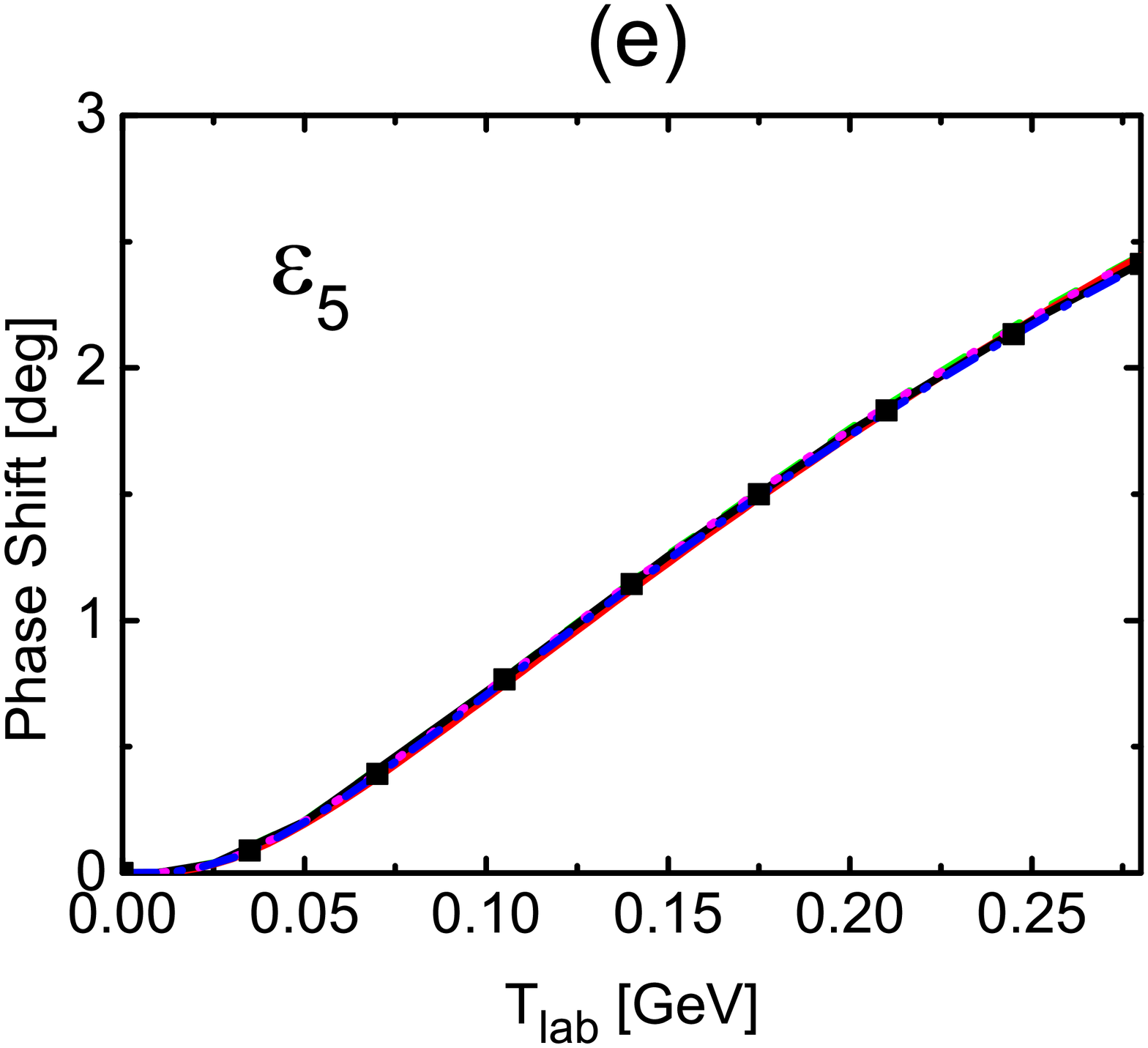}
}
\caption{Same as Fig.~\ref{tb:Dwave}, but for the $H$-wave phase shifts and mixing angle $\epsilon_5$.}
\label{tb:Hwave}
\end{figure}

\begin{figure}[htbp]
\centering
\subfloat{
\includegraphics[width=0.45\textwidth]{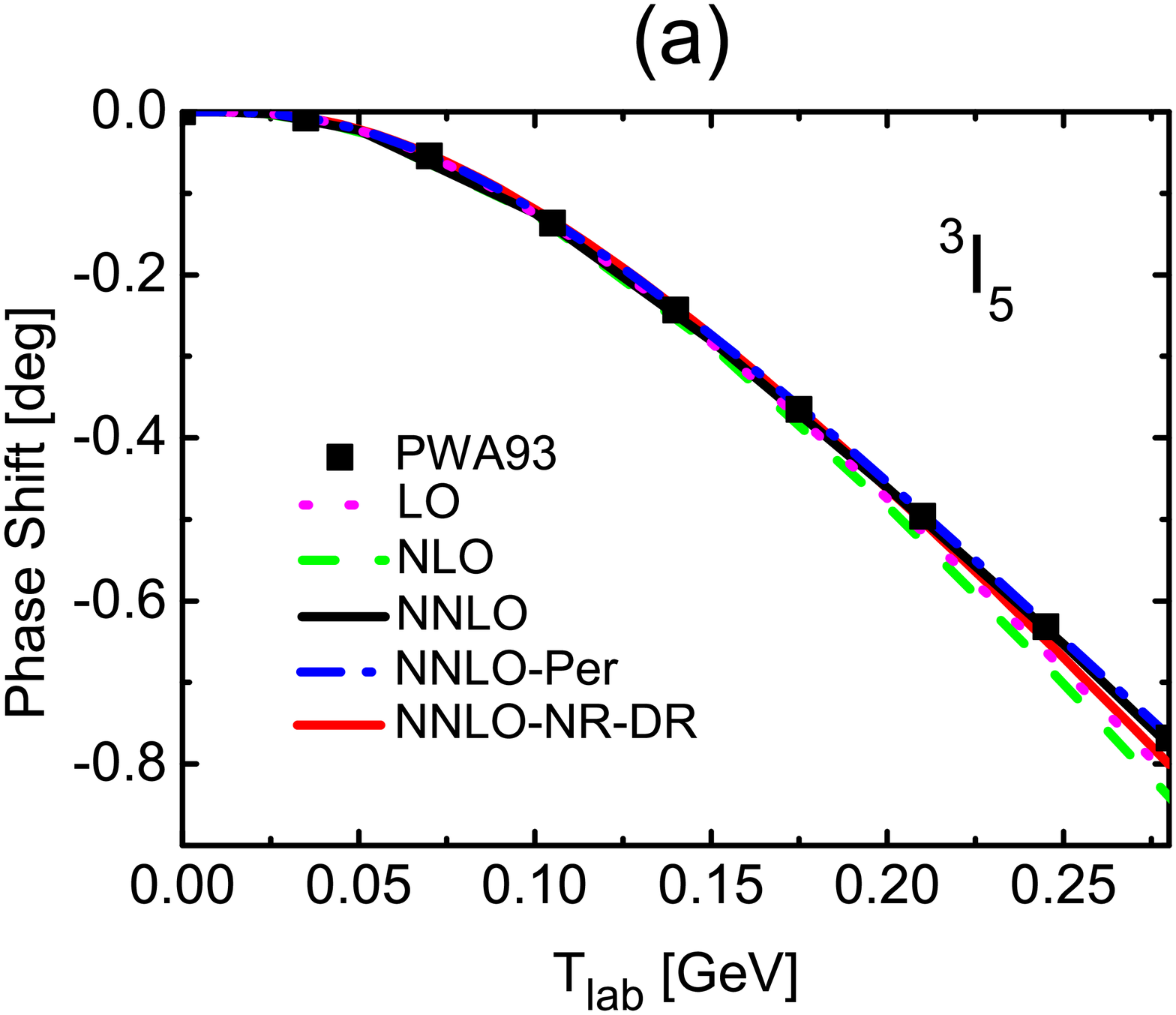}
}\hspace{-7mm}
\subfloat{
\includegraphics[width=0.45\textwidth]{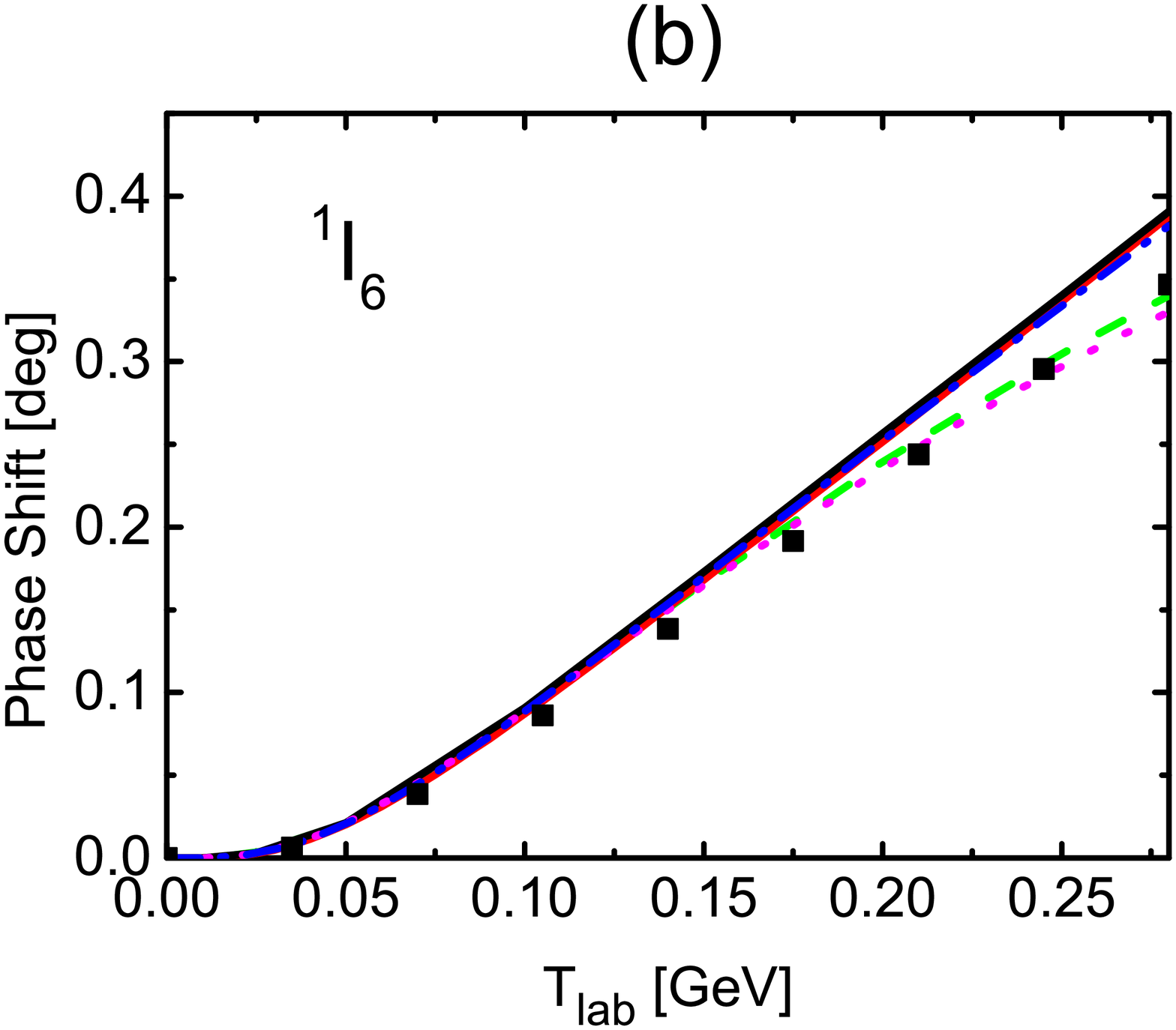}
}\\ \vspace{-2mm}
\subfloat{
\includegraphics[width=0.45\textwidth]{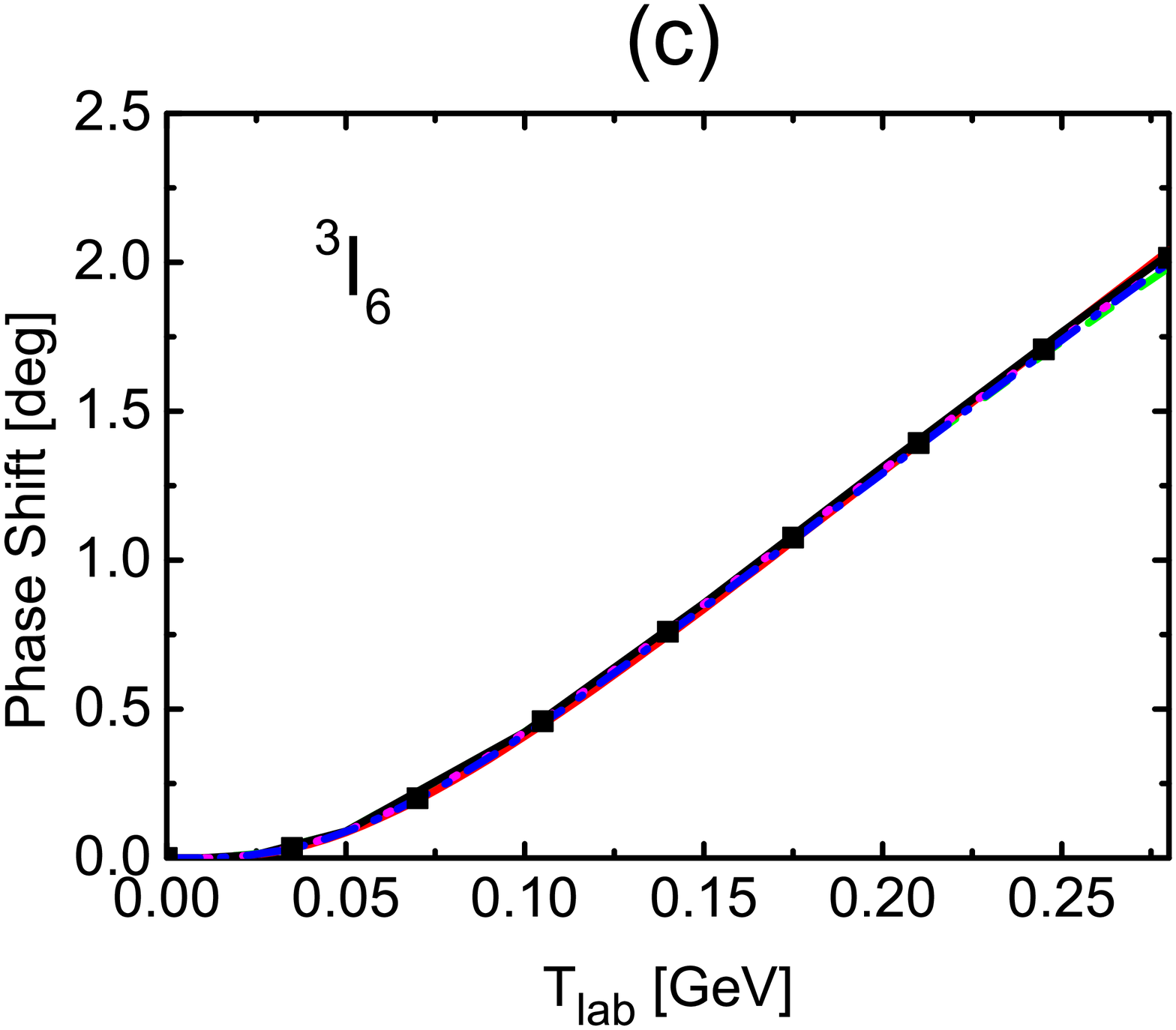}
}\hspace{-7mm}
\subfloat{
\includegraphics[width=0.45\textwidth]{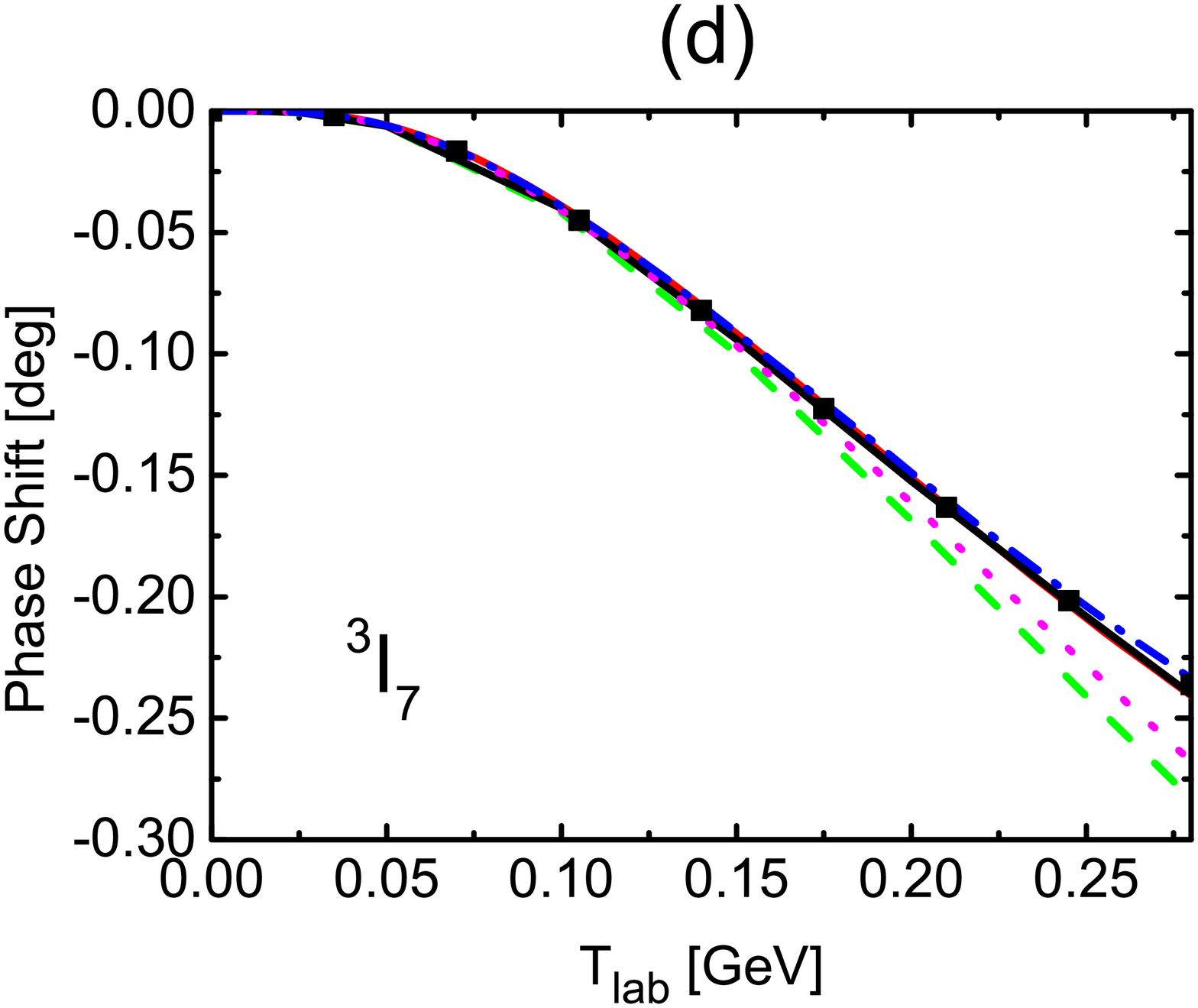}
}\\ \vspace{-2mm}
\subfloat{
\includegraphics[width=0.45\textwidth]{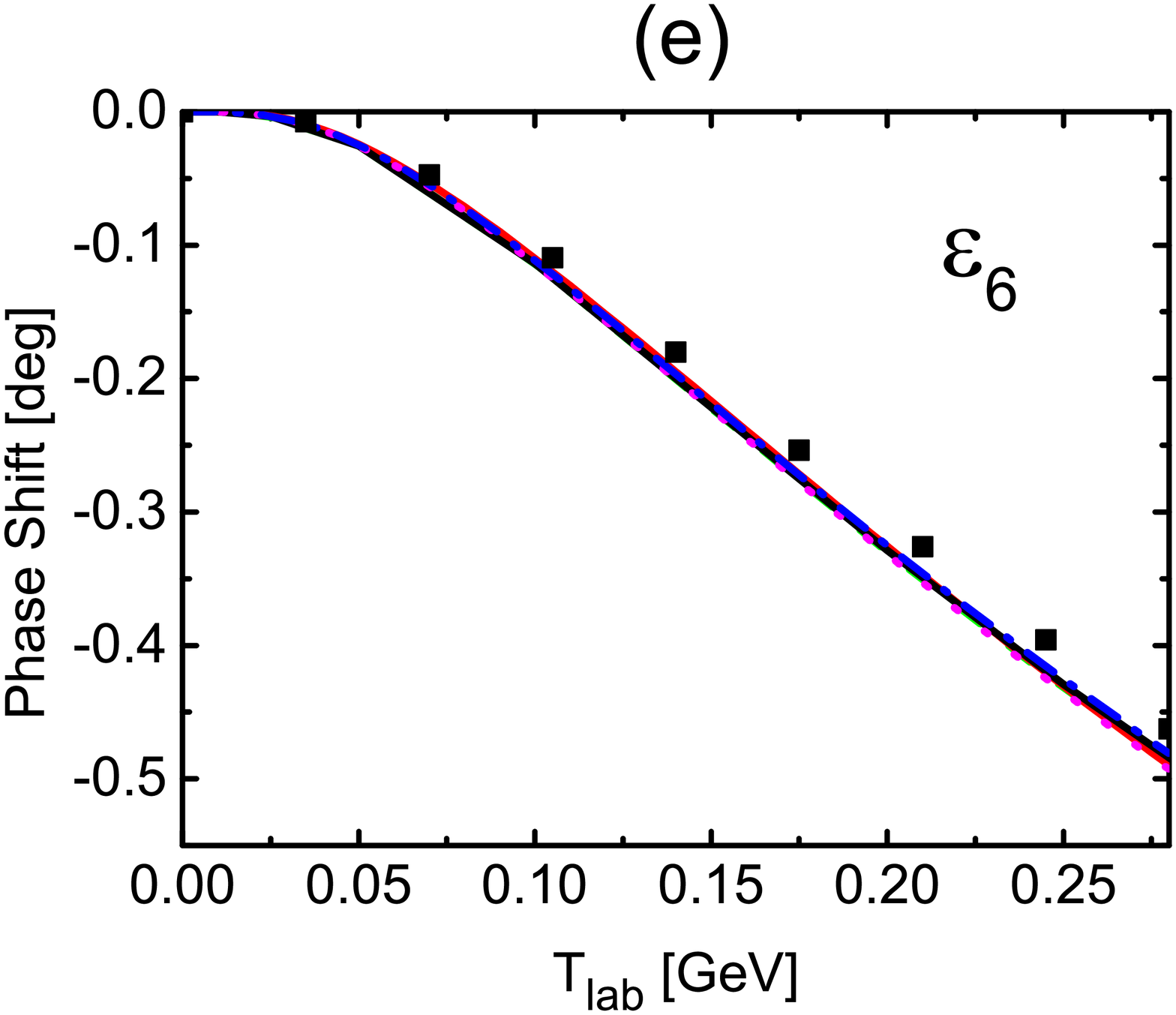}
}
\caption{Same as Fig.~\ref{tb:Dwave}, but for the $I$-wave phase shifts and mixing angle $\epsilon_6$.}
\label{tb:Iwave}
\end{figure}

The $G$-, $H$-, and $I$-wave phase shifts and corresponding mixing angles $\epsilon_4$, $\epsilon_5$, and $\epsilon_6$ are depicted in Figs.~\ref{tb:Gwave},~\ref{tb:Hwave}, and~\ref{tb:Iwave}, respectively.
For partial waves with $L \geq 4$, almost all the NNLO results can describe the PWA93 phase shifts up to $T_{\text{lab}}=200$ MeV equally well. This indicates that neither the nonperturbative effect nor the relativistic effect contributes significantly at NNLO. 

For the $^3{G}_3,$ $^1{G}_4,$ and $^3{G}_4$ partial waves, the relativistic nonperturbative results are slightly worse than the perturbative results while they are better for $^3G_5$. Compared with the nonrelativistic results, the relativistic NNLO  phase shifts are slightly better for $^3{G}_3,$ $^1{G}_4,$ and $^3{G}_5$, while for $^3{G}_4$ and $\varepsilon_4$, the two results almost overlap. For $^3{G}_3,$ $^3{G}_4,$ and $\varepsilon_4$, the OPE alone does already a fairly good job, while for $^1{G}_4$ and $^3{G}_5$, we find a visibly improved description from LO to NNLO, indicating that the subleading TPE is still important for these two partial waves.

For the $H$ wave, the perturbative and nonperturbative $^1{H}_5$ and $^3{H}_5$ phase shifts and mixing angle $\varepsilon_5$ are indistinguishable. The perturbative results are slightly better for $^3{H}_4$ while the nonperturbative results are slightly better for $^3{H}_6$. For the $^3{H}_4,$ $^1{H}_5,$ and $^3{H}_5$ partial waves, the relativistic results are slightly better than the nonrelativistic results. For $\varepsilon_5$, the  phase shifts from different potentials are all very close to the PWA93 results up to $T_{\textrm{lab}} = 280 $ MeV. Only for $^3{H}_6$ does the contribution of the subleading TPE seem to be a bit large for $T_{\text{lab}}\geq150$ MeV. The behavior of the partial waves is quite similar to that of $^3{F}_4$ to which also a strong subleading TPE contributes. 

The $I$ waves are almost in the same situation as the $H$ waves. The nonperturbative phase shifts are  identical to the perturbative phase shifts and nonrelativistic phase shifts, all of which are in good agreement with the PWA93 data. For $^1I_6$, the TPE contribution slightly worsens the description while for $^1I_7$ it slightly improves the description.
For all the $H$- and $I$-wave phase shifts, the OPE potential alone can describe the data reasonably well, which confirm the expectation that for  higher partial waves, the TPE contribution is insignificant.

\iffalse
\begin{table}
\begin{spacing}{1.8}
\begin{center}
\caption{$\tilde{\chi}=\sum\limits_{i=1}^{8}\big{(}\delta(i)-\delta_{\textrm{PWA}}(i)\big{)}^2$ at Tlab(i)=\{1, 5, 10, 25, 50, 100, 150, 200\}MeV for all $3 \leq J $ $\&$  $ L \leq 6$ partial waves}
\setlength{\tabcolsep}{5mm}{
\begin{tabular}{cccccc} \toprule
\hline
\hline
 & \textrm{LO} & \textrm{NLO} & \textrm{NNLO} & \textrm{NNLO-Per} & \textrm{NNLO-NR2000} \\
\midrule[1pt]
  $\tilde{\chi}_{D}$ & $7.69$  & $15.8$ & $2.58\times 10^{-2}$  & $341$  & $1.30$ \\
  $\tilde{\chi}_{F}$ & $2.37$  & $2.23$ & $9.47\times 10^{-1}$  & $9.36\times 10^{-1}$  & $2.79$ \\
  $\tilde{\chi}_{G}$ & $3.00\times 10^{-1}$  & $4.38\times 10^{-1}$ & $1.10\times 10^{-2}$  & $6.19\times 10^{-3}$  & $2.18\times 10^{-2}$ \\
  $\tilde{\chi}_{H}$ & $1.62\times 10^{-2}$  & $1.60\times 10^{-2}$ & $6.54\times 10^{-3}$  & $6.53\times 10^{-3}$  & $5.10\times 10^{-3}$ \\
  $\tilde{\chi}_{I}$ & $2.07\times 10^{-3}$  & $2.75\times 10^{-3}$ & $2.67\times 10^{-3}$  & $2.58\times 10^{-3}$  & $1.97\times 10^{-3}$ \\
 \hline
\hline
\bottomrule
\end{tabular}}
\end{center}
\end{spacing}{1.0}
\end{table}
\fi

\subsection{Quantitative comparisons}
For the purpose of a more quantitative comparison of different theoretical results, in this subsection, we calculate the $\chi^2$-like function $\tilde{\chi}^2$ defined as
\begin{align}\label{Eq:chi2}
\tilde{\chi}^2=\sum\limits_{i=1}^{9}\big{(}\delta(i)-\delta_{\textrm{PWA}}(i)\big{)}^2
\end{align}
at $T_{\textrm{lab}}(i)$ = \{1, 5, 10, 25, 50, 100, 150, 200, 250\} MeV for the LO, NLO, and NNLO relativistic nonperturbative TPE as well as the NNLO-Per and  NNLO-NR-DR results, which are given in Table~\ref{tb:chisquare} in which $\tilde{\chi}^2_D$ only includes the contribution from $^3{D}_3$; $\tilde{\chi}^2_F$ includes those from $^1{F}_3$, $^3{F}_3$, $^3{F}_4$ and $\epsilon_3$; $\tilde{\chi}^2_G$ contains the contributions from all the $G$ waves and mixing angle $\epsilon_4$; $\tilde{\chi}^2_H$ and $\tilde{\chi}^2_I$ are the same as for the $G$ waves. Compared with the relativistic perturbative results, the relativistic nonperturbative results have a much smaller $\tilde{\chi}^2_{D}$ at NNLO but slightly larger $\tilde{\chi}^2_F$ and $\tilde{\chi}^2_G$. Compared with the nonrelativistic results,  our covariant NNLO results lead to a much smaller $\tilde{\chi}^2$ for $D$ wave, $F$ waves, and $G$ waves. For higher partial waves, since the nonperturbative effect and the relativistic effect do not make much difference, $\tilde{\chi}^2_H$ and $\tilde{\chi}^2_I$ are almost the same for the covariant NNLO, NNLO-Per, and  NNLO-NR-DR results. Furthermore, $\tilde{\chi}^2_H$ and $\tilde{\chi}^2_I$ at LO are relatively small and are already comparable with those at NNLO, which indicate the OPE contributions alone can describe the phase shifts well enough. 
\begin{table}
\begin{spacing}{1.8}
\begin{center}
\caption{$\tilde{\chi}^2$ defined in Eq.~(\ref{Eq:chi2}) for all $J \geq 3 $ and $ L \leq 6$ partial waves.}
\label{tb:chisquare}
\setlength{\tabcolsep}{5mm}{
\begin{tabular}{cccccc} \toprule
\hline
\hline
 & \textrm{LO} & \textrm{NLO} & \textrm{NNLO} & \textrm{NNLO-Per} & \textrm{NNLO-NR-DR} \\
\midrule[1pt]
  $\tilde{\chi}_{D}^2$ & $10.9$  & $22.2$ & $0.0129$  & $10.1$  & $2.38$ \\
  $\tilde{\chi}_{F}^2$ & $5.65$  & $5.03$ & $4.28$  & $3.73$  & $14.0$ \\
  $\tilde{\chi}_{G}^2$ & $1.03$  & $1.48$ & $0.0461$  & $0.0144$  & $0.117$ \\
  $\tilde{\chi}_{H}^2$ & $0.0421$  & $0.0420$ & $0.0114$  & $0.0116$  & $0.0116$ \\
  $\tilde{\chi}_{I}^2$ & $0.00458$  & $0.00750$ & $0.00476$  & $0.00449$  & $0.00416$ \\
 \hline
\hline
\bottomrule
\end{tabular}}
\end{center}
\end{spacing}
\end{table}

\subsection{One-pion-exchange in Born approximation}

\begin{figure}[htbp]
\centering
\subfloat{
\includegraphics[width=0.45\textwidth]{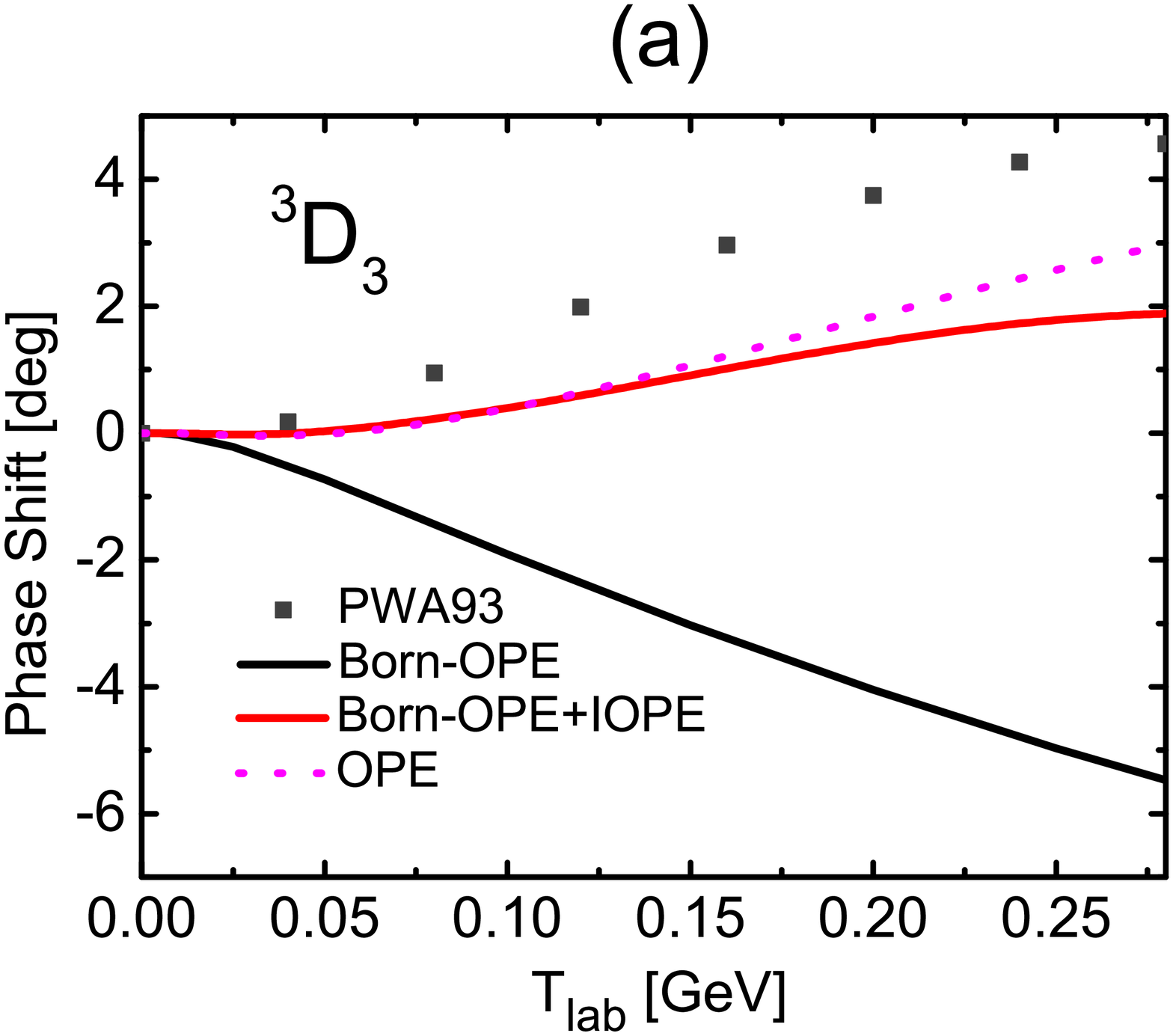}
}\hspace{-7mm}
\subfloat{
\includegraphics[width=0.45\textwidth]{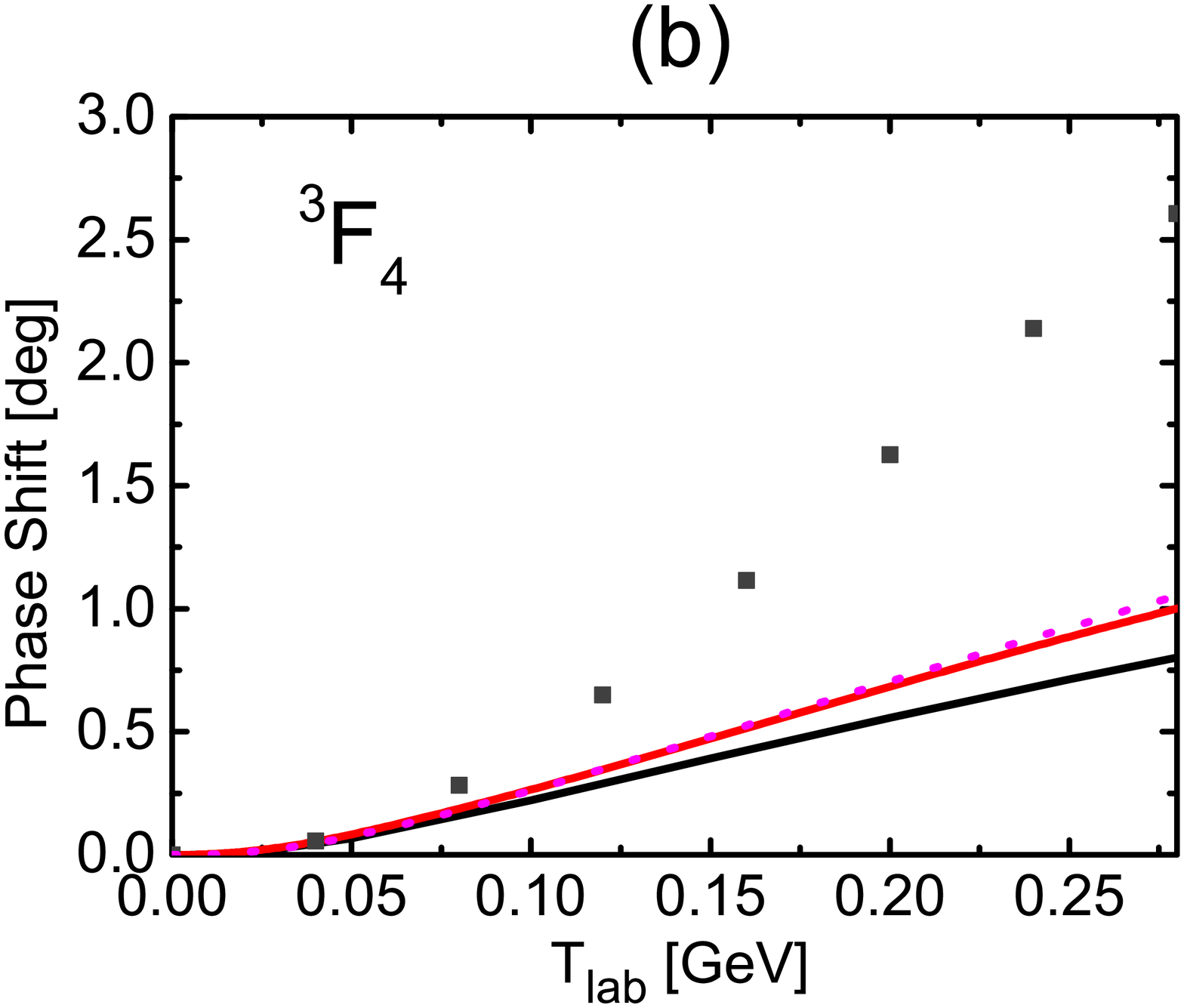}
}\\ \vspace{-6mm}
\subfloat{
\includegraphics[width=0.45\textwidth]{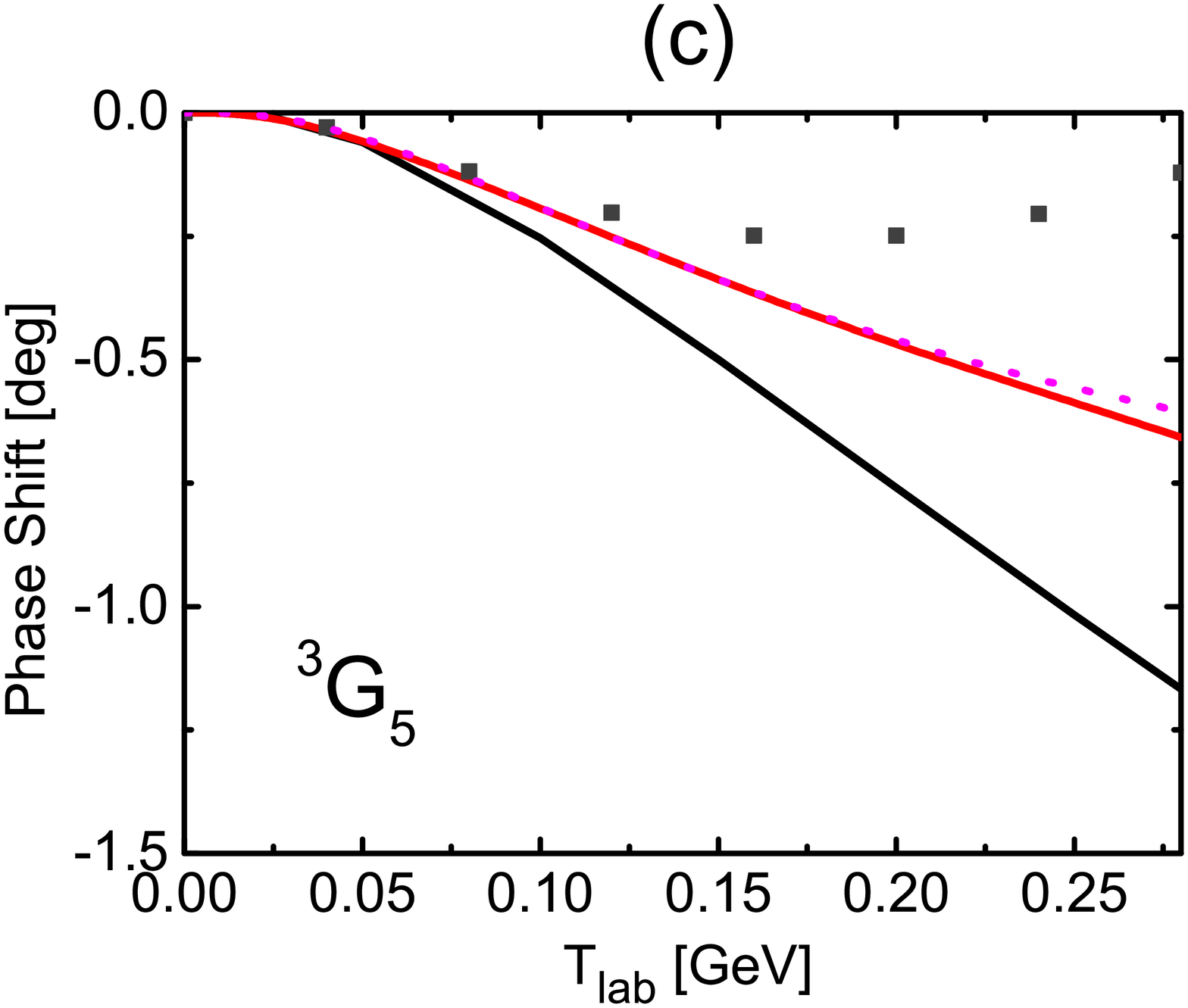}
}
\caption{$^3D_3$, $^3F_4$, and $^3G_5$ phase shifts as a function of $T_{\text{lab}}$. The black dots refer to the PWA93 results. The magenta dotted, black, and red lines represent nonperturbative OPE, OPE in Born approximation, and OPE plus once-iterated OPE in Born approximation, respectively. }
\label{tb:IOPE}
\end{figure}

{In this subsection, we study the OPE in Born approximation. Without loss of generality, we take $^3D_3, ^3F_4$, and $^3G_5$ partial waves as examples and plot the phase shifts in Fig.~\ref{tb:IOPE}. For simplification, we use ``Born-OPE" and ``Born-OPE+IOPE" to denote the phase shifts obtained with OPE in Born approximation and OPE plus once-iterated OPE in Born approximation. For the $^3D_3$ partial wave, we observe a sizable correction of $+7$ deg at $T_{\textrm{lab}} = 250$ MeV by including the contribution of once-iterated OPE and the Born-OPE+IOPE results are  comparable with the nonperturbative OPE results. For the $^3F_4$ partial wave, though the correction of once-iterated OPE is small, and the Born-OPE+IOPE phase shifts are already identical with the nonperturbative OPE results. For the $^3G_5$ partial wave, a correction of $+0.4$ deg at $T_{\textrm{lab}} = 250$ MeV is observed when considering the contribution of once-iterated OPE and the phase shifts are similar to the nonperturbative results up to $T_{\textrm{lab}} \leq 200$ MeV. This feature is similar to what has been known in the nonrelativistic case~\cite{Kaiser:1997mw}.}

\section{Summary and outlook}

Based on the covariant power counting rule proposed in Refs.~\cite{Ren:2016jna,Xiao:2018jot}, we calculated the relativistic TPE potential up to NNLO [$O(p^3)$] and obtained the nonperturbative scattering amplitude by solving the BbS equation with the so-obtained potential. With this $T$ matrix, we further calculated the chiral $NN$ phase shifts for  $J\geq 3$ and $L \leq6$ to which no contact term contributes up to NNLO and then compared our results with the corresponding perturbative  results and the nonrelativistic results. We found that the nonperturbative effect is significant for the $^3{D}_3$ partial wave {(as well as $^1D_2$ and $^3D_2)$} but insignificant for other higher partial waves we studied. Furthermore, a remarkable cutoff dependence is observed for this partial wave, which can be cured by higher order contact interactions. We confirmed that the contributions of relativistic TPE are more moderate than their nonrelativistic counterparts at the nonperturbative level. Moreover, we found that for most partial waves, the TPE contributions play a positive role in improving the descriptions but for the $^3{F}_3$, $^3{F}_4$, and $^3{H}_6$ partial waves, the contribution from subleading TPE is a bit  strong such that the resulting phase shifts become larger than  the data. The nonrelativistic results suffer from the same problem when the dimensional regularization method was applied to treat the ultraviolet divergence of chiral loops~\cite{Epelbaum:1999dj}. The abnormally large contribution of subleading TPE indicates that up to NNLO, the convergence is somehow questionable for these partial waves.

It should be stressed that the  relativistic corrections for the $J\ge3$ and $L\le6$ partial waves are indeed found to improve the description of the $NN$ phase shifts as expected.  Furthermore, although the nonperturbative effects play a minor role for most higher partial waves we studied, we expect that they will play a more important role for lower partial waves of $J\le2$, as found in, e.g., Ref.~\cite{Ren:2016jna}.  In addition, the nonperturbative TPE contributions obtained in the present work set a foundation for the construction of a high-precision relativistic chiral $NN$ interaction. All of these will be left for future studies.

\section{Acknowledgements}
This work is supported in part by the National Natural Science Foundation of China under Grants No.11735003, No.11975041,  and No.11961141004. J.-X. Lu is supported in part
by the National Natural Science Foundation of China under Grants No.12105006.

\section{Appendix}
\begin{figure}[htbp]
\centering
\subfloat{
\includegraphics[width=0.27\textwidth]{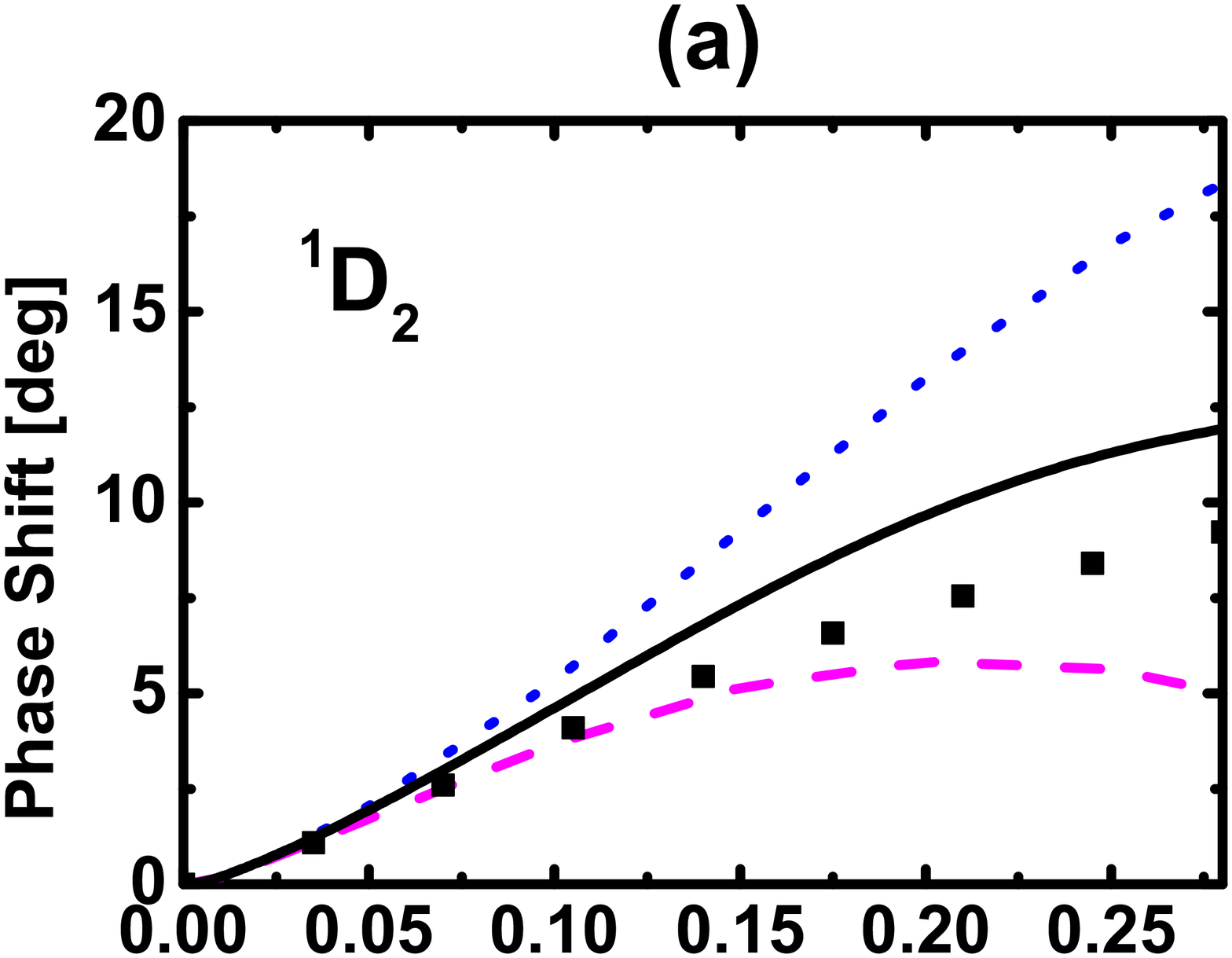}
}\hspace{-11mm}
\subfloat{
\includegraphics[width=0.27\textwidth]{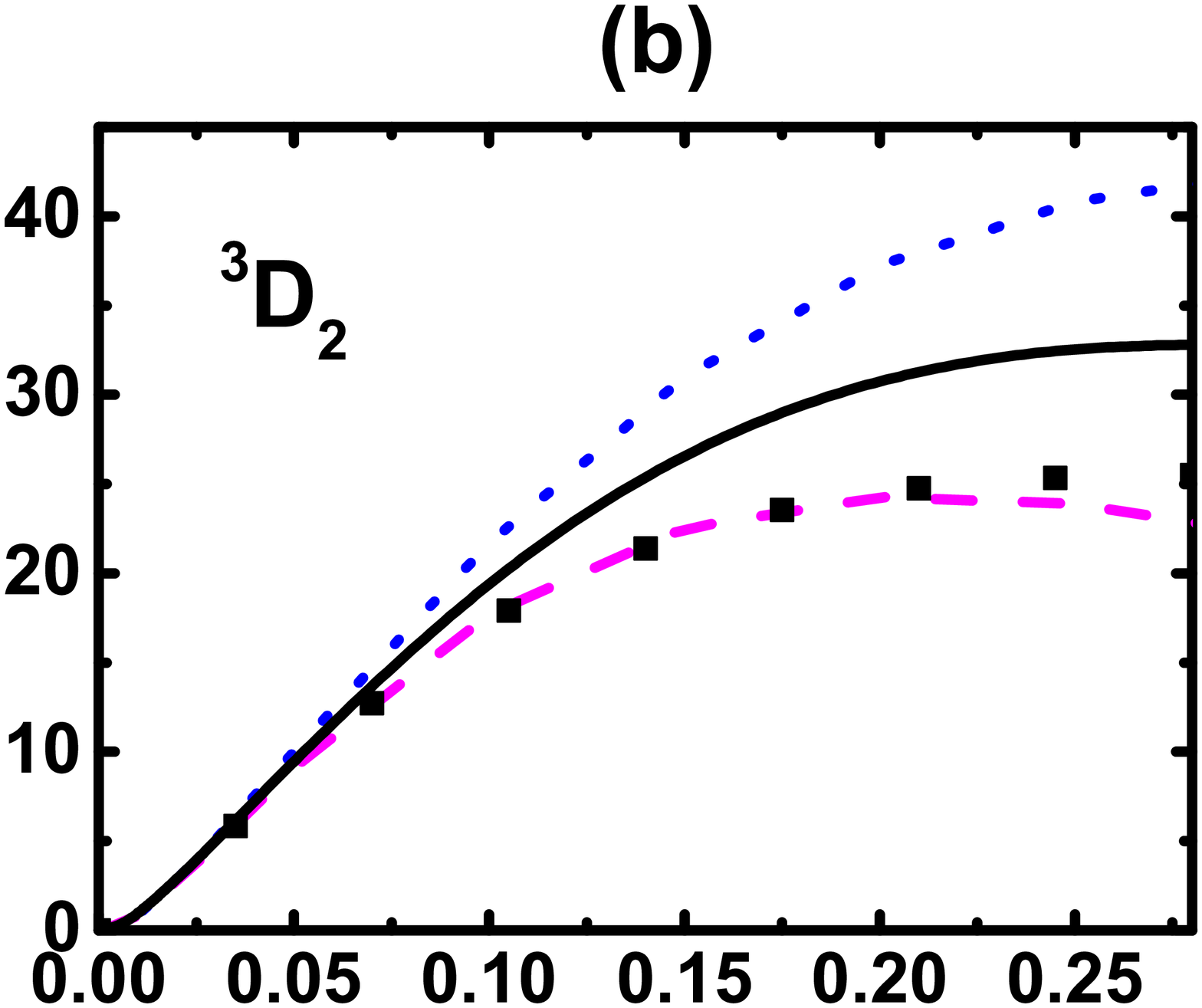}
}\hspace{-11mm}
\subfloat{
\includegraphics[width=0.27\textwidth]{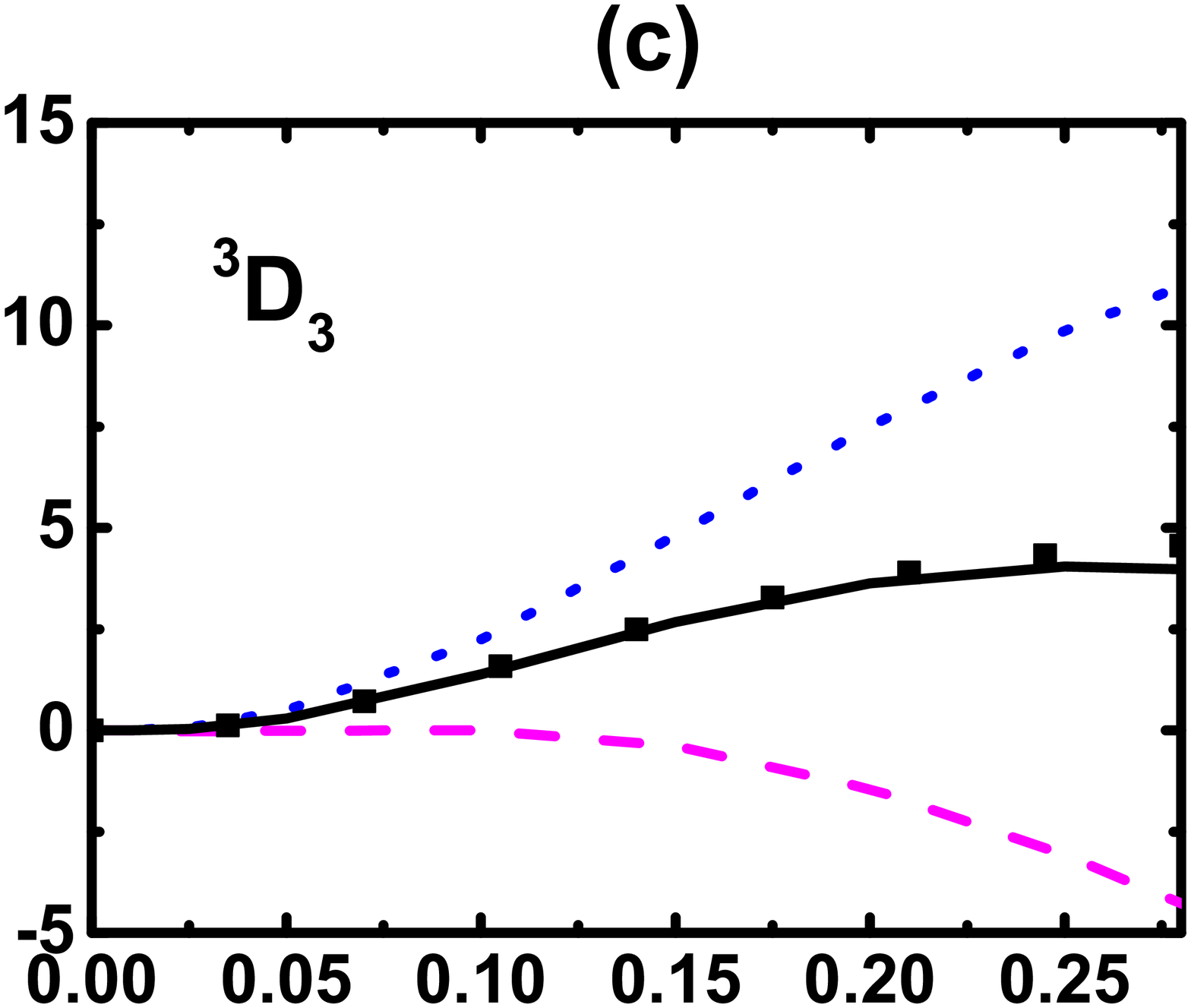}
}\hspace{-11mm}
\subfloat{
\includegraphics[width=0.27\textwidth]{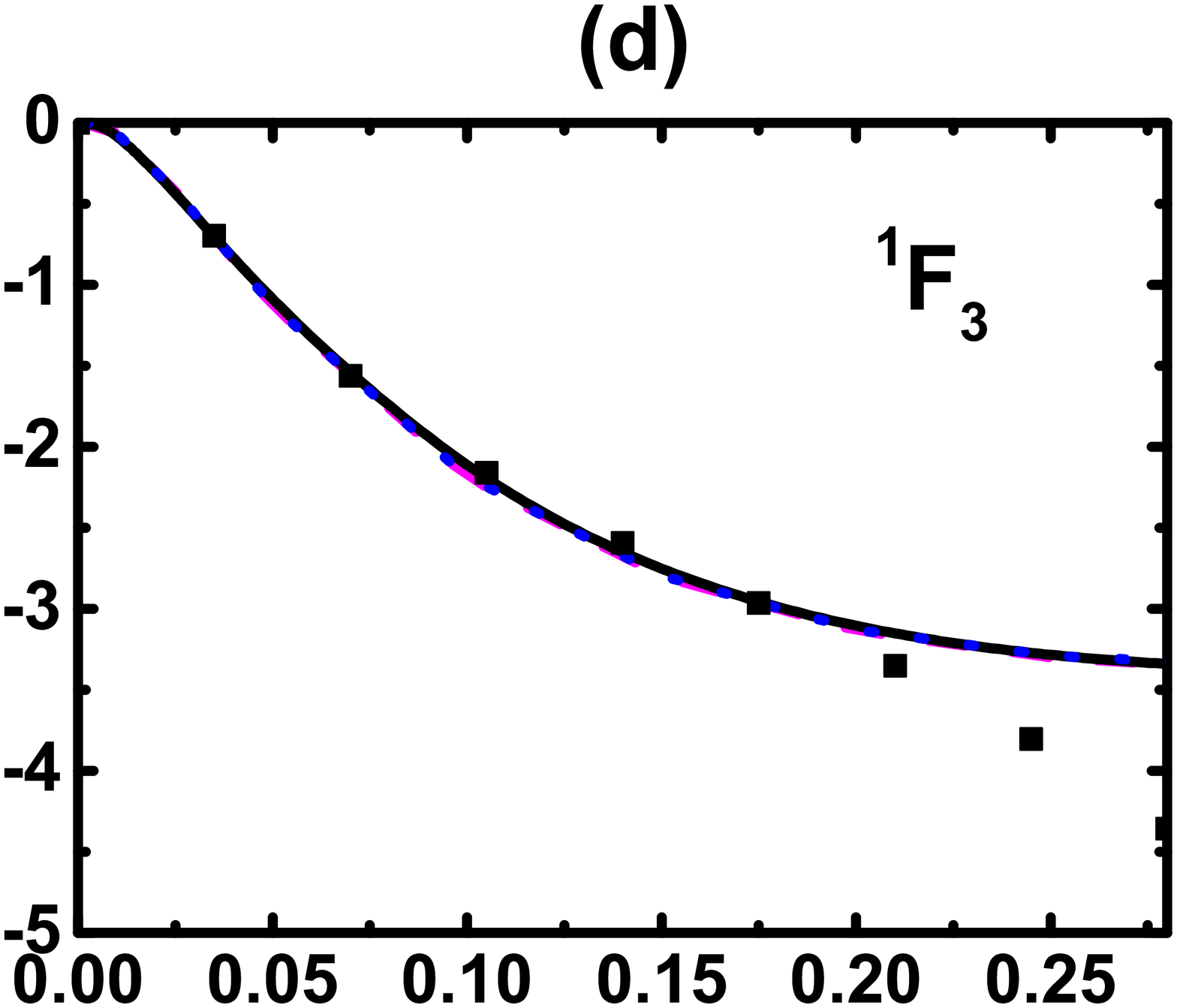}
}\\ \vspace{-6mm}
\subfloat{
\includegraphics[width=0.27\textwidth]{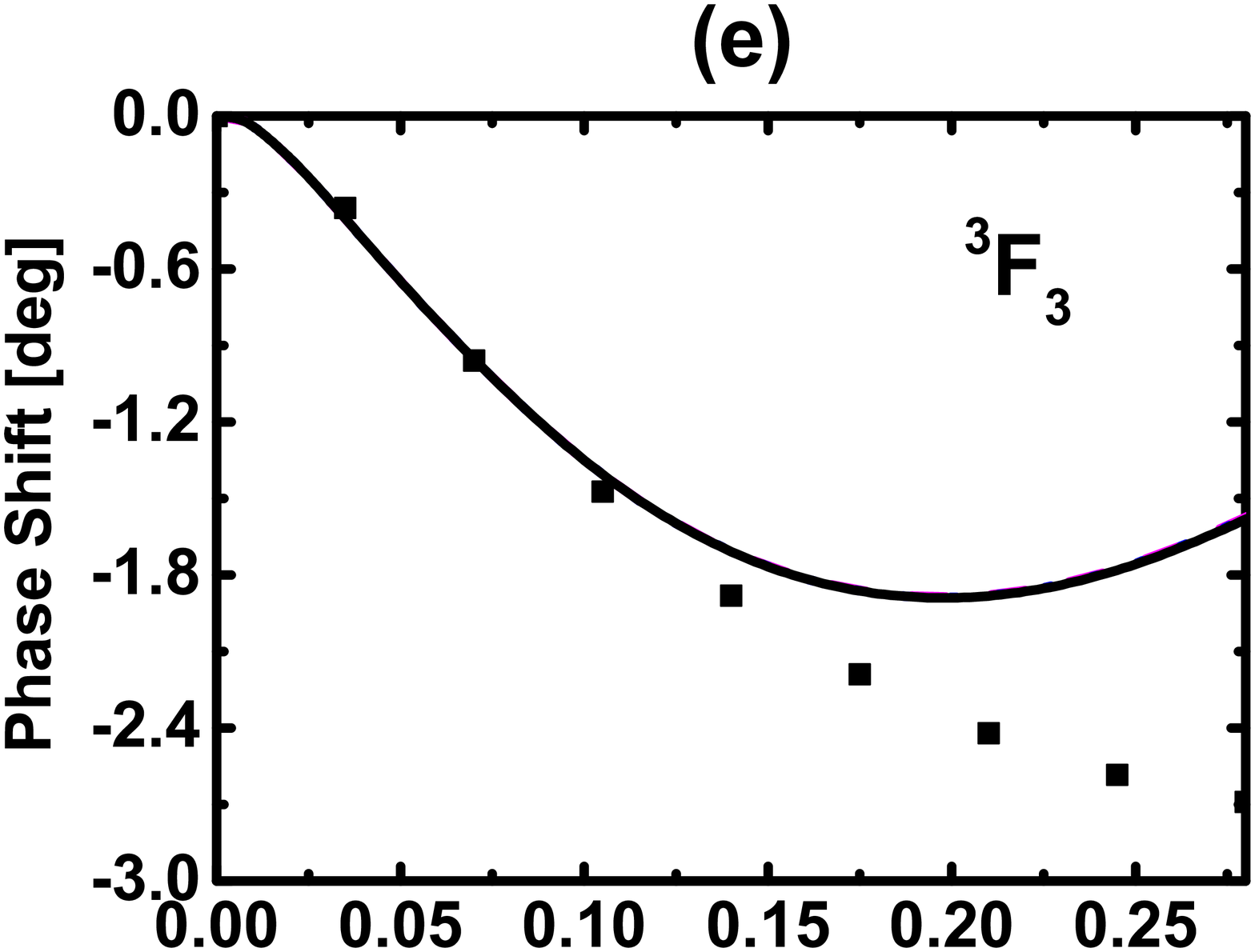}
}\hspace{-11mm}
\subfloat{
\includegraphics[width=0.27\textwidth]{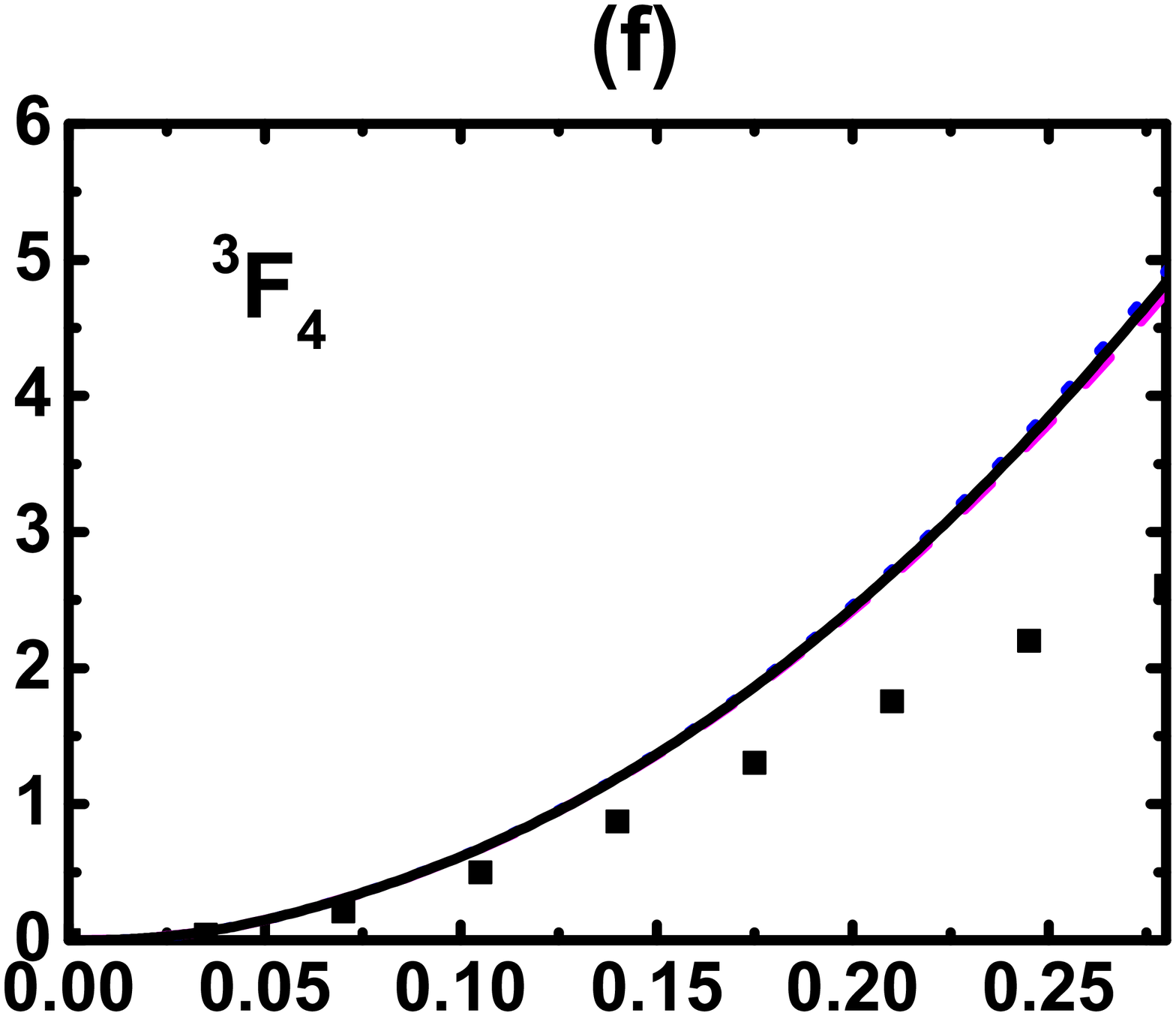}
}\hspace{-11mm}
\subfloat{
\includegraphics[width=0.27\textwidth]{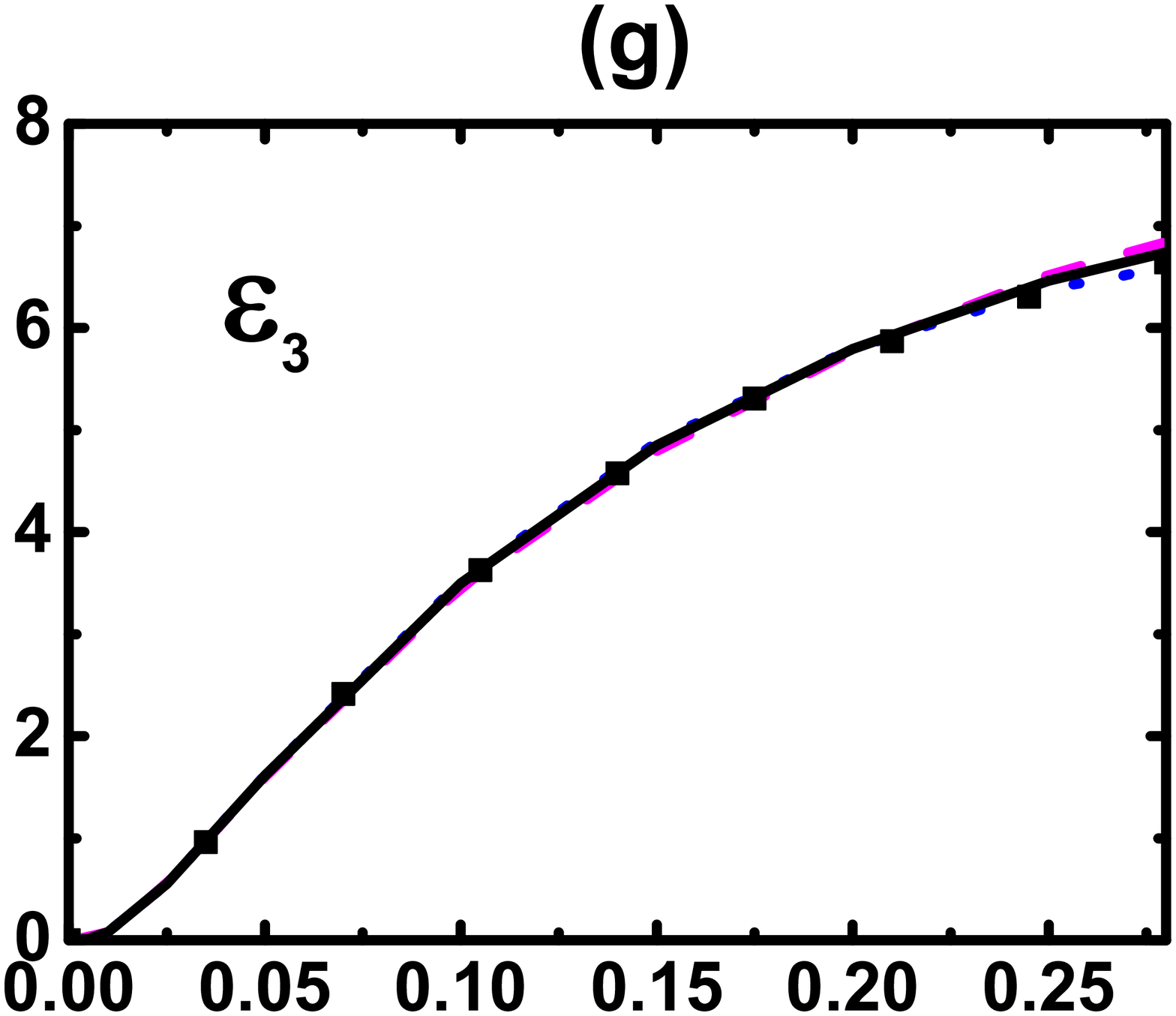}
}\hspace{-11mm}
\subfloat{
\includegraphics[width=0.27\textwidth]{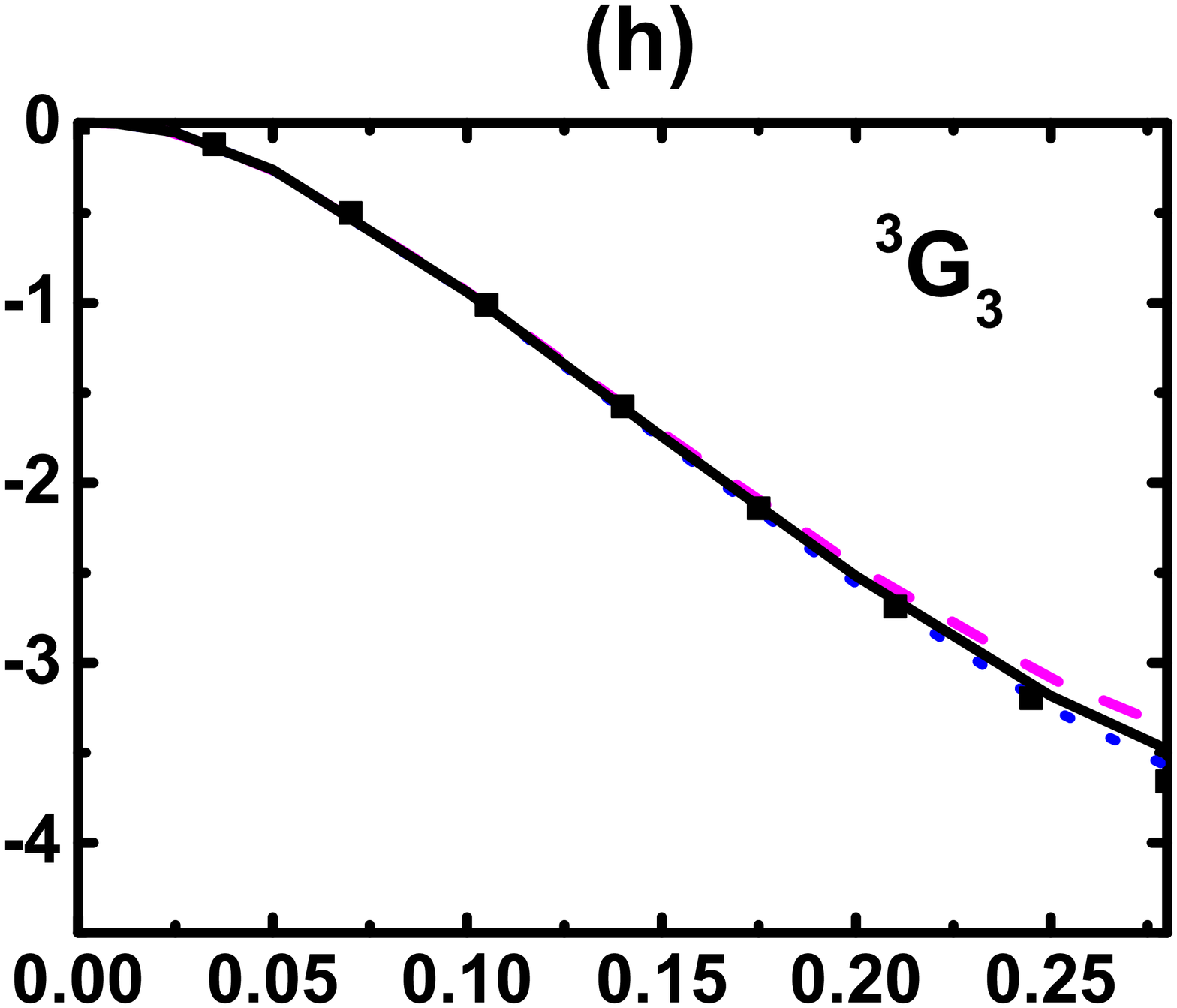}
}\\ \vspace{-6mm}
\subfloat{
\includegraphics[width=0.27\textwidth]{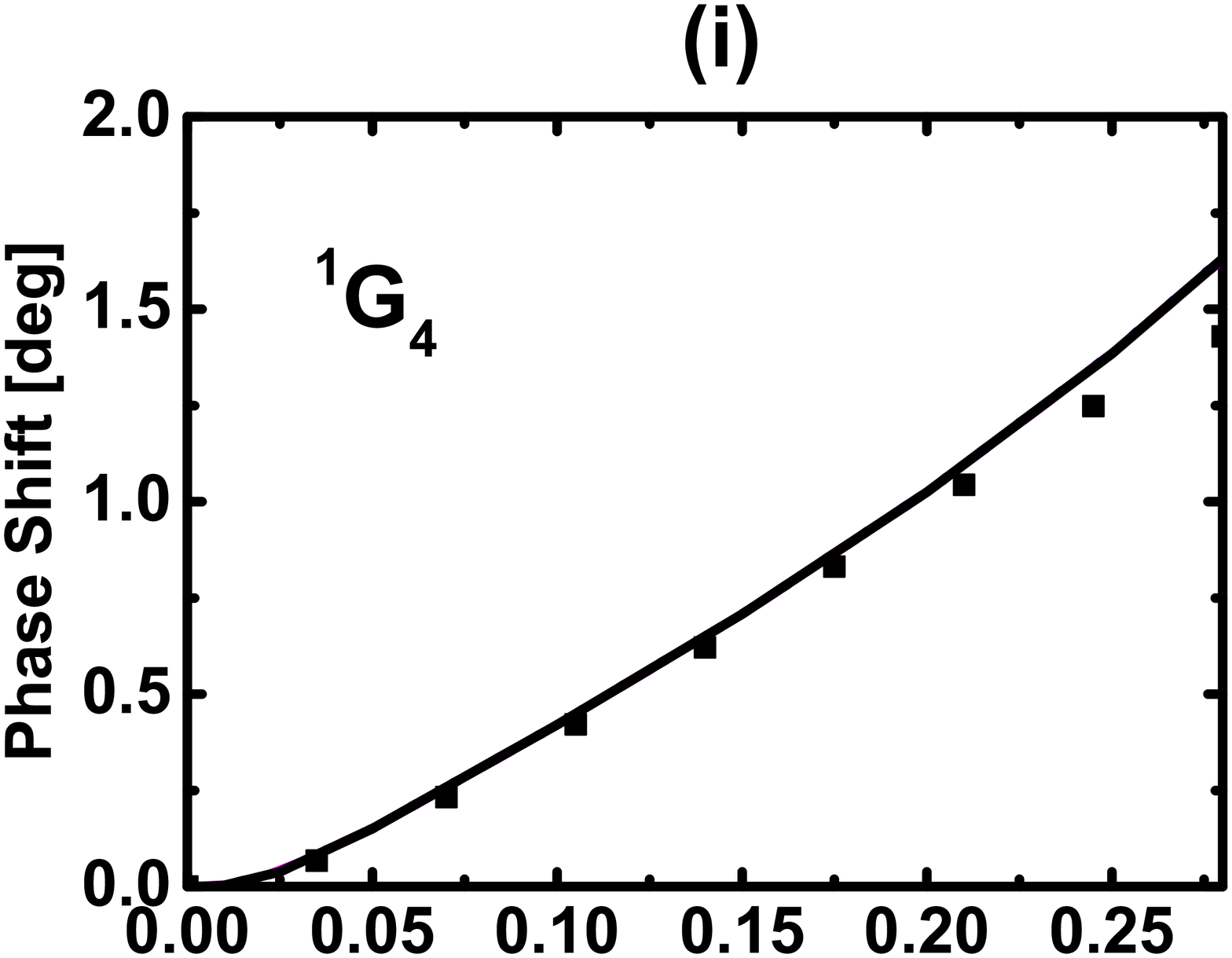}
}\hspace{-11mm}
\subfloat{
\includegraphics[width=0.27\textwidth]{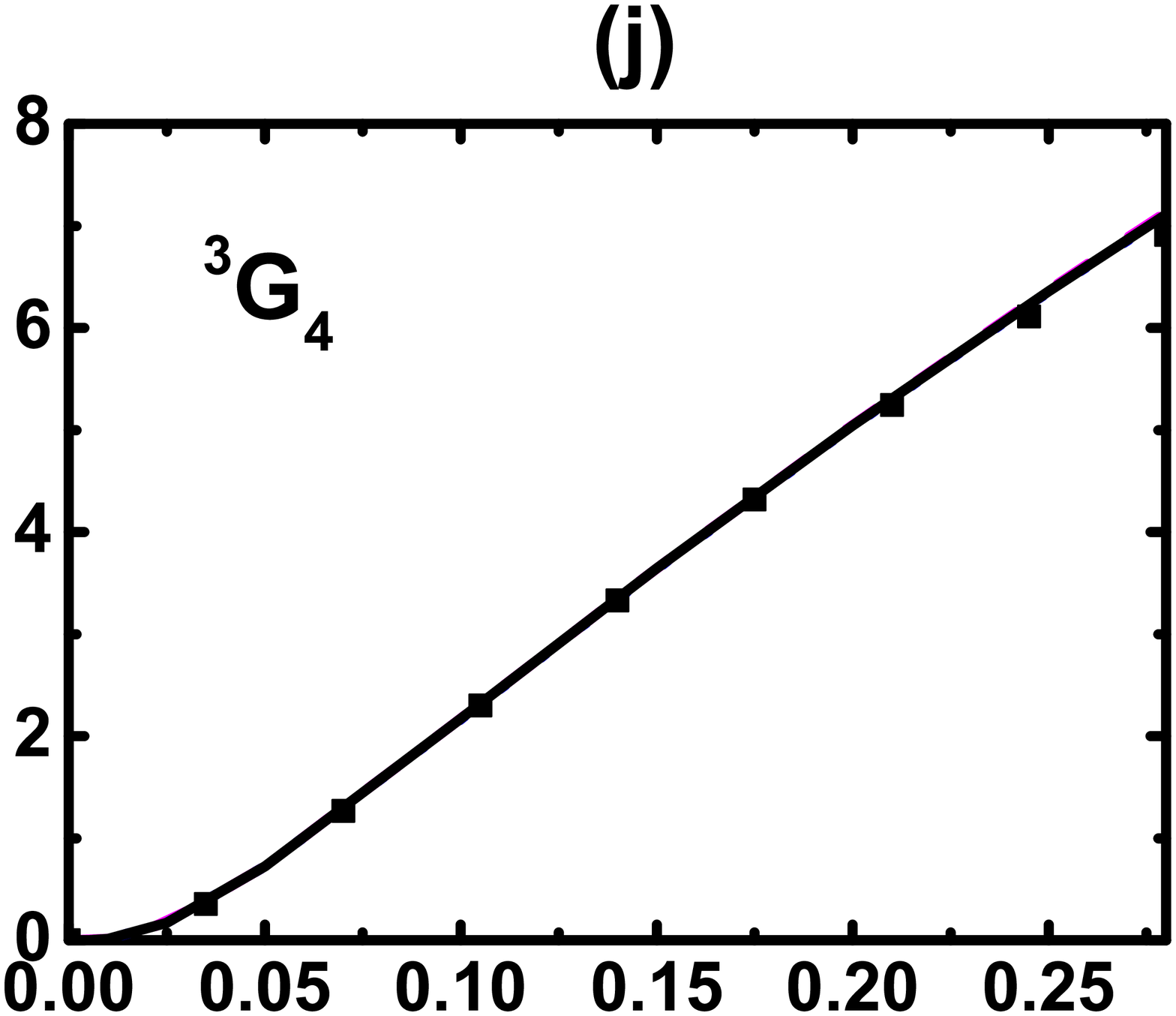}
}\hspace{-11mm}
\subfloat{
\includegraphics[width=0.27\textwidth]{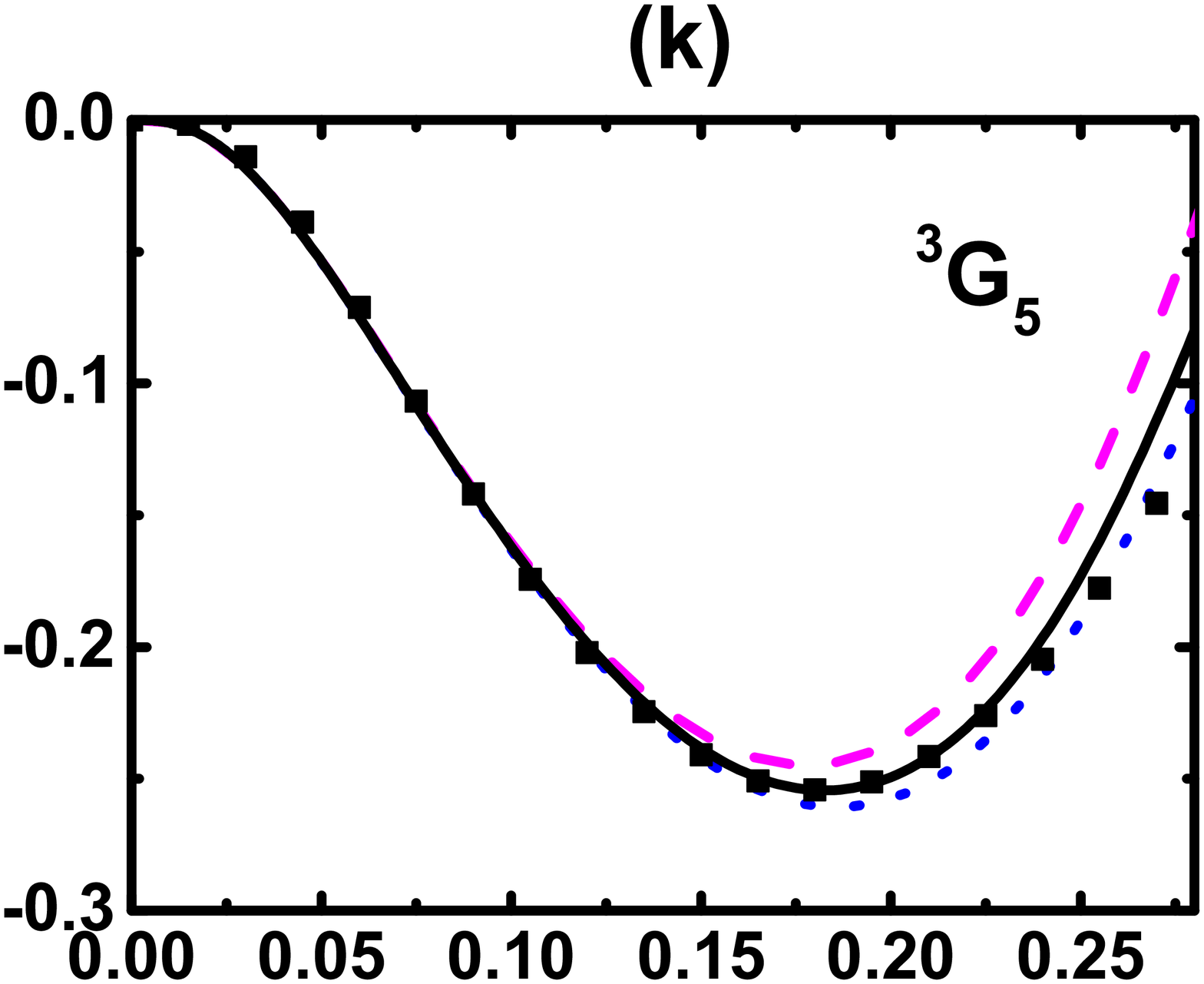}
}\hspace{-11mm}
\subfloat{
\includegraphics[width=0.27\textwidth]{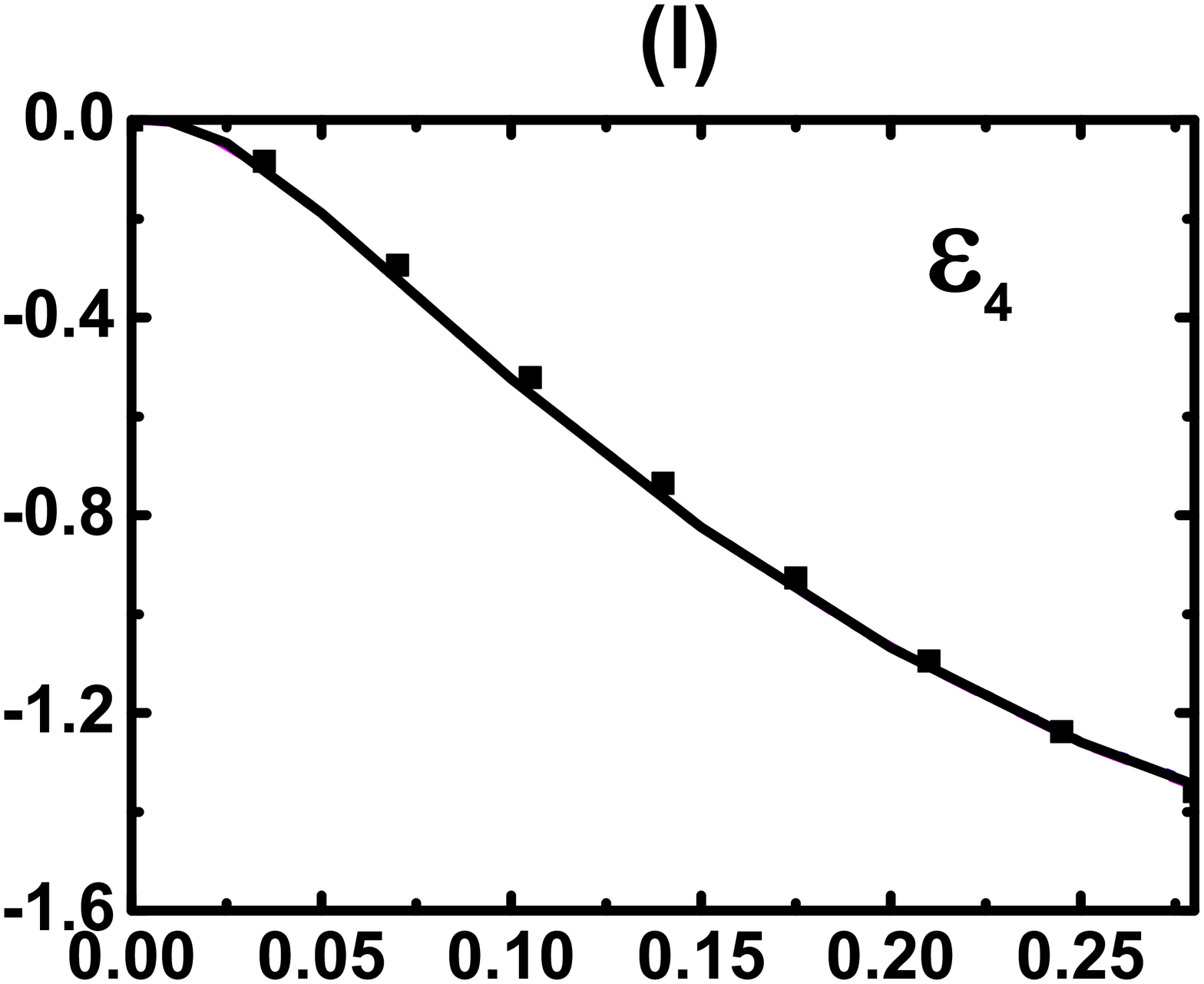}
}\\ \vspace{-6mm}
\subfloat{
\includegraphics[width=0.27\textwidth]{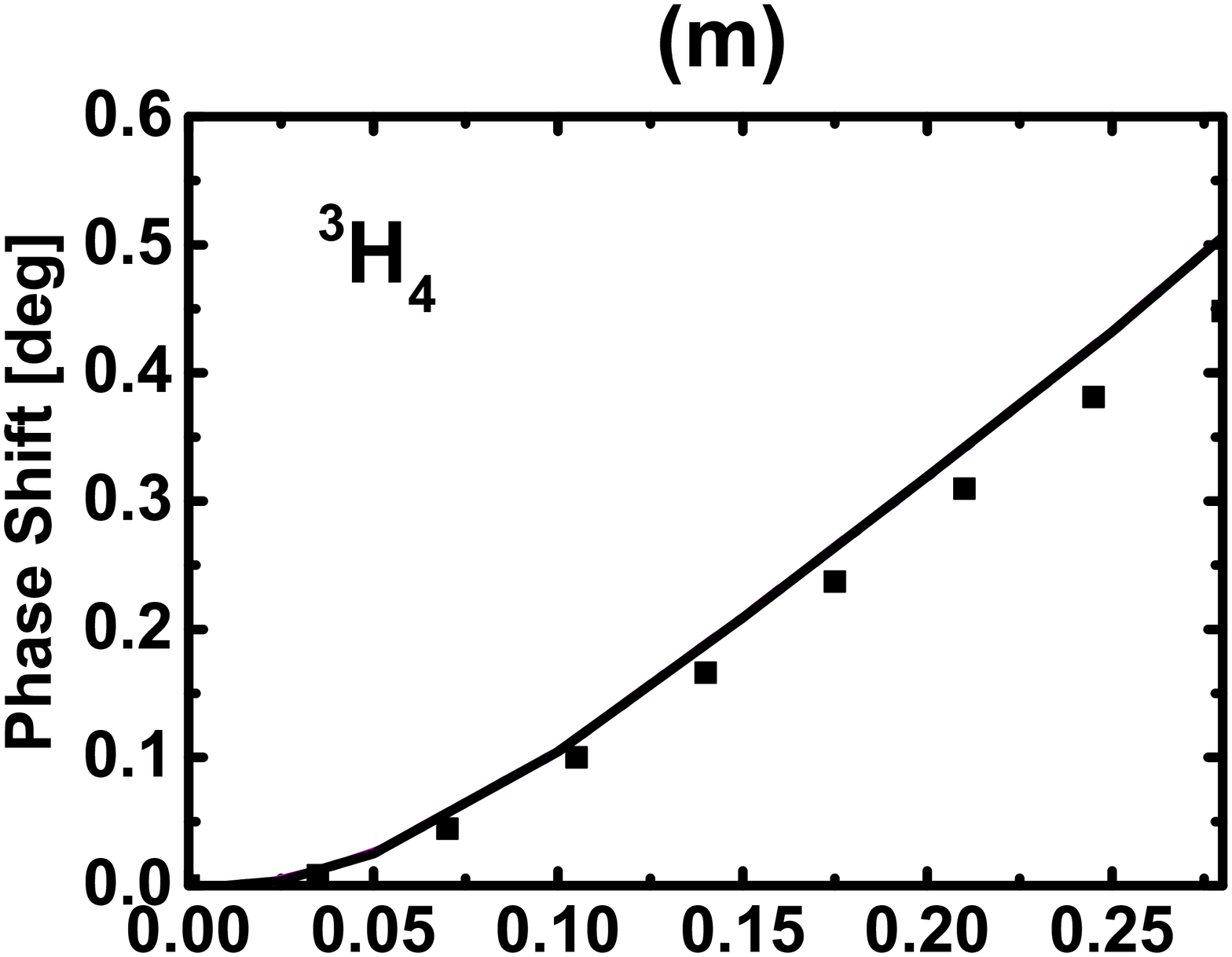}
}\hspace{-11mm}
\subfloat{
\includegraphics[width=0.27\textwidth]{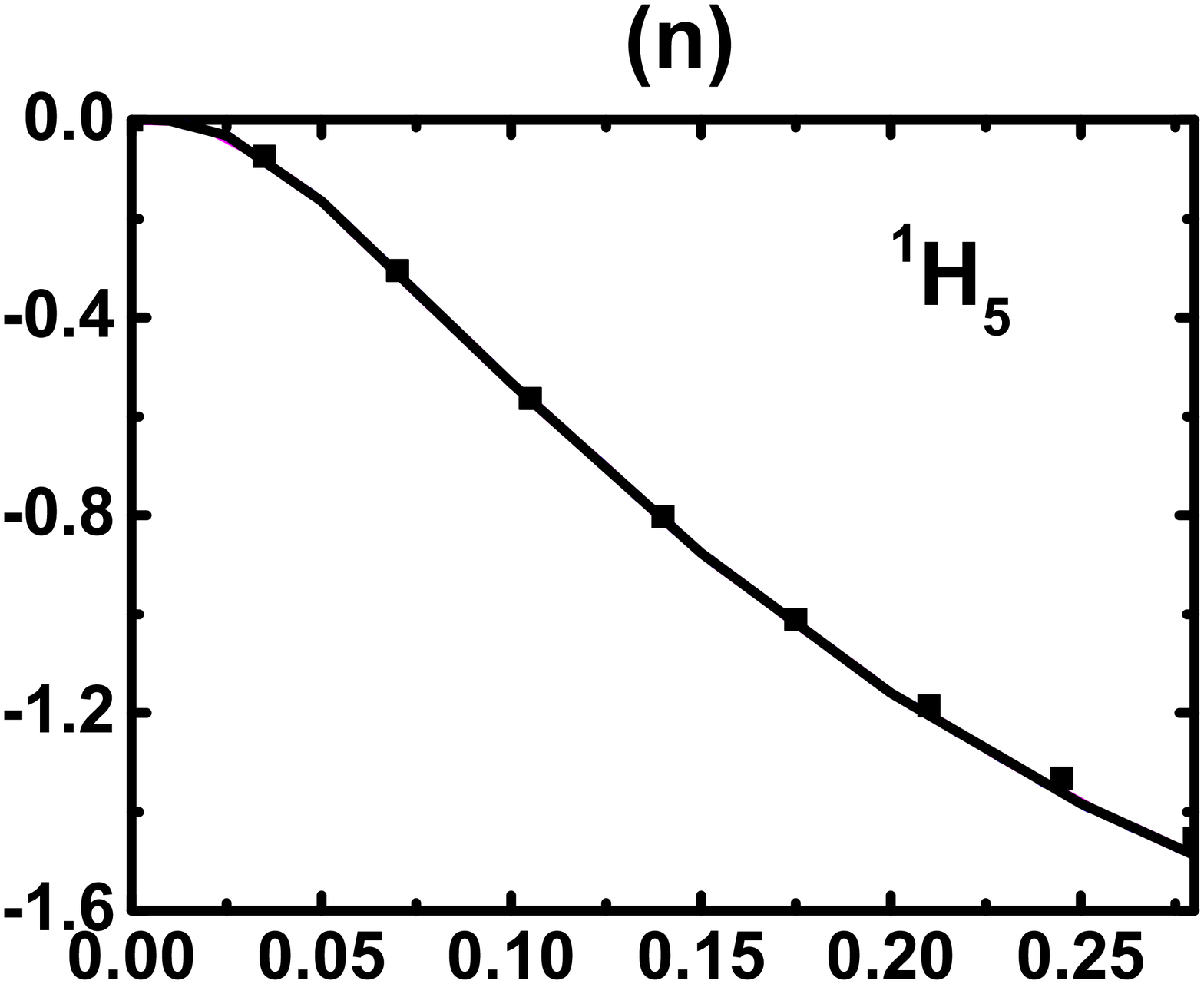}
}\hspace{-11mm}
\subfloat{
\includegraphics[width=0.27\textwidth]{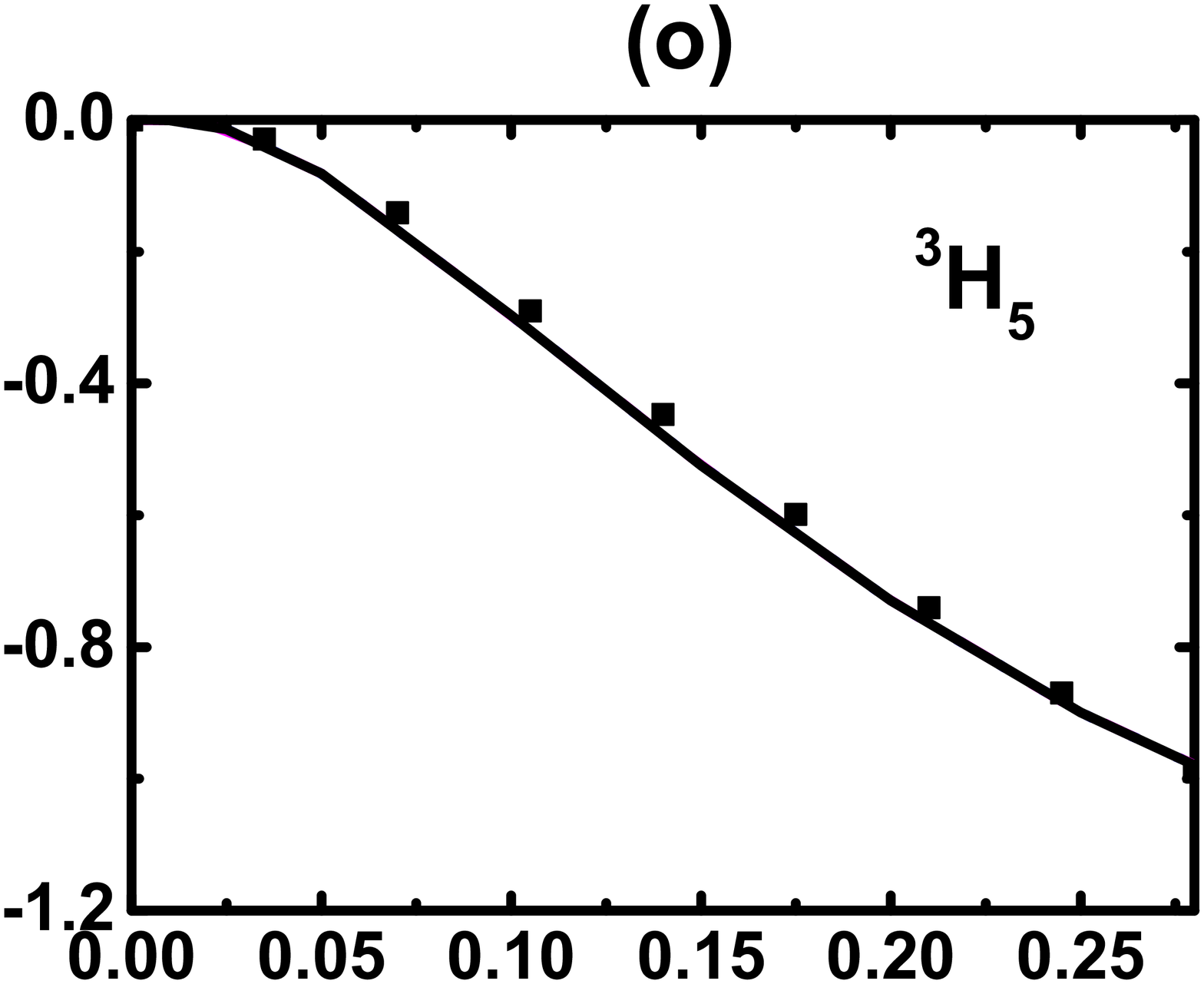}
}\hspace{-11mm}
\subfloat{
\includegraphics[width=0.27\textwidth]{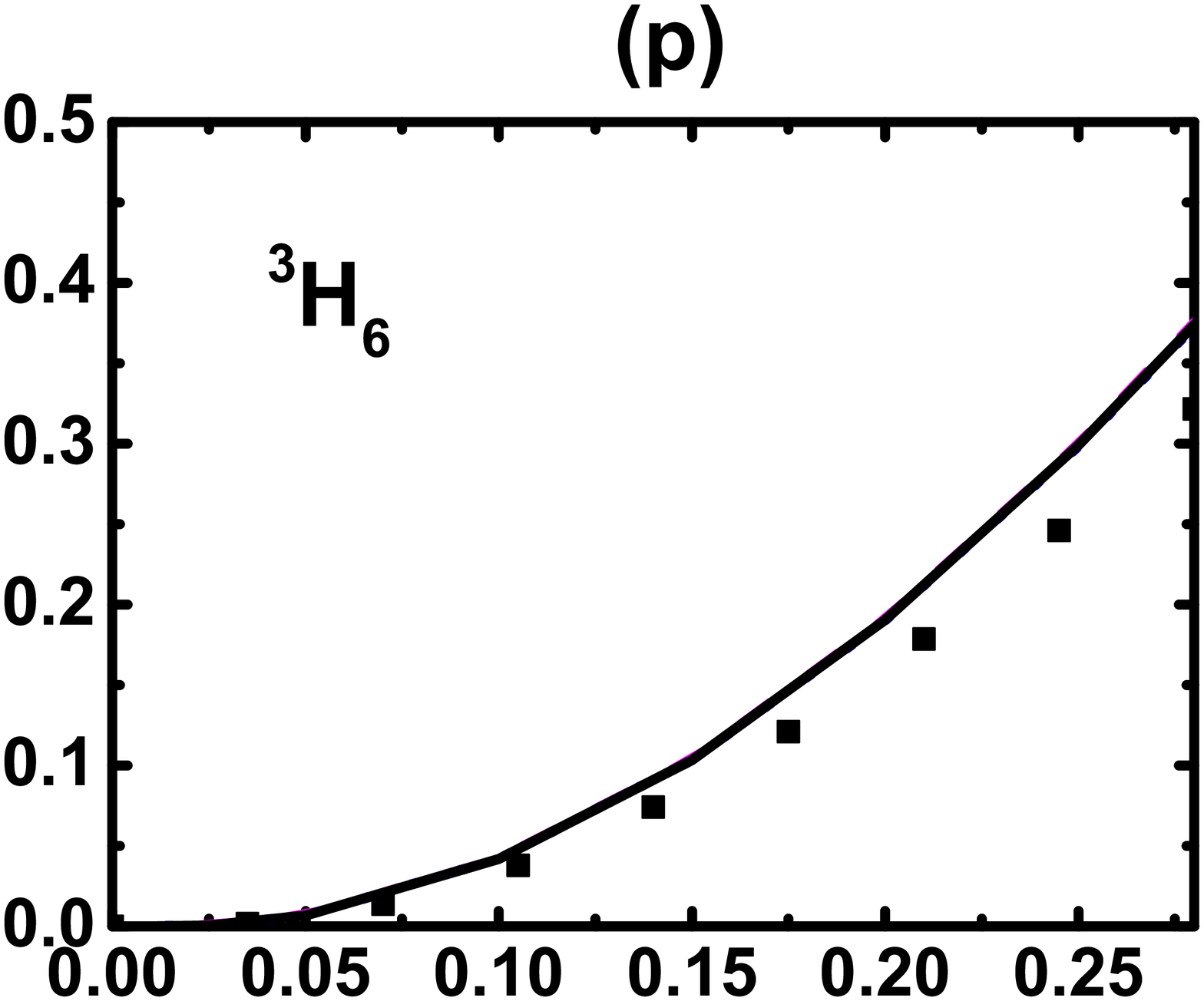}
}\\ \vspace{-6mm}
\subfloat{
\includegraphics[width=0.27\textwidth]{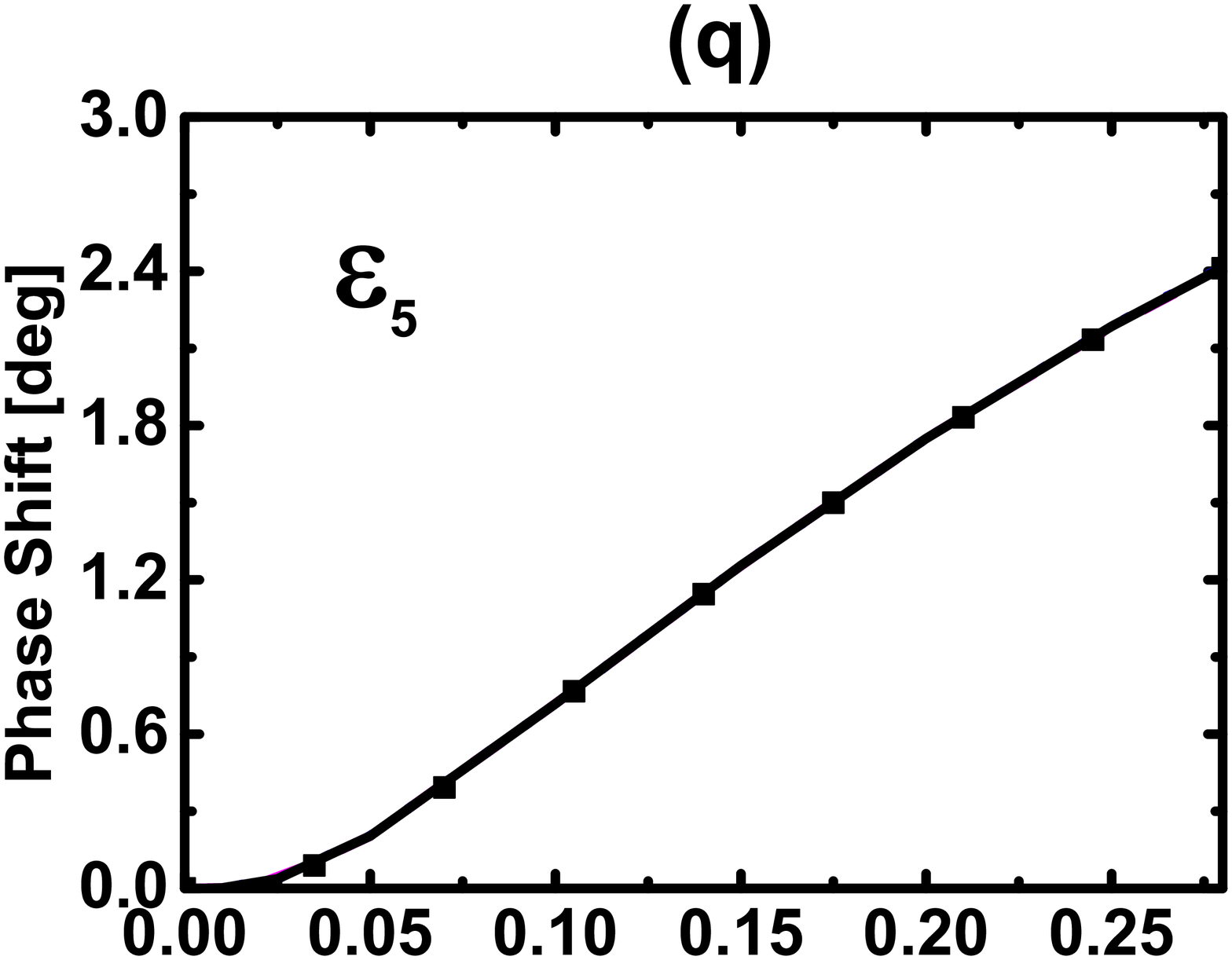}
}\hspace{-11mm}
\subfloat{
\includegraphics[width=0.27\textwidth]{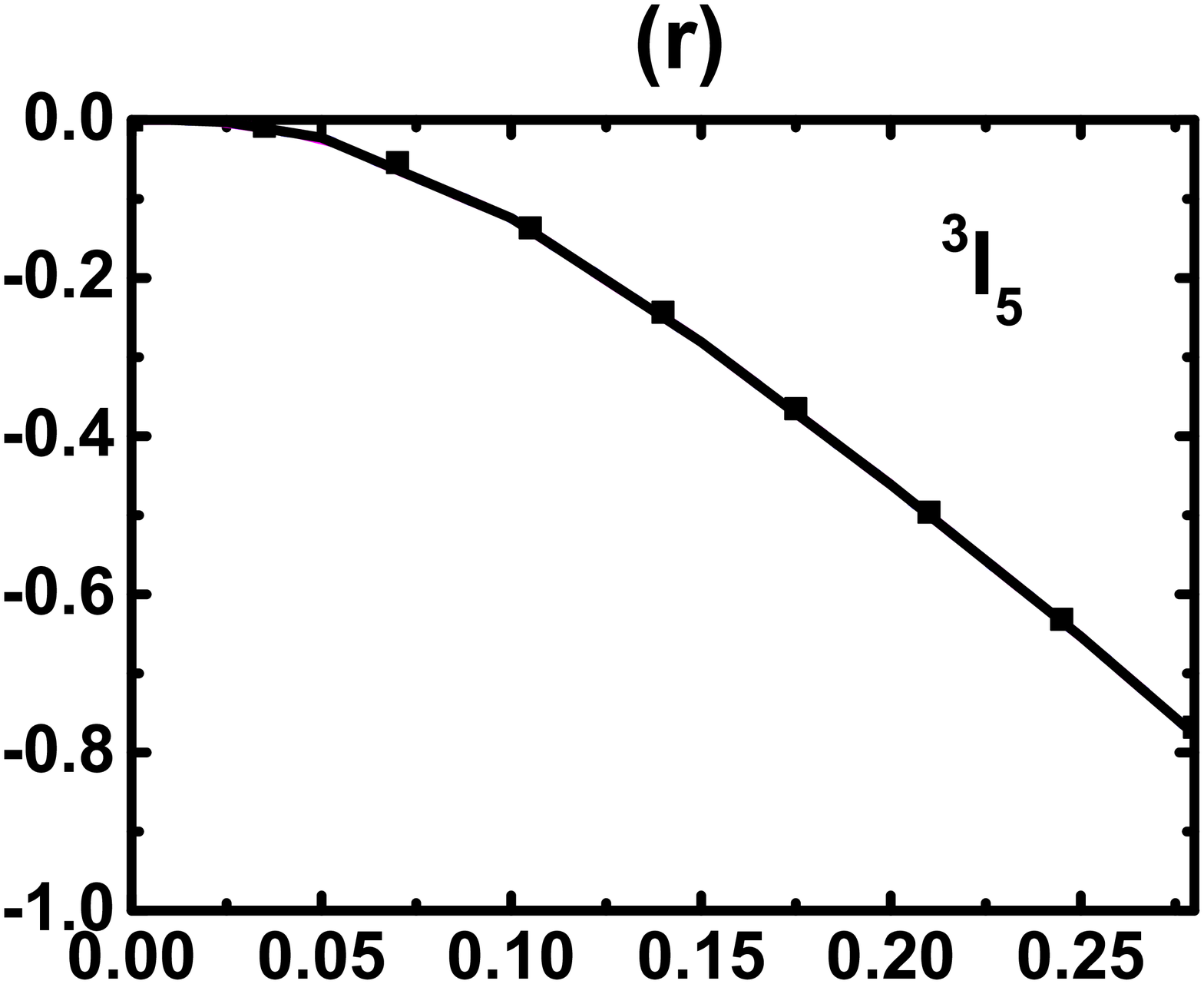}
}\hspace{-11mm}
\subfloat{
\includegraphics[width=0.27\textwidth]{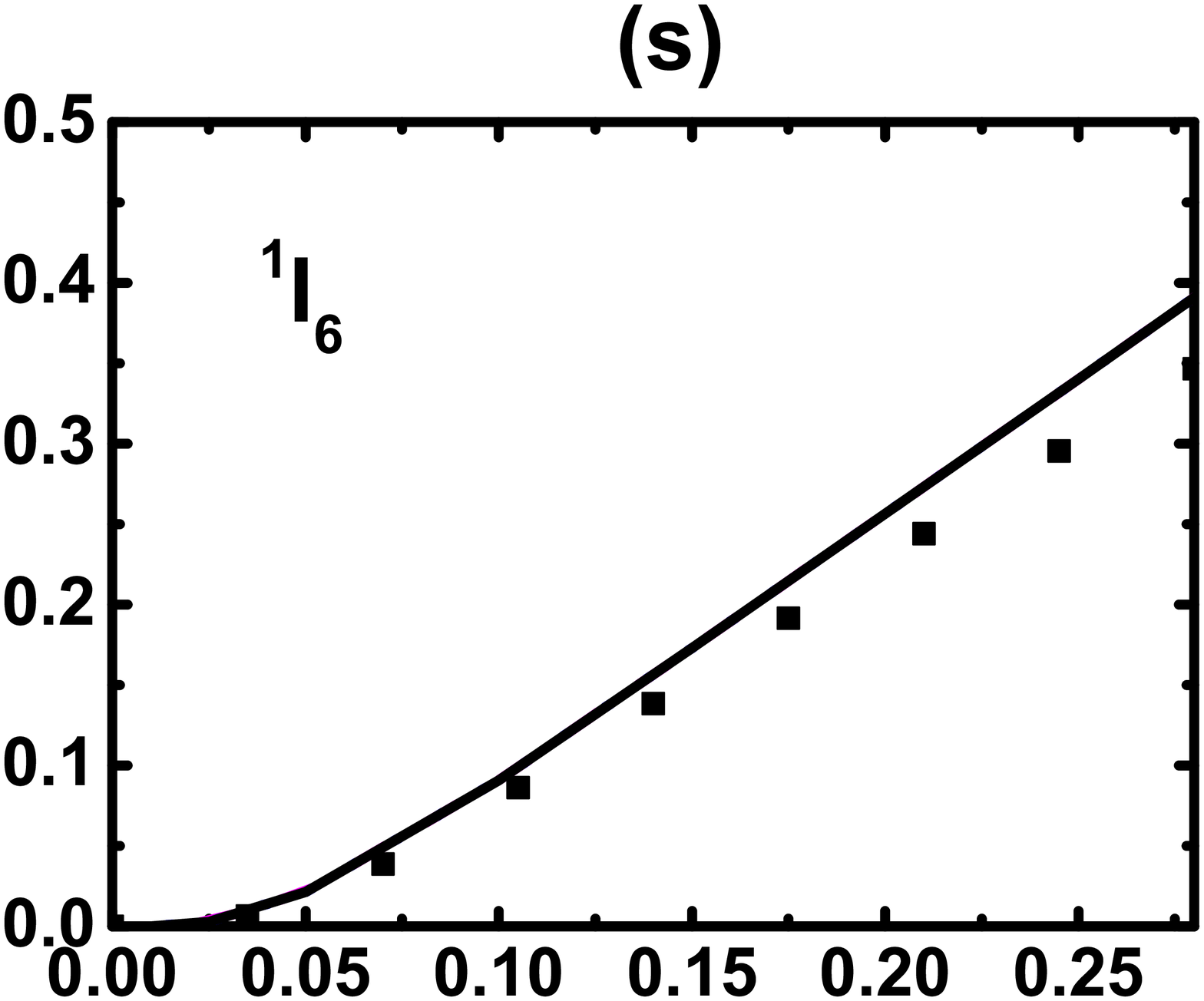}
}\hspace{-11mm}
\subfloat{
\includegraphics[width=0.27\textwidth]{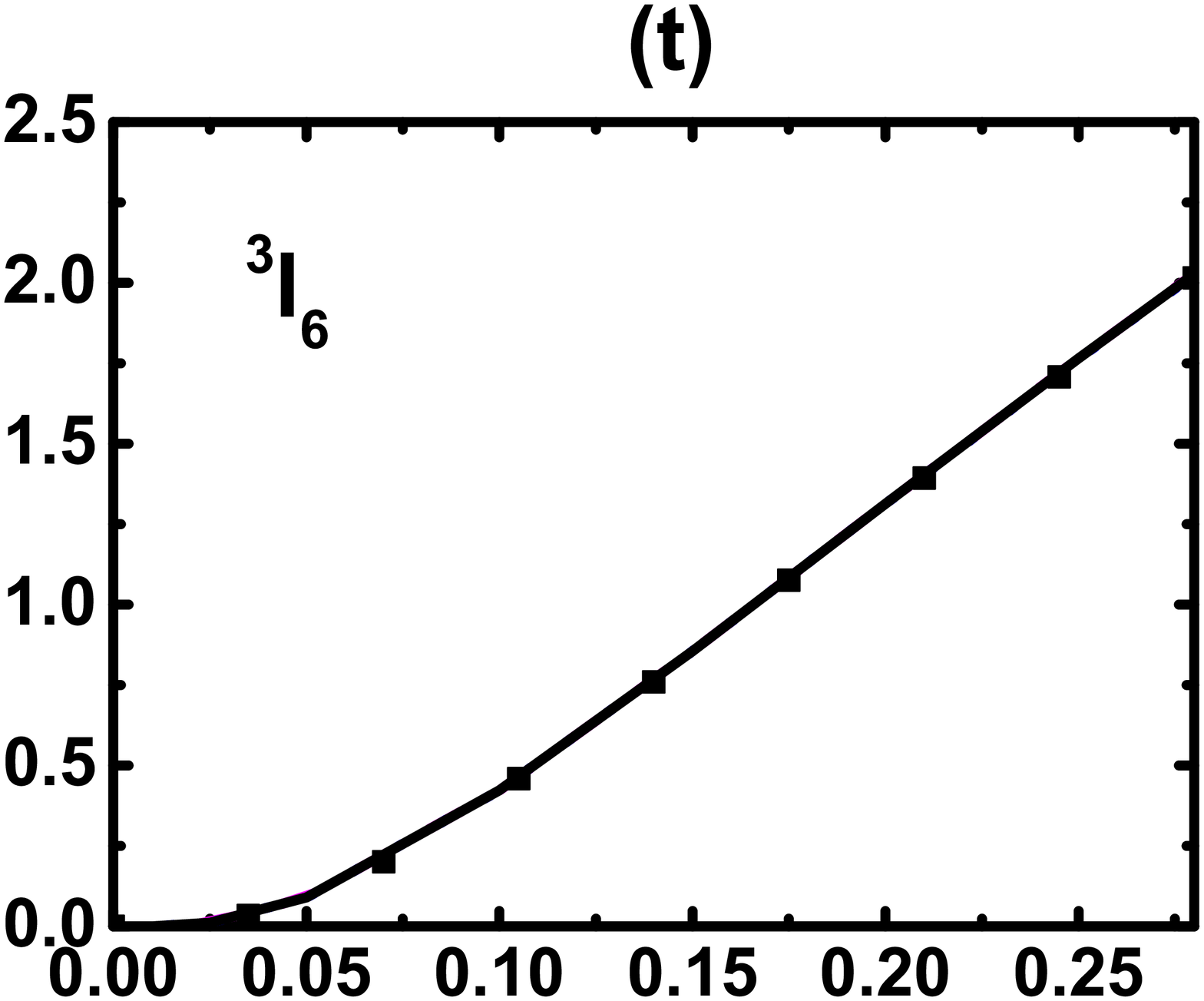}
}\\ \vspace{-6mm}
\subfloat{
\includegraphics[width=0.27\textwidth]{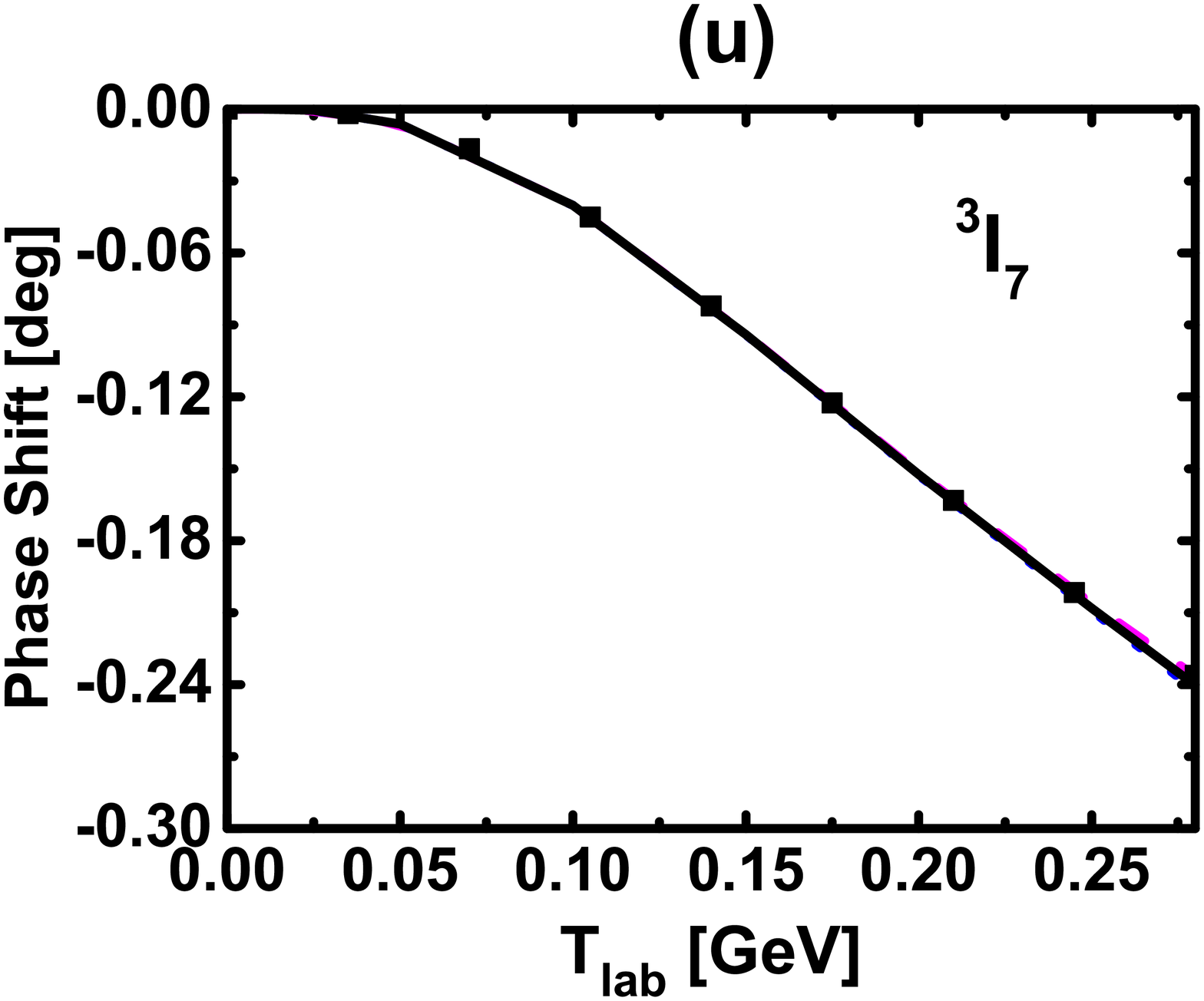}
}\hspace{-11mm}
\subfloat{
\includegraphics[width=0.27\textwidth]{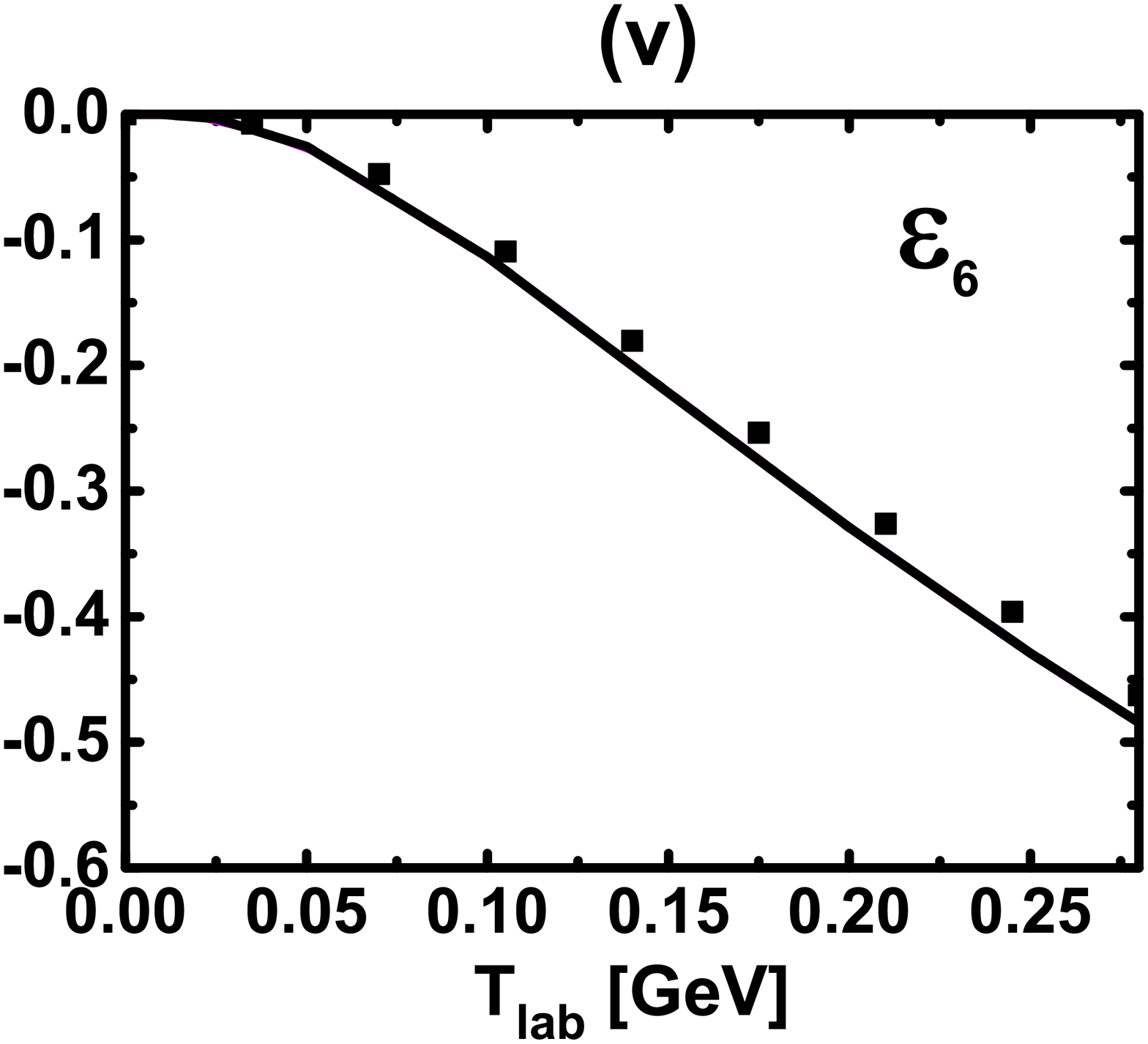}
}
\caption{Cutoff dependence of the covariant OPE and TPE results up to NNLO  for $2 \leq L \leq 6$ partial waves. The black dots denote the Nijmegen partial wave phase shifts~\cite{Stoks:1993tb}. The NNLO relativistic results obtained with cutoff $=800,900,1000$ MeV are depicted with the magenta dotted lines, the black solid lines, and the magenta dashed lines, respectively. }
\label{tb:cutoff-dependence}
\end{figure}

In Fig.~\ref{tb:cutoff-dependence}, we collect the phase shifts for all the partial waves obtained with different cutoffs varying from 800 MeV to 1000 MeV. {The $^1D_2$ and $^3D_2$ partial waves are cutoff dependent since we neglect the corresponding contact interactions. For all the other partial waves except $^3D_3$, we can see that the results are quite stable and almost cutoff independent.}
%We leave a more detailed study of the nenormalizibility of these higher partial waves to a future work.

\bibliography{nuclear-force}

\begin{thebibliography}{62}
\expandafter\ifx\csname natexlab\endcsname\relax\def\natexlab#1{#1}\fi
\expandafter\ifx\csname bibnamefont\endcsname\relax
  \def\bibnamefont#1{#1}\fi
\expandafter\ifx\csname bibfnamefont\endcsname\relax
  \def\bibfnamefont#1{#1}\fi
\expandafter\ifx\csname citenamefont\endcsname\relax
  \def\citenamefont#1{#1}\fi
\expandafter\ifx\csname url\endcsname\relax
  \def\url#1{\texttt{#1}}\fi
\expandafter\ifx\csname urlprefix\endcsname\relax\def\urlprefix{URL }\fi
\providecommand{\bibinfo}[2]{#2}
\providecommand{\eprint}[2][]{\url{#2}}

\bibitem[{\citenamefont{Yukawa}(1935)}]{Yukawa:1935xg}
\bibinfo{author}{\bibfnamefont{H.}~\bibnamefont{Yukawa}},
  \bibinfo{journal}{Proc. Phys. Math. Soc. Jap.} \textbf{\bibinfo{volume}{17}},
  \bibinfo{pages}{48} (\bibinfo{year}{1935}).

\bibitem[{\citenamefont{Stoks et~al.}(1994)\citenamefont{Stoks, Klomp,
  Terheggen, and de~Swart}}]{Stoks:1994wp}
\bibinfo{author}{\bibfnamefont{V.~G.~J.} \bibnamefont{Stoks}},
  \bibinfo{author}{\bibfnamefont{R.~A.~M.} \bibnamefont{Klomp}},
  \bibinfo{author}{\bibfnamefont{C.~P.~F.} \bibnamefont{Terheggen}},
  \bibnamefont{and} \bibinfo{author}{\bibfnamefont{J.~J.}
  \bibnamefont{de~Swart}}, \bibinfo{journal}{Phys. Rev. C}
  \textbf{\bibinfo{volume}{49}}, \bibinfo{pages}{2950} (\bibinfo{year}{1994}),
  \eprint{nucl-th/9406039}.

\bibitem[{\citenamefont{Wiringa et~al.}(1995)\citenamefont{Wiringa, Stoks, and
  Schiavilla}}]{Wiringa:1994wb}
\bibinfo{author}{\bibfnamefont{R.~B.} \bibnamefont{Wiringa}},
  \bibinfo{author}{\bibfnamefont{V.~G.~J.} \bibnamefont{Stoks}},
  \bibnamefont{and}
  \bibinfo{author}{\bibfnamefont{R.}~\bibnamefont{Schiavilla}},
  \bibinfo{journal}{Phys. Rev. C} \textbf{\bibinfo{volume}{51}},
  \bibinfo{pages}{38} (\bibinfo{year}{1995}), \eprint{nucl-th/9408016}.

\bibitem[{\citenamefont{Machleidt}(2001)}]{Machleidt:2000ge}
\bibinfo{author}{\bibfnamefont{R.}~\bibnamefont{Machleidt}},
  \bibinfo{journal}{Phys. Rev. C} \textbf{\bibinfo{volume}{63}},
  \bibinfo{pages}{024001} (\bibinfo{year}{2001}), \eprint{nucl-th/0006014}.

\bibitem[{\citenamefont{Ishii et~al.}(2007)\citenamefont{Ishii, Aoki, and
  Hatsuda}}]{Ishii:2006ec}
\bibinfo{author}{\bibfnamefont{N.}~\bibnamefont{Ishii}},
  \bibinfo{author}{\bibfnamefont{S.}~\bibnamefont{Aoki}}, \bibnamefont{and}
  \bibinfo{author}{\bibfnamefont{T.}~\bibnamefont{Hatsuda}},
  \bibinfo{journal}{Phys. Rev. Lett.} \textbf{\bibinfo{volume}{99}},
  \bibinfo{pages}{022001} (\bibinfo{year}{2007}), \eprint{nucl-th/0611096}.

\bibitem[{\citenamefont{Aoki et~al.}(2010)\citenamefont{Aoki, Hatsuda, and
  Ishii}}]{Aoki:2009ji}
\bibinfo{author}{\bibfnamefont{S.}~\bibnamefont{Aoki}},
  \bibinfo{author}{\bibfnamefont{T.}~\bibnamefont{Hatsuda}}, \bibnamefont{and}
  \bibinfo{author}{\bibfnamefont{N.}~\bibnamefont{Ishii}},
  \bibinfo{journal}{Prog. Theor. Phys.} \textbf{\bibinfo{volume}{123}},
  \bibinfo{pages}{89} (\bibinfo{year}{2010}), \eprint{0909.5585}.

\bibitem[{\citenamefont{Beane et~al.}(2011)\citenamefont{Beane, Detmold,
  Orginos, and Savage}}]{Beane:2010em}
\bibinfo{author}{\bibfnamefont{S.~R.} \bibnamefont{Beane}},
  \bibinfo{author}{\bibfnamefont{W.}~\bibnamefont{Detmold}},
  \bibinfo{author}{\bibfnamefont{K.}~\bibnamefont{Orginos}}, \bibnamefont{and}
  \bibinfo{author}{\bibfnamefont{M.~J.} \bibnamefont{Savage}},
  \bibinfo{journal}{Prog. Part. Nucl. Phys.} \textbf{\bibinfo{volume}{66}},
  \bibinfo{pages}{1} (\bibinfo{year}{2011}), \eprint{1004.2935}.

\bibitem[{\citenamefont{Barnea et~al.}(2015)\citenamefont{Barnea, Contessi,
  Gazit, Pederiva, and van Kolck}}]{Barnea:2013uqa}
\bibinfo{author}{\bibfnamefont{N.}~\bibnamefont{Barnea}},
  \bibinfo{author}{\bibfnamefont{L.}~\bibnamefont{Contessi}},
  \bibinfo{author}{\bibfnamefont{D.}~\bibnamefont{Gazit}},
  \bibinfo{author}{\bibfnamefont{F.}~\bibnamefont{Pederiva}}, \bibnamefont{and}
  \bibinfo{author}{\bibfnamefont{U.}~\bibnamefont{van Kolck}},
  \bibinfo{journal}{Phys. Rev. Lett.} \textbf{\bibinfo{volume}{114}},
  \bibinfo{pages}{052501} (\bibinfo{year}{2015}), \eprint{1311.4966}.

\bibitem[{\citenamefont{Drischler et~al.}(2021)\citenamefont{Drischler, Haxton,
  McElvain, Mereghetti, Nicholson, Vranas, and
  Walker-Loud}}]{Drischler:2019xuo}
\bibinfo{author}{\bibfnamefont{C.}~\bibnamefont{Drischler}},
  \bibinfo{author}{\bibfnamefont{W.}~\bibnamefont{Haxton}},
  \bibinfo{author}{\bibfnamefont{K.}~\bibnamefont{McElvain}},
  \bibinfo{author}{\bibfnamefont{E.}~\bibnamefont{Mereghetti}},
  \bibinfo{author}{\bibfnamefont{A.}~\bibnamefont{Nicholson}},
  \bibinfo{author}{\bibfnamefont{P.}~\bibnamefont{Vranas}}, \bibnamefont{and}
  \bibinfo{author}{\bibfnamefont{A.}~\bibnamefont{Walker-Loud}},
  \bibinfo{journal}{Prog. Part. Nucl. Phys.} \textbf{\bibinfo{volume}{121}},
  \bibinfo{pages}{103888} (\bibinfo{year}{2021}), \eprint{1910.07961}.

\bibitem[{\citenamefont{Aoki and Doi}(2020)}]{Aoki:2020bew}
\bibinfo{author}{\bibfnamefont{S.}~\bibnamefont{Aoki}} \bibnamefont{and}
  \bibinfo{author}{\bibfnamefont{T.}~\bibnamefont{Doi}},
  \bibinfo{journal}{Front. in Phys.} \textbf{\bibinfo{volume}{8}},
  \bibinfo{pages}{307} (\bibinfo{year}{2020}), \eprint{2003.10730}.

\bibitem[{\citenamefont{Illa et~al.}(2021)}]{NPLQCD:2020lxg}
\bibinfo{author}{\bibfnamefont{M.}~\bibnamefont{Illa}} \bibnamefont{et~al.}
  (\bibinfo{collaboration}{NPLQCD}), \bibinfo{journal}{Phys. Rev. D}
  \textbf{\bibinfo{volume}{103}}, \bibinfo{pages}{054508}
  (\bibinfo{year}{2021}), \eprint{2009.12357}.

\bibitem[{\citenamefont{McIlroy et~al.}(2018)\citenamefont{McIlroy, Barbieri,
  Inoue, Doi, and Hatsuda}}]{McIlroy:2017ssf}
\bibinfo{author}{\bibfnamefont{C.}~\bibnamefont{McIlroy}},
  \bibinfo{author}{\bibfnamefont{C.}~\bibnamefont{Barbieri}},
  \bibinfo{author}{\bibfnamefont{T.}~\bibnamefont{Inoue}},
  \bibinfo{author}{\bibfnamefont{T.}~\bibnamefont{Doi}}, \bibnamefont{and}
  \bibinfo{author}{\bibfnamefont{T.}~\bibnamefont{Hatsuda}},
  \bibinfo{journal}{Phys. Rev. C} \textbf{\bibinfo{volume}{97}},
  \bibinfo{pages}{021303} (\bibinfo{year}{2018}), \eprint{1701.02607}.

\bibitem[{\citenamefont{Weinberg}(1990)}]{Weinberg:1990rz}
\bibinfo{author}{\bibfnamefont{S.}~\bibnamefont{Weinberg}},
  \bibinfo{journal}{Phys. Lett. B} \textbf{\bibinfo{volume}{251}},
  \bibinfo{pages}{288} (\bibinfo{year}{1990}).

\bibitem[{\citenamefont{Weinberg}(1991)}]{Weinberg:1991um}
\bibinfo{author}{\bibfnamefont{S.}~\bibnamefont{Weinberg}},
  \bibinfo{journal}{Nucl. Phys. B} \textbf{\bibinfo{volume}{363}},
  \bibinfo{pages}{3} (\bibinfo{year}{1991}).

\bibitem[{\citenamefont{Epelbaum et~al.}(2015)\citenamefont{Epelbaum, Krebs,
  and Mei\ss{}ner}}]{Epelbaum:2014sza}
\bibinfo{author}{\bibfnamefont{E.}~\bibnamefont{Epelbaum}},
  \bibinfo{author}{\bibfnamefont{H.}~\bibnamefont{Krebs}}, \bibnamefont{and}
  \bibinfo{author}{\bibfnamefont{U.~G.} \bibnamefont{Mei\ss{}ner}},
  \bibinfo{journal}{Phys. Rev. Lett.} \textbf{\bibinfo{volume}{115}},
  \bibinfo{pages}{122301} (\bibinfo{year}{2015}), \eprint{1412.4623}.

\bibitem[{\citenamefont{Reinert et~al.}(2018)\citenamefont{Reinert, Krebs, and
  Epelbaum}}]{Reinert:2017usi}
\bibinfo{author}{\bibfnamefont{P.}~\bibnamefont{Reinert}},
  \bibinfo{author}{\bibfnamefont{H.}~\bibnamefont{Krebs}}, \bibnamefont{and}
  \bibinfo{author}{\bibfnamefont{E.}~\bibnamefont{Epelbaum}},
  \bibinfo{journal}{Eur. Phys. J. A} \textbf{\bibinfo{volume}{54}},
  \bibinfo{pages}{86} (\bibinfo{year}{2018}), \eprint{1711.08821}.

\bibitem[{\citenamefont{Entem et~al.}(2017)\citenamefont{Entem, Machleidt, and
  Nosyk}}]{Entem:2017gor}
\bibinfo{author}{\bibfnamefont{D.~R.} \bibnamefont{Entem}},
  \bibinfo{author}{\bibfnamefont{R.}~\bibnamefont{Machleidt}},
  \bibnamefont{and} \bibinfo{author}{\bibfnamefont{Y.}~\bibnamefont{Nosyk}},
  \bibinfo{journal}{Phys. Rev. C} \textbf{\bibinfo{volume}{96}},
  \bibinfo{pages}{024004} (\bibinfo{year}{2017}), \eprint{1703.05454}.

\bibitem[{\citenamefont{Epelbaum et~al.}(2009)\citenamefont{Epelbaum, Hammer,
  and Meissner}}]{Epelbaum:2008ga}
\bibinfo{author}{\bibfnamefont{E.}~\bibnamefont{Epelbaum}},
  \bibinfo{author}{\bibfnamefont{H.-W.} \bibnamefont{Hammer}},
  \bibnamefont{and} \bibinfo{author}{\bibfnamefont{U.-G.}
  \bibnamefont{Meissner}}, \bibinfo{journal}{Rev. Mod. Phys.}
  \textbf{\bibinfo{volume}{81}}, \bibinfo{pages}{1773} (\bibinfo{year}{2009}),
  \eprint{0811.1338}.

\bibitem[{\citenamefont{Machleidt and Entem}(2011)}]{Machleidt:2011zz}
\bibinfo{author}{\bibfnamefont{R.}~\bibnamefont{Machleidt}} \bibnamefont{and}
  \bibinfo{author}{\bibfnamefont{D.~R.} \bibnamefont{Entem}},
  \bibinfo{journal}{Phys. Rept.} \textbf{\bibinfo{volume}{503}},
  \bibinfo{pages}{1} (\bibinfo{year}{2011}), \eprint{1105.2919}.

\bibitem[{\citenamefont{Hammer et~al.}(2020)\citenamefont{Hammer, K\"onig, and
  van Kolck}}]{Hammer:2019poc}
\bibinfo{author}{\bibfnamefont{H.~W.} \bibnamefont{Hammer}},
  \bibinfo{author}{\bibfnamefont{S.}~\bibnamefont{K\"onig}}, \bibnamefont{and}
  \bibinfo{author}{\bibfnamefont{U.}~\bibnamefont{van Kolck}},
  \bibinfo{journal}{Rev. Mod. Phys.} \textbf{\bibinfo{volume}{92}},
  \bibinfo{pages}{025004} (\bibinfo{year}{2020}), \eprint{1906.12122}.

\bibitem[{\citenamefont{Tews et~al.}(2020)\citenamefont{Tews, Davoudi,
  Ekstr\"om, Holt, and Lynn}}]{Tews:2020hgp}
\bibinfo{author}{\bibfnamefont{I.}~\bibnamefont{Tews}},
  \bibinfo{author}{\bibfnamefont{Z.}~\bibnamefont{Davoudi}},
  \bibinfo{author}{\bibfnamefont{A.}~\bibnamefont{Ekstr\"om}},
  \bibinfo{author}{\bibfnamefont{J.~D.} \bibnamefont{Holt}}, \bibnamefont{and}
  \bibinfo{author}{\bibfnamefont{J.~E.} \bibnamefont{Lynn}},
  \bibinfo{journal}{J. Phys. G} \textbf{\bibinfo{volume}{47}},
  \bibinfo{pages}{103001} (\bibinfo{year}{2020}), \eprint{2001.03334}.

\bibitem[{\citenamefont{Ren et~al.}(2018)\citenamefont{Ren, Li, Geng, Long,
  Ring, and Meng}}]{Ren:2016jna}
\bibinfo{author}{\bibfnamefont{X.-L.} \bibnamefont{Ren}},
  \bibinfo{author}{\bibfnamefont{K.-W.} \bibnamefont{Li}},
  \bibinfo{author}{\bibfnamefont{L.-S.} \bibnamefont{Geng}},
  \bibinfo{author}{\bibfnamefont{B.-W.} \bibnamefont{Long}},
  \bibinfo{author}{\bibfnamefont{P.}~\bibnamefont{Ring}}, \bibnamefont{and}
  \bibinfo{author}{\bibfnamefont{J.}~\bibnamefont{Meng}},
  \bibinfo{journal}{Chin. Phys. C} \textbf{\bibinfo{volume}{42}},
  \bibinfo{pages}{014103} (\bibinfo{year}{2018}), \eprint{1611.08475}.

\bibitem[{\citenamefont{Xiao et~al.}(2019)\citenamefont{Xiao, Geng, and
  Ren}}]{Xiao:2018jot}
\bibinfo{author}{\bibfnamefont{Y.}~\bibnamefont{Xiao}},
  \bibinfo{author}{\bibfnamefont{L.-S.} \bibnamefont{Geng}}, \bibnamefont{and}
  \bibinfo{author}{\bibfnamefont{X.-L.} \bibnamefont{Ren}},
  \bibinfo{journal}{Phys. Rev. C} \textbf{\bibinfo{volume}{99}},
  \bibinfo{pages}{024004} (\bibinfo{year}{2019}), \eprint{1812.03005}.

\bibitem[{\citenamefont{S\'anchez~S\'anchez
  et~al.}(2018)\citenamefont{S\'anchez~S\'anchez, Yang, Long, and van
  Kolck}}]{SanchezSanchez:2017tws}
\bibinfo{author}{\bibfnamefont{M.}~\bibnamefont{S\'anchez~S\'anchez}},
  \bibinfo{author}{\bibfnamefont{C.~J.} \bibnamefont{Yang}},
  \bibinfo{author}{\bibfnamefont{B.}~\bibnamefont{Long}}, \bibnamefont{and}
  \bibinfo{author}{\bibfnamefont{U.}~\bibnamefont{van Kolck}},
  \bibinfo{journal}{Phys. Rev. C} \textbf{\bibinfo{volume}{97}},
  \bibinfo{pages}{024001} (\bibinfo{year}{2018}), \eprint{1704.08524}.

\bibitem[{\citenamefont{Ren et~al.}(2021)\citenamefont{Ren, Wang, Li, Geng, and
  Meng}}]{Ren:2017yvw}
\bibinfo{author}{\bibfnamefont{X.-L.} \bibnamefont{Ren}},
  \bibinfo{author}{\bibfnamefont{C.-X.} \bibnamefont{Wang}},
  \bibinfo{author}{\bibfnamefont{K.-W.} \bibnamefont{Li}},
  \bibinfo{author}{\bibfnamefont{L.-S.} \bibnamefont{Geng}}, \bibnamefont{and}
  \bibinfo{author}{\bibfnamefont{J.}~\bibnamefont{Meng}},
  \bibinfo{journal}{Chin. Phys. Lett.} \textbf{\bibinfo{volume}{38}},
  \bibinfo{pages}{062101} (\bibinfo{year}{2021}), \eprint{1712.10083}.

\bibitem[{\citenamefont{Bai et~al.}(2020)\citenamefont{Bai, Wang, Xiao, and
  Geng}}]{Bai:2020yml}
\bibinfo{author}{\bibfnamefont{Q.-Q.} \bibnamefont{Bai}},
  \bibinfo{author}{\bibfnamefont{C.-X.} \bibnamefont{Wang}},
  \bibinfo{author}{\bibfnamefont{Y.}~\bibnamefont{Xiao}}, \bibnamefont{and}
  \bibinfo{author}{\bibfnamefont{L.-S.} \bibnamefont{Geng}},
  \bibinfo{journal}{Phys. Lett.} \textbf{\bibinfo{volume}{B}},
  \bibinfo{pages}{135745} (\bibinfo{year}{2020}), \eprint{2007.01638}.

\bibitem[{\citenamefont{Bai et~al.}(2021)\citenamefont{Bai, Wang, Xiao, and
  Geng}}]{Bai:2021uim}
\bibinfo{author}{\bibfnamefont{Q.-Q.} \bibnamefont{Bai}},
  \bibinfo{author}{\bibfnamefont{C.-X.} \bibnamefont{Wang}},
  \bibinfo{author}{\bibfnamefont{Y.}~\bibnamefont{Xiao}}, \bibnamefont{and}
  \bibinfo{author}{\bibfnamefont{L.-S.} \bibnamefont{Geng}}
  (\bibinfo{year}{2021}), \eprint{2105.06113}.

\bibitem[{\citenamefont{Li et~al.}(2016)\citenamefont{Li, Ren, Geng, and
  Long}}]{Li:2016paq}
\bibinfo{author}{\bibfnamefont{K.-W.} \bibnamefont{Li}},
  \bibinfo{author}{\bibfnamefont{X.-L.} \bibnamefont{Ren}},
  \bibinfo{author}{\bibfnamefont{L.-S.} \bibnamefont{Geng}}, \bibnamefont{and}
  \bibinfo{author}{\bibfnamefont{B.}~\bibnamefont{Long}},
  \bibinfo{journal}{Phys. Rev. D} \textbf{\bibinfo{volume}{94}},
  \bibinfo{pages}{014029} (\bibinfo{year}{2016}), \eprint{1603.07802}.

\bibitem[{\citenamefont{Li et~al.}(2018{\natexlab{a}})\citenamefont{Li, Ren,
  Geng, and Long}}]{Li:2016mln}
\bibinfo{author}{\bibfnamefont{K.-W.} \bibnamefont{Li}},
  \bibinfo{author}{\bibfnamefont{X.-L.} \bibnamefont{Ren}},
  \bibinfo{author}{\bibfnamefont{L.-S.} \bibnamefont{Geng}}, \bibnamefont{and}
  \bibinfo{author}{\bibfnamefont{B.-W.} \bibnamefont{Long}},
  \bibinfo{journal}{Chin. Phys. C} \textbf{\bibinfo{volume}{42}},
  \bibinfo{pages}{014105} (\bibinfo{year}{2018}{\natexlab{a}}),
  \eprint{1612.08482}.

\bibitem[{\citenamefont{Li et~al.}(2018{\natexlab{b}})\citenamefont{Li, Hyodo,
  and Geng}}]{Li:2018tbt}
\bibinfo{author}{\bibfnamefont{K.-W.} \bibnamefont{Li}},
  \bibinfo{author}{\bibfnamefont{T.}~\bibnamefont{Hyodo}}, \bibnamefont{and}
  \bibinfo{author}{\bibfnamefont{L.-S.} \bibnamefont{Geng}},
  \bibinfo{journal}{Phys. Rev. C} \textbf{\bibinfo{volume}{98}},
  \bibinfo{pages}{065203} (\bibinfo{year}{2018}{\natexlab{b}}),
  \eprint{1809.03199}.

\bibitem[{\citenamefont{Song et~al.}(2018)\citenamefont{Song, Li, and
  Geng}}]{Song:2018qqm}
\bibinfo{author}{\bibfnamefont{J.}~\bibnamefont{Song}},
  \bibinfo{author}{\bibfnamefont{K.-W.} \bibnamefont{Li}}, \bibnamefont{and}
  \bibinfo{author}{\bibfnamefont{L.-S.} \bibnamefont{Geng}},
  \bibinfo{journal}{Phys. Rev. C} \textbf{\bibinfo{volume}{97}},
  \bibinfo{pages}{065201} (\bibinfo{year}{2018}), \eprint{1802.04433}.

\bibitem[{\citenamefont{Liu et~al.}(2021)\citenamefont{Liu, Song, Li, and
  Geng}}]{Liu:2020uxi}
\bibinfo{author}{\bibfnamefont{Z.-W.} \bibnamefont{Liu}},
  \bibinfo{author}{\bibfnamefont{J.}~\bibnamefont{Song}},
  \bibinfo{author}{\bibfnamefont{K.-W.} \bibnamefont{Li}}, \bibnamefont{and}
  \bibinfo{author}{\bibfnamefont{L.-S.} \bibnamefont{Geng}},
  \bibinfo{journal}{Phys. Rev. C} \textbf{\bibinfo{volume}{103}},
  \bibinfo{pages}{025201} (\bibinfo{year}{2021}), \eprint{2011.05510}.

\bibitem[{\citenamefont{Song et~al.}(2021{\natexlab{a}})\citenamefont{Song,
  Liu, Li, and Geng}}]{Song:2021yab}
\bibinfo{author}{\bibfnamefont{J.}~\bibnamefont{Song}},
  \bibinfo{author}{\bibfnamefont{Z.-W.} \bibnamefont{Liu}},
  \bibinfo{author}{\bibfnamefont{K.-W.} \bibnamefont{Li}}, \bibnamefont{and}
  \bibinfo{author}{\bibfnamefont{L.-S.} \bibnamefont{Geng}}
  (\bibinfo{year}{2021}{\natexlab{a}}), \eprint{2107.04742}.

\bibitem[{\citenamefont{Song et~al.}(2020)\citenamefont{Song, Xiao, Liu, Wang,
  Li, and Geng}}]{Song:2020isu}
\bibinfo{author}{\bibfnamefont{J.}~\bibnamefont{Song}},
  \bibinfo{author}{\bibfnamefont{Y.}~\bibnamefont{Xiao}},
  \bibinfo{author}{\bibfnamefont{Z.-W.} \bibnamefont{Liu}},
  \bibinfo{author}{\bibfnamefont{C.-X.} \bibnamefont{Wang}},
  \bibinfo{author}{\bibfnamefont{K.-W.} \bibnamefont{Li}}, \bibnamefont{and}
  \bibinfo{author}{\bibfnamefont{L.-S.} \bibnamefont{Geng}},
  \bibinfo{journal}{Phys. Rev. C} \textbf{\bibinfo{volume}{102}},
  \bibinfo{pages}{065208} (\bibinfo{year}{2020}), \eprint{2010.06916}.

\bibitem[{\citenamefont{Song et~al.}(2021{\natexlab{b}})\citenamefont{Song,
  Xiao, Liu, Li, and Geng}}]{Song:2021war}
\bibinfo{author}{\bibfnamefont{J.}~\bibnamefont{Song}},
  \bibinfo{author}{\bibfnamefont{Y.}~\bibnamefont{Xiao}},
  \bibinfo{author}{\bibfnamefont{Z.-W.} \bibnamefont{Liu}},
  \bibinfo{author}{\bibfnamefont{K.-W.} \bibnamefont{Li}}, \bibnamefont{and}
  \bibinfo{author}{\bibfnamefont{L.-S.} \bibnamefont{Geng}}
  (\bibinfo{year}{2021}{\natexlab{b}}), \eprint{2104.02380}.

\bibitem[{\citenamefont{Wang et~al.}(2021)\citenamefont{Wang, Geng, and
  Long}}]{Wang:2020myr}
\bibinfo{author}{\bibfnamefont{C.-X.} \bibnamefont{Wang}},
  \bibinfo{author}{\bibfnamefont{L.-S.} \bibnamefont{Geng}}, \bibnamefont{and}
  \bibinfo{author}{\bibfnamefont{B.}~\bibnamefont{Long}},
  \bibinfo{journal}{Chin. Phys. C} \textbf{\bibinfo{volume}{45}},
  \bibinfo{pages}{054101} (\bibinfo{year}{2021}), \eprint{2001.08483}.

\bibitem[{\citenamefont{Xiao et~al.}(2020)\citenamefont{Xiao, Wang, Lu, and
  Geng}}]{Xiao:2020ozd}
\bibinfo{author}{\bibfnamefont{Y.}~\bibnamefont{Xiao}},
  \bibinfo{author}{\bibfnamefont{C.-X.} \bibnamefont{Wang}},
  \bibinfo{author}{\bibfnamefont{J.-X.} \bibnamefont{Lu}}, \bibnamefont{and}
  \bibinfo{author}{\bibfnamefont{L.-S.} \bibnamefont{Geng}},
  \bibinfo{journal}{Phys. Rev. C} \textbf{\bibinfo{volume}{102}},
  \bibinfo{pages}{054001} (\bibinfo{year}{2020}), \eprint{2007.13675}.

\bibitem[{\citenamefont{Fettes et~al.}(2000)\citenamefont{Fettes, Meissner,
  Mojzis, and Steininger}}]{Fettes:2000gb}
\bibinfo{author}{\bibfnamefont{N.}~\bibnamefont{Fettes}},
  \bibinfo{author}{\bibfnamefont{U.-G.} \bibnamefont{Meissner}},
  \bibinfo{author}{\bibfnamefont{M.}~\bibnamefont{Mojzis}}, \bibnamefont{and}
  \bibinfo{author}{\bibfnamefont{S.}~\bibnamefont{Steininger}},
  \bibinfo{journal}{Annals Phys.} \textbf{\bibinfo{volume}{283}},
  \bibinfo{pages}{273} (\bibinfo{year}{2000}), \bibinfo{note}{[Erratum: Annals
  Phys. 288, 249--250 (2001)]}, \eprint{hep-ph/0001308}.

\bibitem[{\citenamefont{Blankenbecler and Sugar}(1966)}]{Blankenbecler:1965gx}
\bibinfo{author}{\bibfnamefont{R.}~\bibnamefont{Blankenbecler}}
  \bibnamefont{and} \bibinfo{author}{\bibfnamefont{R.}~\bibnamefont{Sugar}},
  \bibinfo{journal}{Phys. Rev.} \textbf{\bibinfo{volume}{142}},
  \bibinfo{pages}{1051} (\bibinfo{year}{1966}).

\bibitem[{\citenamefont{Epelbaum et~al.}(2000)\citenamefont{Epelbaum, Gloeckle,
  and Meissner}}]{Epelbaum:1999dj}
\bibinfo{author}{\bibfnamefont{E.}~\bibnamefont{Epelbaum}},
  \bibinfo{author}{\bibfnamefont{W.}~\bibnamefont{Gloeckle}}, \bibnamefont{and}
  \bibinfo{author}{\bibfnamefont{U.-G.} \bibnamefont{Meissner}},
  \bibinfo{journal}{Nucl. Phys. A} \textbf{\bibinfo{volume}{671}},
  \bibinfo{pages}{295} (\bibinfo{year}{2000}), \eprint{nucl-th/9910064}.

\bibitem[{\citenamefont{Tanabashi et~al.}(2018)}]{ParticleDataGroup:2018ovx}
\bibinfo{author}{\bibfnamefont{M.}~\bibnamefont{Tanabashi}}
  \bibnamefont{et~al.} (\bibinfo{collaboration}{Particle Data Group}),
  \bibinfo{journal}{Phys. Rev. D} \textbf{\bibinfo{volume}{98}},
  \bibinfo{pages}{030001} (\bibinfo{year}{2018}).

\bibitem[{\citenamefont{Chen et~al.}(2013)\citenamefont{Chen, Yao, and
  Zheng}}]{Chen:2012nx}
\bibinfo{author}{\bibfnamefont{Y.-H.} \bibnamefont{Chen}},
  \bibinfo{author}{\bibfnamefont{D.-L.} \bibnamefont{Yao}}, \bibnamefont{and}
  \bibinfo{author}{\bibfnamefont{H.~Q.} \bibnamefont{Zheng}},
  \bibinfo{journal}{Phys. Rev. D} \textbf{\bibinfo{volume}{87}},
  \bibinfo{pages}{054019} (\bibinfo{year}{2013}), \eprint{1212.1893}.

\bibitem[{\citenamefont{Gegelia and Japaridze}(1999)}]{Gegelia:1999gf}
\bibinfo{author}{\bibfnamefont{J.}~\bibnamefont{Gegelia}} \bibnamefont{and}
  \bibinfo{author}{\bibfnamefont{G.}~\bibnamefont{Japaridze}},
  \bibinfo{journal}{Phys. Rev. D} \textbf{\bibinfo{volume}{60}},
  \bibinfo{pages}{114038} (\bibinfo{year}{1999}), \eprint{hep-ph/9908377}.

\bibitem[{\citenamefont{Fuchs et~al.}(2003)\citenamefont{Fuchs, Gegelia,
  Japaridze, and Scherer}}]{Fuchs:2003qc}
\bibinfo{author}{\bibfnamefont{T.}~\bibnamefont{Fuchs}},
  \bibinfo{author}{\bibfnamefont{J.}~\bibnamefont{Gegelia}},
  \bibinfo{author}{\bibfnamefont{G.}~\bibnamefont{Japaridze}},
  \bibnamefont{and} \bibinfo{author}{\bibfnamefont{S.}~\bibnamefont{Scherer}},
  \bibinfo{journal}{Phys. Rev. D} \textbf{\bibinfo{volume}{68}},
  \bibinfo{pages}{056005} (\bibinfo{year}{2003}), \eprint{hep-ph/0302117}.

\bibitem[{\citenamefont{Geng}(2013)}]{Geng:2013xn}
\bibinfo{author}{\bibfnamefont{L.}~\bibnamefont{Geng}},
  \bibinfo{journal}{Front. Phys. (Beijing)} \textbf{\bibinfo{volume}{8}},
  \bibinfo{pages}{328} (\bibinfo{year}{2013}), \eprint{1301.6815}.

\bibitem[{\citenamefont{Lu et~al.}(2019)\citenamefont{Lu, Geng, Ren, and
  Du}}]{Lu:2018zof}
\bibinfo{author}{\bibfnamefont{J.-X.} \bibnamefont{Lu}},
  \bibinfo{author}{\bibfnamefont{L.-S.} \bibnamefont{Geng}},
  \bibinfo{author}{\bibfnamefont{X.-L.} \bibnamefont{Ren}}, \bibnamefont{and}
  \bibinfo{author}{\bibfnamefont{M.-L.} \bibnamefont{Du}},
  \bibinfo{journal}{Phys. Rev. D} \textbf{\bibinfo{volume}{99}},
  \bibinfo{pages}{054024} (\bibinfo{year}{2019}), \eprint{1812.03799}.

\bibitem[{\citenamefont{Shtabovenko et~al.}(2020)\citenamefont{Shtabovenko,
  Mertig, and Orellana}}]{Shtabovenko:2020gxv}
\bibinfo{author}{\bibfnamefont{V.}~\bibnamefont{Shtabovenko}},
  \bibinfo{author}{\bibfnamefont{R.}~\bibnamefont{Mertig}}, \bibnamefont{and}
  \bibinfo{author}{\bibfnamefont{F.}~\bibnamefont{Orellana}},
  \bibinfo{journal}{Comput. Phys. Commun.} \textbf{\bibinfo{volume}{256}},
  \bibinfo{pages}{107478} (\bibinfo{year}{2020}), \eprint{2001.04407}.

\bibitem[{\citenamefont{Shtabovenko et~al.}(2016)\citenamefont{Shtabovenko,
  Mertig, and Orellana}}]{Shtabovenko:2016sxi}
\bibinfo{author}{\bibfnamefont{V.}~\bibnamefont{Shtabovenko}},
  \bibinfo{author}{\bibfnamefont{R.}~\bibnamefont{Mertig}}, \bibnamefont{and}
  \bibinfo{author}{\bibfnamefont{F.}~\bibnamefont{Orellana}},
  \bibinfo{journal}{Comput. Phys. Commun.} \textbf{\bibinfo{volume}{207}},
  \bibinfo{pages}{432} (\bibinfo{year}{2016}), \eprint{1601.01167}.

\bibitem[{\citenamefont{Mertig et~al.}(1991)\citenamefont{Mertig, Bohm, and
  Denner}}]{Mertig:1990an}
\bibinfo{author}{\bibfnamefont{R.}~\bibnamefont{Mertig}},
  \bibinfo{author}{\bibfnamefont{M.}~\bibnamefont{Bohm}}, \bibnamefont{and}
  \bibinfo{author}{\bibfnamefont{A.}~\bibnamefont{Denner}},
  \bibinfo{journal}{Comput. Phys. Commun.} \textbf{\bibinfo{volume}{64}},
  \bibinfo{pages}{345} (\bibinfo{year}{1991}).

\bibitem[{\citenamefont{van Hameren et~al.}(2009)\citenamefont{van Hameren,
  Papadopoulos, and Pittau}}]{vanHameren:2009dr}
\bibinfo{author}{\bibfnamefont{A.}~\bibnamefont{van Hameren}},
  \bibinfo{author}{\bibfnamefont{C.~G.} \bibnamefont{Papadopoulos}},
  \bibnamefont{and} \bibinfo{author}{\bibfnamefont{R.}~\bibnamefont{Pittau}},
  \bibinfo{journal}{JHEP} \textbf{\bibinfo{volume}{09}}, \bibinfo{pages}{106}
  (\bibinfo{year}{2009}), \eprint{0903.4665}.

\bibitem[{\citenamefont{van Hameren}(2011)}]{vanHameren:2010cp}
\bibinfo{author}{\bibfnamefont{A.}~\bibnamefont{van Hameren}},
  \bibinfo{journal}{Comput. Phys. Commun.} \textbf{\bibinfo{volume}{182}},
  \bibinfo{pages}{2427} (\bibinfo{year}{2011}), \eprint{1007.4716}.

\bibitem[{\citenamefont{Woloshyn and Jackson}(1973)}]{Woloshyn:1973mce}
\bibinfo{author}{\bibfnamefont{R.~M.} \bibnamefont{Woloshyn}} \bibnamefont{and}
  \bibinfo{author}{\bibfnamefont{A.~D.} \bibnamefont{Jackson}},
  \bibinfo{journal}{Nucl. Phys. B} \textbf{\bibinfo{volume}{64}},
  \bibinfo{pages}{269} (\bibinfo{year}{1973}).

\bibitem[{\citenamefont{Epelbaum
  et~al.}(2004{\natexlab{a}})\citenamefont{Epelbaum, Gloeckle, and
  Meissner}}]{Epelbaum:2003xx}
\bibinfo{author}{\bibfnamefont{E.}~\bibnamefont{Epelbaum}},
  \bibinfo{author}{\bibfnamefont{W.}~\bibnamefont{Gloeckle}}, \bibnamefont{and}
  \bibinfo{author}{\bibfnamefont{U.-G.} \bibnamefont{Meissner}},
  \bibinfo{journal}{Eur. Phys. J. A} \textbf{\bibinfo{volume}{19}},
  \bibinfo{pages}{401} (\bibinfo{year}{2004}{\natexlab{a}}),
  \eprint{nucl-th/0308010}.

\bibitem[{\citenamefont{Entem and
  Machleidt}(2002{\natexlab{a}})}]{Entem:2001cg}
\bibinfo{author}{\bibfnamefont{D.~R.} \bibnamefont{Entem}} \bibnamefont{and}
  \bibinfo{author}{\bibfnamefont{R.}~\bibnamefont{Machleidt}},
  \bibinfo{journal}{Phys. Lett. B} \textbf{\bibinfo{volume}{524}},
  \bibinfo{pages}{93} (\bibinfo{year}{2002}{\natexlab{a}}),
  \eprint{nucl-th/0108057}.

\bibitem[{\citenamefont{Epelbaum et~al.}(2005)\citenamefont{Epelbaum, Glockle,
  and Meissner}}]{Epelbaum:2004fk}
\bibinfo{author}{\bibfnamefont{E.}~\bibnamefont{Epelbaum}},
  \bibinfo{author}{\bibfnamefont{W.}~\bibnamefont{Glockle}}, \bibnamefont{and}
  \bibinfo{author}{\bibfnamefont{U.-G.} \bibnamefont{Meissner}},
  \bibinfo{journal}{Nucl. Phys. A} \textbf{\bibinfo{volume}{747}},
  \bibinfo{pages}{362} (\bibinfo{year}{2005}), \eprint{nucl-th/0405048}.

\bibitem[{\citenamefont{Stapp et~al.}(1957)\citenamefont{Stapp, Ypsilantis, and
  Metropolis}}]{Stapp:1956mz}
\bibinfo{author}{\bibfnamefont{H.~P.} \bibnamefont{Stapp}},
  \bibinfo{author}{\bibfnamefont{T.~J.} \bibnamefont{Ypsilantis}},
  \bibnamefont{and}
  \bibinfo{author}{\bibfnamefont{N.}~\bibnamefont{Metropolis}},
  \bibinfo{journal}{Phys. Rev.} \textbf{\bibinfo{volume}{105}},
  \bibinfo{pages}{302} (\bibinfo{year}{1957}).

\bibitem[{\citenamefont{Stoks et~al.}(1993)\citenamefont{Stoks, Klomp,
  Rentmeester, and de~Swart}}]{Stoks:1993tb}
\bibinfo{author}{\bibfnamefont{V.~G.~J.} \bibnamefont{Stoks}},
  \bibinfo{author}{\bibfnamefont{R.~A.~M.} \bibnamefont{Klomp}},
  \bibinfo{author}{\bibfnamefont{M.~C.~M.} \bibnamefont{Rentmeester}},
  \bibnamefont{and} \bibinfo{author}{\bibfnamefont{J.~J.}
  \bibnamefont{de~Swart}}, \bibinfo{journal}{Phys. Rev. C}
  \textbf{\bibinfo{volume}{48}}, \bibinfo{pages}{792} (\bibinfo{year}{1993}).

\bibitem[{\citenamefont{Kaiser et~al.}(1997)\citenamefont{Kaiser, Brockmann,
  and Weise}}]{Kaiser:1997mw}
\bibinfo{author}{\bibfnamefont{N.}~\bibnamefont{Kaiser}},
  \bibinfo{author}{\bibfnamefont{R.}~\bibnamefont{Brockmann}},
  \bibnamefont{and} \bibinfo{author}{\bibfnamefont{W.}~\bibnamefont{Weise}},
  \bibinfo{journal}{Nucl. Phys. A} \textbf{\bibinfo{volume}{625}},
  \bibinfo{pages}{758} (\bibinfo{year}{1997}), \eprint{nucl-th/9706045}.

\bibitem[{\citenamefont{Epelbaum
  et~al.}(2004{\natexlab{b}})\citenamefont{Epelbaum, Gloeckle, and
  Meissner}}]{Epelbaum:2003gr}
\bibinfo{author}{\bibfnamefont{E.}~\bibnamefont{Epelbaum}},
  \bibinfo{author}{\bibfnamefont{W.}~\bibnamefont{Gloeckle}}, \bibnamefont{and}
  \bibinfo{author}{\bibfnamefont{U.-G.} \bibnamefont{Meissner}},
  \bibinfo{journal}{Eur. Phys. J. A} \textbf{\bibinfo{volume}{19}},
  \bibinfo{pages}{125} (\bibinfo{year}{2004}{\natexlab{b}}),
  \eprint{nucl-th/0304037}.

\bibitem[{\citenamefont{Entem and
  Machleidt}(2002{\natexlab{b}})}]{Entem:2002sf}
\bibinfo{author}{\bibfnamefont{D.~R.} \bibnamefont{Entem}} \bibnamefont{and}
  \bibinfo{author}{\bibfnamefont{R.}~\bibnamefont{Machleidt}},
  \bibinfo{journal}{Phys. Rev. C} \textbf{\bibinfo{volume}{66}},
  \bibinfo{pages}{014002} (\bibinfo{year}{2002}{\natexlab{b}}),
  \eprint{nucl-th/0202039}.

\bibitem[{\citenamefont{Entem and Machleidt}(2003)}]{Entem:2003ft}
\bibinfo{author}{\bibfnamefont{D.~R.} \bibnamefont{Entem}} \bibnamefont{and}
  \bibinfo{author}{\bibfnamefont{R.}~\bibnamefont{Machleidt}},
  \bibinfo{journal}{Phys. Rev. C} \textbf{\bibinfo{volume}{68}},
  \bibinfo{pages}{041001} (\bibinfo{year}{2003}), \eprint{nucl-th/0304018}.

\bibitem[{\citenamefont{Lu et~al.}(2021)\citenamefont{Lu, Wang, Xiao, Geng,
  Meng, and Ring}}]{Lu:2021gsb}
\bibinfo{author}{\bibfnamefont{J.-X.} \bibnamefont{Lu}},
  \bibinfo{author}{\bibfnamefont{C.-X.} \bibnamefont{Wang}},
  \bibinfo{author}{\bibfnamefont{Y.}~\bibnamefont{Xiao}},
  \bibinfo{author}{\bibfnamefont{L.-S.} \bibnamefont{Geng}},
  \bibinfo{author}{\bibfnamefont{J.}~\bibnamefont{Meng}}, \bibnamefont{and}
  \bibinfo{author}{\bibfnamefont{P.}~\bibnamefont{Ring}}
  (\bibinfo{year}{2021}), \eprint{2111.07766}.

\end{thebibliography}

\end{document}